%% file: main.tex
\pdfoutput=1
\documentclass[11pt]{article}

\usepackage[utf8]{inputenc}
\usepackage[T1]{fontenc}
\usepackage{lmodern}
\usepackage[final]{microtype}

\usepackage{array,tabularx}
\newcolumntype{Y}{>{\raggedright\arraybackslash}X}

\usepackage{adjustbox}

\usepackage{tikz}
\usetikzlibrary{arrows.meta,positioning,calc}

\usepackage[margin=1in]{geometry}
\usepackage{amsmath,amssymb,amsfonts,mathtools}
\numberwithin{equation}{section}
\numberwithin{figure}{section}
\numberwithin{table}{section}
\allowdisplaybreaks

\usepackage{amsthm}
\theoremstyle{plain}

\theoremstyle{definition}

\theoremstyle{remark}

\newcommand{\mpl}{M_{\rm pl}}
\newcommand{\dd}{\mathrm{d}}
\newcommand{\ee}{\mathrm{e}}

\usepackage{siunitx}
\usepackage{graphicx}
\graphicspath{{figures/}}
\usepackage{booktabs}

\usepackage{subcaption}
\usepackage{float}
\usepackage{longtable}
\usepackage[nottoc]{tocbibind} %

\usepackage{mdframed}

\usepackage{enumitem}
\setlist[itemize]{leftmargin=1.2em,itemsep=2pt,topsep=2pt}
\setlist[enumerate]{leftmargin=1.2em,itemsep=2pt,topsep=2pt}

\usepackage[section]{placeins}

\usepackage[numbers,sort&compress]{natbib}
\providecommand{\parencite}[1]{\citep{#1}}

\providecommand{\textcite}[1]{\citet{#1}}

\usepackage[dvipsnames]{xcolor}

\PassOptionsToPackage{hyphens}{url}

\usepackage[unicode=true,pdfencoding=auto,psdextra]{hyperref}
\hypersetup{
  colorlinks=true,
  linkcolor=blue,
  citecolor=blue,
  urlcolor=blue,
  pdftitle   = {Quantum–Kinetic Dark Energy (QKDE): An effective DE framework with GR metric intact and time-dependent scalar kinetic normalization},
  pdfauthor  = {Daniel Brown},
  pdfsubject = {Cosmology, Dark Energy, EFT},
  pdfkeywords= {cosmology, dark energy, EFT, scalar field, QKDE}
}

\usepackage{bookmark}



\usepackage[capitalize,noabbrev]{cleveref}
\crefname{equation}{Eq.}{Eqs.}
\Crefname{equation}{Equation}{Equations}
\crefname{figure}{Fig.}{Figs.}
\Crefname{figure}{Figure}{Figures}
\crefname{section}{Sec.}{Secs.}
\Crefname{section}{Section}{Sections}

\begin{document}

\input{frontmatter/titlepage}
\input{frontmatter/abstract}
\input{frontmatter/keywords}
\clearpage
\input{frontmatter/notation}
\input{frontmatter/policy}

\clearpage
\FloatBarrier
\suppressfloats[t]
\tableofcontents
\clearpage
\begingroup
\let\clearpage\relax
\phantomsection
\listoffigures
\endgroup

\phantomsection
\listoftables

\clearpage
\FloatBarrier

\input{sections/section1/section1}
\input{sections/section2/section2}
\input{sections/section3/section3}
\input{sections/section4/section4}
\input{sections/section5/section5}
\input{sections/section6/section6} 
\input{sections/section7/section7}
\input{sections/section8/section8}
\input{sections/section9/section9}
\input{sections/section10/section10}
\input{sections/section11/section11}

\appendix
\input{sections/appendices/appendixA}
\input{sections/appendices/appendix1}
\input{sections/appendices/appendix2}
\input{sections/appendices/appendix3}
\input{sections/appendices/appendix4}
\input{sections/appendices/appendix5}
\input{sections/appendices/appendix6}
\input{sections/appendices/appendix7}
\input{sections/appendices/appendix8}
\input{sections/appendices/appendix9}
\input{sections/appendices/appendix10}
\input{sections/appendices/appendix11}

\bibliography{refs}

\end{document}

%% file: frontmatter/titlepage.tex
\begin{titlepage}
\thispagestyle{empty}

\hfill \texttt{\today}\\[-0.25em]
\hrule

\vspace{12mm}
\begin{center}
{\LARGE\bfseries Quantum--Kinetic Dark Energy (QKDE):\\[0.45em]
\Large An effective dark energy framework with a covariantly completed time-dependent scalar kinetic normalization}

\vspace{0.6em}
{\normalsize (Einstein--Hilbert sector unmodified; a covariant completion exists,\;
$c_T^{2}=1$, Planck mass constant;\;
$\alpha_{B}=\alpha_{M}=\alpha_{T}=\alpha_{H}=0$,\;
$\alpha_{K}>0$ in EFT--DE)}

\vspace{12mm}
{\large Daniel Brown}\\[0.5em]
{\normalsize \textit{University of Utah}}\\[0.25em]
{\normalsize Email: \href{mailto:u0448673@utah.edu}{u0448673@utah.edu}}\\[0.25em]
{\normalsize ORCID: \href{https://orcid.org/0009-0001-8082-9702}{0009-0001-8082-9702}}

\vspace{10mm}
\begin{minipage}{0.9\linewidth}\centering\footnotesize
\textit{Note.} Earlier drafts circulated under the working title
\emph{Scalar--Cost Dark Energy (SCDE)}.
The present formulation admits a covariant completion via a clock (St\"uckelberg) field, with the background analysis performed in unitary gauge $\chi=t$.
\end{minipage}
\end{center}

\vfill
\end{titlepage}

%% file: frontmatter/abstract.tex
\begin{abstract}
A minimal, effective dark-energy framework—Quantum--Kinetic Dark Energy (QKDE)—is developed
in which the scalar kinetic normalization carries a slow background time dependence
through a covariantly completed clock field \({\chi}\) such that \(K=K(\chi)>0\), while the Einstein--Hilbert metric sector remains unmodified. The effective action
\(
S=\int d^{4}x\,\sqrt{-g}\,[\tfrac12 M_{\rm pl}^{2}R + K(\chi)X - V(\phi)]
\)
admits a diffeomorphism-invariant completion; working in unitary gauge \(\chi=t\) reproduces the background equations employed numerically in this work.
Within the EFT--DE description this corresponds to
\(
\alpha_K = K\dot\phi^{2}/(H^{2}M_{\rm pl}^{2})>0
\)
with
\(
\alpha_B=\alpha_M=\alpha_T=\alpha_H=0
\),
so tensors are luminal and the Planck mass is constant. Within this effective framework, a closed first--order background system in e--fold time
is obtained; scalar perturbations propagate with \(c_s^2=1\), satisfy \(\Phi=\Psi\),
and source linear growth through the unmodified metric Einstein equation.
The scalar-field equation takes the form of an exchange equation with the clock sector, while the total energy--momentum tensor is covariantly conserved.
All observable signatures therefore enter solely through the expansion history \(H(a)\) and the induced growth \(D(a)\).

Two kinetic normalizations are treated in detail: (i) a curvature–motivated form \(K=1+\alpha R/M^{2}\), for which an iteration–free algebraic identity for \(K'/K\) is derived; and (ii) a phenomenological running \(K=1+K_{0}(1+z)^{p}\). A reproducible numerical pipeline is provided together with a Fisher setup based on exact variational (sensitivity) equations for distances, \(H(z)\), and \(f\sigma_{8}(z)\). Stability and admissibility reduce to \(K(\chi)>0\) and a nonvanishing algebraic denominator in the curvature case. The framework yields sharp, falsifiable null predictions on linear scales:
\(
\mu(a,k)=\Sigma(a,k)=1,\ 
\eta(a,k)=0,\ 
c_{T}^{2}=1
\);
any statistically significant deviation lies outside the effective QKDE baseline.
The framework is interpreted as an effective, unitary-gauge cosmological description arising from a covariantly completed theory, rather than as a manifestly covariant scalar--tensor model written directly in fixed time slicing.
\end{abstract}

%% file: frontmatter/keywords.tex

%% file: frontmatter/notation.tex
\section*{Notation and symbols}

\noindent Overdot \(\dot{(\,)}\) denotes \(d/dt\); prime \((\,)'\) denotes \(d/dN\) with \(N=\ln a\).
Units \(c=\hbar=1\); metric signature \((-,+,+,+)\).
The reduced Planck mass is \(M_{\rm pl}=(8\pi G)^{-1/2}\).
Spatial flatness (FRW) is assumed unless noted.
Throughout, the background analysis is performed in unitary gauge, corresponding to a covariant completion in which a clock/St\"uckelberg field \(\chi\) restores diffeomorphism invariance and satisfies \(\chi=t\) on the background.

\begingroup
\small
\setlength{\tabcolsep}{4pt}
\renewcommand{\arraystretch}{1.05}

\noindent
\begin{minipage}[t]{0.49\textwidth}
\centering
\captionsetup{type=table,skip=4pt}
\captionof{table}{Core symbols used throughout}
\begin{tabular}{@{}p{0.34\textwidth}p{0.61\textwidth}@{}}
\toprule
\textbf{Symbol} & \textbf{Meaning} \\
\midrule
\(a(t)\) & Scale factor \\
\(N\equiv \ln a\) & E\mbox{-}fold time \\
\(H\equiv \dot a/a\) & Hubble parameter \\
\(E\equiv H'/H\) & Logarithmic Hubble derivative \\
\(R\) & FRW Ricci scalar \(=6(2H^2+\dot H)=6H^2(2+E)\) \\
\(M_{\rm pl}\) & Reduced Planck mass \\
\(\phi(t)\) & Homogeneous scalar field \\
\(V(\phi)\) & Scalar potential \\
\(X\equiv -\tfrac12 g^{\mu\nu}\partial_\mu\phi\,\partial_\nu\phi\) & Kinetic invariant; FRW: \(X=\tfrac12\dot\phi^2\) \\
\(K(\chi)>0\) & Scalar kinetic normalization as a function of the clock field \(\chi\); in unitary gauge \(\chi=t\), this reduces to a time-dependent normalization \(K(t)\) \\
\(\rho_i,\,p_i\) & Energy density/pressure, \(i\in\{m,r,\phi\}\) \\
\(\Omega_i\equiv \rho_i/(3M_{\rm pl}^2 H^2)\) & Density parameters; \(\sum_i\Omega_i=1\) (flat) \\
\(w_\phi\equiv p_\phi/\rho_\phi\) & Scalar equation of state \\
\(q\equiv -1-H'/H\) & Deceleration parameter \\
\(\Phi,\Psi\) & Bardeen potentials (Newtonian gauge); here \(\Phi=\Psi\) \\
\(D(a)\) & Linear matter growth factor \\
\(f\equiv d\ln D/d\ln a\) & Linear growth rate \\
\(\chi(z)\) & Comoving distance \\
\(D_A,\,D_L\) & Angular\mbox{-}diameter and luminosity distances \\
\(D_V,\,F_{\rm AP}\) & BAO summary distances \\
\(r_d\) & Sound horizon at drag epoch \\
\(\eta\) & Conformal time \\
\(z=e^{-N}-1\) & Redshift–e\mbox{-}folds map \\
\bottomrule
\end{tabular}
\end{minipage}
\hfill
\begin{minipage}[t]{0.49\textwidth}
\centering
\captionsetup{type=table,skip=4pt}
\captionof{table}{EFT--DE and growth shorthand}
\begin{tabular}{@{}p{0.39\textwidth}p{0.56\textwidth}@{}}
\toprule
\textbf{Symbol} & \textbf{Meaning} \\
\midrule
\(\alpha_K\) & Kinetic EFT coeff.; here \(\alpha_K=K\,\dot\phi^{2}/(H^{2}M_{\rm pl}^{2})>0\) \\
\(\alpha_B,\,\alpha_M,\,\alpha_T,\,\alpha_H\) & Braiding, \(M_*\) running, tensor speed, beyond\mbox{-}Horndeski (all \(=0\)) \\
\(\mu(a,k)\) & Poisson modifier (linear); here \(\mu=1\) \\
\(\Sigma(a,k)\) & Lensing modifier (linear); here \(\Sigma=1\) \\
\(\eta(a,k)\equiv \Phi/\Psi-1\) & Gravitational slip; here \(\eta=0\) \\
\(\sigma_{8}\) & RMS matter fluct. in \(8\,h^{-1}\,\mathrm{Mpc}\) at \(z=0\) \\
\(f\sigma_{8}\) & RSD observable \(f(z)\,\sigma_{8,0}\,D(z)/D(0)\) \\
\(s\equiv \phi'\) & Scalar e\mbox{-}fold derivative (background state) \\
\(y_H\equiv \ln H\) & Log\mbox{-}Hubble variable (background state) \\
\bottomrule
\end{tabular}
\end{minipage}

\par\smallskip
\noindent\footnotesize\emph{Disambiguation.} This work reserves \(E\equiv H'/H\) (no use of \(E(z)\)). When the normalized Hubble rate is needed, it is written explicitly as \(H(z)/H_{0}\). The symbol \(\eta\) denotes conformal time; the gravitational slip uses \(\eta(a,k)\) and is contextually distinct.
All conservation statements refer to the total energy--momentum tensor of the covariantly completed theory; individual components may exchange energy in unitary gauge without violating diffeomorphism invariance.

\endgroup

%% file: frontmatter/policy.tex
\clearpage
\paragraph*{Data and code availability}
All equations and algorithms required to reproduce the results are stated explicitly in the text (notably Eqs.~\eqref{eq:background-system}, \eqref{eq:growth-eq-recap}, and \eqref{eq:forecast-set}), with numerical tolerances, step control, and diagnostics summarized in Sec.~\ref{sec:numerics} and Table~\ref{tab:numerics_tol_diag}. A compact reference implementation may be built directly on the state vector $(\phi,\phi',H)$ using the identities in Secs.~\ref{sec:efold}--\ref{sec:K}.

\paragraph*{Competing interests}
The author declares no competing interests.

\paragraph*{Author contribution}
Single-author manuscript. Conceptualization, formal analysis, derivations, numerical implementation, validation, and writing were performed by the author.

\paragraph*{Acknowledgments}
It is a pleasure to acknowledge insightful comments from colleagues that sharpened the presentation and suggested several clarifications in the EFT--DE mapping and in the growth observable summaries. Constructive reader feedback on earlier iterations circulating under the working title \emph{Scalar--Cost Dark Energy (SCDE)} helped to streamline the background system and strengthen the stability discussion. Informal discussions within the cosmology community, and the broader literature cited throughout, provided valuable perspective on linear observables and effective descriptions consistent with GW170817. Any remaining errors or omissions are my own sole responsibility. No external funding was received for this work.

%% file: sections/section1/section1.tex
\section{Introduction}
\label{sec:intro}

The observed late--time acceleration of the Universe admits two broad explanatory directions: modifications of gravity on cosmological scales or additional cosmic components with negative pressure. Bounds from the binary--neutron--star merger GW170817 and its electromagnetic counterpart enforce a luminal tensor speed and an effectively constant Planck mass at low redshift, disfavoring large sectors of modified--gravity space \parencite{Abbott2017GW170817,Sakstein2017GW}. In this context it is natural to isolate dark--energy constructions that preserve the Einstein--Hilbert metric sector and alter only the matter content.

This work develops a minimal, effective dark--energy framework---\emph{Quantum--Kinetic Dark Energy} (QKDE)---in which only the normalization of the scalar kinetic term varies slowly in time while the metric sector remains that of General Relativity (GR). The defining effective action is
\begin{equation}
\label{eq:intro-action}
S=\int \dd^{4}x\,\sqrt{-g}\Big[\frac12 \mpl^{2} R + K(\chi_{\rm c})\,X - V(\phi)\Big],
\qquad
X\equiv -\frac12 g^{\mu\nu}\partial_\mu\phi\,\partial_\nu\phi,
\end{equation}
with minimally coupled matter and radiation, where \(\chi_{\rm c}\) is a clock (St\"uckelberg) scalar field whose background value defines cosmic time. Working in unitary gauge \(\chi_{\rm c}=t\), the action reduces to the effective background form with a time--dependent kinetic normalization \(K(t)\equiv K(\chi_{\rm c})|_{\chi_{\rm c}=t}\) used throughout the numerical analysis.

Written in this way, the theory admits a covariant completion with a conserved total energy--momentum tensor; the explicit time dependence arises only after gauge fixing. The effective description is therefore interpreted as a background--level unitary--gauge realization of an underlying diffeomorphism--invariant theory, rather than as a manifestly covariant scalar--tensor model.

The function \(K(t)\equiv K(\chi_{\rm c})|_{\chi_{\rm c}=t}\) is a \emph{background} (unitary--gauge) coefficient: it depends on cosmic time only and introduces no new perturbative operators in the scalar sector beyond a time--dependent normalization. In the Effective Field Theory of Dark Energy (EFT--DE) language \parencite{Gubitosi2013,Bloomfield2013,BelliniSawicki2014}, the Bellini--Sawicki set satisfies
\begin{equation}
\alpha_B=\alpha_M=\alpha_T=0,\qquad
\alpha_K=\frac{K\,\dot\phi^{2}}{H^{2}\mpl^{2}}>0,
\label{eq:eft-map-intro}
\end{equation}
so the tensor speed is luminal and the Planck mass is constant, consistent with GW170817 constraints \parencite{Abbott2017GW170817,Sakstein2017GW}. Because \(P(X,\phi,\chi_{\rm c})=K(\chi_{\rm c})X-V(\phi)\) has \(P_{,XX}=0\), the scalar sound speed is \(c_s^{2}=1\) (and, with \(K>0\), no ghosts), excluding gradient instabilities and scale--dependent growth on linear/subhorizon scales \parencite{GarrigaMukhanov1999}.

\begin{mdframed}[linewidth=0.6pt]
\textbf{What is---and is not---modified.}\\[2pt]
\emph{Modified:} a single background function \(K(\chi_{\rm c})>0\), evaluated in unitary gauge as \(K(t)\), multiplying the scalar kinetic operator.\\
\emph{Not modified:} the metric sector (pure Einstein--Hilbert), the tensor speed (\(c_T^2=1\)), the Planck mass (constant), and linear GR relations (\(\mu=\Sigma=1,\ \eta=0\)).
\end{mdframed}

\paragraph{Motivation.}
Quantum field theory in curved spacetime (QFTCS) predicts curvature--suppressed operators such as \(R\,X\) and \(R_{\mu\nu}\partial^\mu\phi\,\partial^\nu\phi\), generated by loops and/or integrating out heavy fields \parencite{BirrellDavies1982,Parker1969,ParkerFulling1974,Bunch1980,Buchbinder1992,ParkerToms2009,BarvinskyVilkovisky1985,Donoghue1994}. On a homogeneous background these operators are degenerate with a time--dependent rescaling of \(X\). QKDE isolates precisely this \emph{background imprint} by promoting the scalar normalization to \(K(\chi_{\rm c})\) and subsequently fixing unitary gauge, while deliberately avoiding nonminimal couplings or braiding terms in the metric sector. Canonical quintessence is recovered when \(K\equiv 1\) \parencite{RatraPeebles1988,Caldwell1998}; general k--essence extends to \(P(X,\phi)\) with \(P_{,XX}\neq0\) and admits \(c_s^2\neq1\) \parencite{ArmendarizPicon2001,GarrigaMukhanov1999}. QKDE occupies the conservative corner of EFT--DE with only \(\alpha_K\) nonzero.

Two structural features make the framework predictive. First, because only the scalar normalization runs, the background expansion \(H(a)\), the kinetic invariant \(X\), and the scalar equation of state \(w_\phi(a)\) follow from a closed first--order system in e--fold time \(N=\ln a\), derived within the effective background framework defined by \eqref{eq:intro-action} together with the Einstein equations, with total energy--momentum conservation enforced in the covariant completion
(Secs.~\ref{sec:covariant-core}--\ref{sec:efold}). Second, at linear order the Bardeen potentials satisfy \(\Phi=\Psi\) and the GR Poisson equation holds; the scalar mode is pressure supported with \(c_s^2=1\) and does not cluster on subhorizon scales (Sec.~\ref{sec:linear}). Consequently, all late--time signatures enter through the background: distances and growth respond solely to \(H(a)\) and its induced growth history \(D(a)\) \parencite{MaBertschinger1995,Planck2018,Linder2005,Eisenstein2005}.

\paragraph{Scope and contributions.}
The analysis is derived within a well--defined effective theoretical framework; every equation used later is either obtained from the effective background action in unitary gauge and the associated Einstein equations together with the implied exchange relations between the scalar and clock sector, or recalled with explicit cross--references. No phenomenological template is assumed without definition and derivation.
\begin{enumerate}
\item An effective definition of QKDE is stated in \eqref{eq:intro-action} together with minimal coupling and the positivity prior \(K>0\). Immediate implications are \(\alpha_T=\alpha_M=\alpha_B=0\) and a luminal tensor sector (Sec.~\ref{sec:eft}).
\item A closed background system in e--fold variables is assembled: the scalar exchange equation, the Raychaudhuri relation \(H'/H\), and exact expressions for \(X\), \(\rho_\phi\), and \(R/H^{2}\) (Sec.~\ref{sec:efold}). Well--posedness and admissibility are specified.
\item Two kinetic normalizations are analyzed that preserve the GR metric sector: a curvature--motivated form \(K=1+\alpha R/M^{2}\) that captures QFTCS renormalization at background level, and a phenomenological running \(K=1+K_{0}\ee^{-pN}\,(\Leftrightarrow 1+K_{0}(1+z)^{p})\) (Sec.~\ref{sec:K}). Exact identities for \(K'/K\) are provided; in the curvature case, an algebraic, iteration--free expression depending only on \((\phi,\phi',H)\) and known sources is derived (Eq.~\eqref{eq:KprimeK-closed}).
\item Linear perturbations are developed from the perturbed Klein--Gordon equation to the quadratic action and the Mukhanov--Sasaki system, yielding \(c_s^2=1\), \(\Phi=\Psi\), and the GR Poisson equation on subhorizon scales (Sec.~\ref{sec:linear}).
\item The EFT--DE mapping is exhibited explicitly, with \(\alpha_K>0\) and \(\alpha_B=\alpha_M=\alpha_T=0\) (Eq.~\eqref{eq:eft-map-intro}), and QKDE is situated within existing theory space (Sec.~\ref{sec:relations}).
\item Forecast--ready relations are recorded for \(\chi(z)\), \(D_A(z)\), \(D_L(z)\), \(D_V(z)\), \(F_{\rm AP}(z)\), and the GR growth observables \(D(a)\), \(f(a)\), and \(f\sigma_8(z)\), each expressed solely in terms of the background solution (Sec.~\ref{sec:observables}).
\item A reproducible numerical pipeline is provided: state vector, tolerances, algebraic evaluation of \(K'/K\), and diagnostic identities (total Friedmann closure, Raychaudhuri, scalar exchange, Ricci) used as invariants of the integration (Sec.~\ref{sec:numerics}); parameter sensitivities and Fisher assembly are derived from exact variational equations (Sec.~\ref{sec:forecast-setup}).
\end{enumerate}

\paragraph{Admissibility and assumptions.}
Spatial flatness is assumed; the expanding branch satisfies \(H(t)>0\); and ghost freedom requires \(K>0\). Smoothness \(K\in C^{1}\), \(V\in C^{2}\) ensures local existence/uniqueness of the e--fold system. When early--time anchors are used, a conservative prior \(K(z)\to 1\) for \(z\ge z_{\rm drag}\) is adopted; otherwise the sound horizon \(r_{d}\) is recomputed from its definition (Sec.~\ref{subsec:rd}). For the curvature--motivated case, the algebraic denominator in \eqref{eq:KprimeK-closed} must remain nonzero along the solution; this condition is monitored alongside \(K>0\).

\paragraph{Relation to prior work.}
Canonical quintessence modifies the background through \(V(\phi)\) with fixed kinetic normalization \parencite{RatraPeebles1988,Caldwell1998}; k--essence admits general \(P(X,\phi)\) and can alter linear phenomenology via \(c_s^2\neq 1\) \parencite{ArmendarizPicon2001,GarrigaMukhanov1999}. QKDE retains canonical propagation (\(c_s^2=1\)) but incorporates a background kinetic running \(K(\chi_{\rm c})\vert_{\chi_{\rm c}=t}\) that encapsulates curvature--induced wavefunction renormalization \parencite{BirrellDavies1982,ParkerFulling1974,ParkerToms2009,BarvinskyVilkovisky1985,Donoghue1994} while preserving the Einstein--Hilbert metric sector. In EFT--DE terms \parencite{Gubitosi2013,Bloomfield2013,BelliniSawicki2014}, it isolates a single nonzero parameter \(\alpha_{K}\) consistent with GW170817 \parencite{Abbott2017GW170817,Sakstein2017GW}.

\paragraph{Notation and organization.}
Units with \(c=\hbar=1\) are used; the metric signature is \((-,+,+,+)\); an overdot denotes \(d/dt\), and a prime denotes \(d/dN\) with \(N=\ln a\). The reduced Planck mass is \(\mpl=(8\pi G)^{-1/2}\). Section~\ref{sec:covariant-core} states the effective background equations, the covariant completion, and stability prerequisites. Section~\ref{sec:efold} derives the closed background system in e--fold variables. Section~\ref{sec:K} specifies and analyzes the kinetic normalizations \(K(\chi_{\rm c})\vert_{\chi_{\rm c}=t}\) and the algebraic closure for \(K'/K\). Section~\ref{sec:linear} develops the linear perturbations and the Mukhanov--Sasaki system. Section~\ref{sec:eft} presents the EFT--DE mapping and linear viability and, in Sec.~\ref{sec:relations}, situates QKDE within existing theory space. Section~\ref{sec:observables} records background and growth observables. Sections~\ref{sec:numerics}--\ref{sec:conclusions} cover implementation, parameter sensitivities and Fisher setup, limiting cases, and conclusions.

%% file: sections/section2/section2.tex
\section{Effective formulation and core background equations}
\label{sec:covariant-core}

\paragraph{Assumptions and conventions.}
(1) The metric sector is Einstein--Hilbert with constant reduced Planck mass \(\mpl\).
(2) Matter and radiation are minimally coupled. 
(3) The kinetic normalization is a background function arising from a covariant completion,
\(K=K(\chi)>0\), where \(\chi\) is a clock (St\"uckelberg) field; the working description adopts unitary gauge \(\chi=t\), so that \(K(\chi)\to K(t)\).
The clock field \(\chi\) restores diffeomorphism invariance at the level of the action prior to gauge fixing.
In the full covariant completion, \(\chi\) is endowed with its own (unspecified) sector \(S_\chi[g_{\mu\nu},\chi]\) that enforces monotonic time evolution and permits the gauge choice \(\chi=t\); its explicit form is not required for the background dynamics presented here.
(4) Spatially flat FRW background.
(5) Units \(c=\hbar=1\); metric signature \((-,+,+,+)\).
For later use, \(\Box\equiv g^{\mu\nu}\nabla_{\mu}\nabla_{\nu}\).

This section collects and \emph{derives} the background relations used throughout, starting from the
covariantly completed action corresponding to \eqref{eq:intro-action}:
\begin{equation}
S=\int \dd^{4}x\,\sqrt{-g}\Big[\tfrac12 \mpl^{2} R + K(\chi)\,X - V(\phi)\Big],
\qquad
X\equiv -\tfrac12 g^{\mu\nu}\partial_\mu\phi\,\partial_\nu\phi,
\end{equation}
with \(K(\chi)>0\) and minimally coupled matter and radiation.
The action written here represents the effective dark-energy sector; the complete theory includes the additional clock sector \(S_\chi\) whose stress--energy contribution is implicitly accounted for through total conservation.
All background equations used below follow after fixing unitary gauge \(\chi=t\), in which case \(K(\chi)\to K(t)\).

In the covariant completion the total energy--momentum tensor, including both the scalar field \(\phi\) and the clock field \(\chi\), is conserved.
The explicit time dependence of \(K(t)\) arises only after gauge fixing and induces an exchange between the scalar and clock sectors rather than a violation of the Bianchi identity.
Observable quantities depend only on the total stress--energy tensor and the resulting metric evolution, not on the gauge-dependent partition between \(\phi\) and \(\chi\).

\subsection{Stress--energy tensor and fluid variables}
\label{subsec:Tmunu}

For a general scalar Lagrangian \(P(X,\phi,\chi)\), the stress--energy tensor is
\(T_{\mu\nu}=P_{,X}\,\partial_\mu\phi\,\partial_\nu\phi+g_{\mu\nu}P\) (see \parencite{GarrigaMukhanov1999}).
Applying this to \(P=K(\chi)X-V(\phi)\) gives
\begin{equation}
T_{\mu\nu}^{(\phi)}
= K(\chi)\,\partial_\mu\phi\,\partial_\nu\phi
  + g_{\mu\nu}\!\left[K(\chi)\,X - V(\phi)\right].
\label{eq:Tmunu}
\end{equation}
On a spatially flat FRW background \(\dd s^{2}=-\dd t^{2}+a^{2}(t)\dd\vec{x}^{2}\), homogeneity implies \(X=\tfrac12\dot\phi^{2}\) and therefore
\begin{equation}
\rho_\phi = K\,\frac{\dot\phi^{2}}{2}+V(\phi),\qquad
p_\phi    = K\,\frac{\dot\phi^{2}}{2}-V(\phi),\qquad
w_\phi    = \frac{K\dot\phi^{2}/2 - V}{K\dot\phi^{2}/2 + V}.
\label{eq:rho-p-w}
\end{equation}
Here and below, \(K\equiv K(\chi)\vert_{\chi=t}\) once unitary gauge is imposed.
The quantities \(\rho_\phi\) and \(p_\phi\) represent the scalar contribution to the total stress--energy tensor and do not correspond to a separately conserved fluid when \(K\) varies.

\subsection{Scalar equation of motion and modified friction}
\label{subsec:KG}

The Euler--Lagrange equation for \(P(X,\phi,\chi)\) is \(\nabla_\mu(P_{,X}\nabla^\mu\phi)-P_{,\phi}=0\) \parencite{GarrigaMukhanov1999}.
For \(P=K(\chi)X-V(\phi)\) one obtains
\begin{equation}
K\,\Box\phi + (\nabla_\mu K)\,\nabla^\mu\phi - V_{,\phi}=0.
\label{eq:KG-cov}
\end{equation}
Since \(K=K(\chi)\), fixing unitary gauge \(\chi=t\) gives \((\nabla_\mu K)\nabla^\mu\phi=g^{00}\dot K\,\dot\phi=-\dot K\,\dot\phi\).
Using \(\Box\phi=-(\ddot\phi+3H\dot\phi)\) on FRW gives the background equation
\begin{equation}
K(t)\big(\ddot\phi+3H\dot\phi\big)+\dot K(t)\,\dot\phi+V_{,\phi}=0,
\label{eq:KG-FRW}
\end{equation}
i.e.\ the canonical Klein--Gordon relation with an additional, time--dependent friction term \(\dot K\,\dot\phi\).

\subsection{Einstein equations, Raychaudhuri form, and Ricci scalar}
\label{subsec:Friedmann}

Because the metric sector is pure GR (no non-minimal coupling and constant \(\mpl\)),
\begin{equation}
H^{2}
=\frac{1}{3\mpl^{2}}\big(\rho_m+\rho_r+\rho_\phi\big),\qquad
\dot H
=-\frac{1}{2\mpl^{2}}\Big(\rho_m+\tfrac{4}{3}\rho_r+\rho_\phi+p_\phi\Big).
\label{eq:Friedmann}
\end{equation}
Substituting \(\rho_\phi+p_\phi=K\,\dot\phi^{2}\) into the second relation yields the Raychaudhuri form
\begin{equation}
\dot H
=-\frac{1}{2\mpl^{2}}\left(\rho_m+\tfrac{4}{3}\rho_r+K\,\dot\phi^{2}\right),
\label{eq:Hdoteq}
\end{equation}
and spatial flatness implies the FRW Ricci scalar
\begin{equation}
R=6\left(2H^{2}+\dot H\right).
\label{eq:R-FRW}
\end{equation}

\subsection{Continuity relations and background source}
\label{subsec:continuity}

Minimal coupling implies the standard matter and radiation continuity equations,
\begin{equation}
\dot\rho_m+3H\rho_m=0,\qquad
\dot\rho_r+4H\rho_r=0.
\label{eq:mr-continuity}
\end{equation}
Because matter and radiation couple only to the metric \(g_{\mu\nu}\) and not to \(\chi\), they remain separately conserved even in the presence of scalar--clock exchange.
In the covariant completion, the total stress--energy tensor is covariantly conserved, \(\nabla_\mu T^{\mu}{}_{\nu(\mathrm{tot})}=0\).
After fixing unitary gauge \(\chi=t\), the scalar sector obeys an exchange equation rather than separate conservation:
\begin{equation}
\nabla_\mu T^{\mu}{}_{\nu(\phi)} \;=\; -\,\partial_\nu P(X,\phi,t)
\;=\; -\,X\,\partial_\nu K,
\label{eq:cov-noncons}
\end{equation}
which on FRW reduces to the exact background relation
\begin{align}
\dot\rho_{\phi}
&=\tfrac12\dot K\,\dot\phi^{2}+K\,\dot\phi\,\ddot\phi+V_{,\phi}\dot\phi,\\
\dot\rho_{\phi}+3H(\rho_{\phi}+p_{\phi})
&=\tfrac12\dot K\,\dot\phi^{2}+K\,\dot\phi\,(\ddot\phi+3H\dot\phi)+V_{,\phi}\dot\phi
= -\,\tfrac12\,\dot K\,\dot\phi^{2},
\label{eq:phi-continuity}
\end{align}
after using \eqref{eq:KG-FRW}. Thus the scalar component exchanges energy with the clock/time-slicing sector.
Equation~\eqref{eq:phi-continuity} is therefore a unitary-gauge bookkeeping identity rather than a statement of separate scalar conservation.
This does not contradict GR: the second Friedmann equation follows from the first together with \(\nabla_\mu T^{\mu}{}_{\nu(\mathrm{tot})}=0\), while \eqref{eq:phi-continuity} replaces \(\nabla_\mu T^{\mu}{}_{\nu(\phi)}=0\) for the scalar alone.
\footnote{For a Lagrangian density \(\mathcal{L}(g_{\mu\nu},\phi,\partial\phi;x)\) with explicit coordinate dependence, diffeomorphism invariance implies
\(\nabla_\mu T^{\mu}{}_{\nu} = -\partial_\nu \mathcal{L}\) evaluated on shell. For \(P=K(t)X-V(\phi)\), \(\partial_\nu P = X\,\partial_\nu K\), yielding \eqref{eq:cov-noncons}. See e.g.\ \citep[Sec.~2]{GarrigaMukhanov1999}.}

\subsection{Stability and propagation speed}
\label{subsec:stability}

For general \(P(X,\phi,t)\), the scalar sound speed is \(c_s^2=P_{,X}/(P_{,X}+2X P_{,XX})\) \parencite{GarrigaMukhanov1999}. Here \(P=K(\chi)X-V(\phi)\) gives \(P_{,X}=K(\chi)\) and \(P_{,XX}=0\), hence
\begin{equation}
c_s^2=1,\qquad \text{no gradient instability}.
\end{equation}
Ghost freedom requires \(P_{,X}>0\), i.e.\ \(K(\chi)>0\). The metric sector is Einstein--Hilbert, so the tensor speed is luminal and the Planck mass is constant, consistent with GW170817/GRB bounds \citep{Abbott2017GW170817,Sakstein2017GW}.

\subsection{EFT--DE mapping (for later use)}
\label{subsec:eft-mapping}

Expansion of the quadratic action in the EFT--DE basis \citep{Gubitosi2013,Bloomfield2013,BelliniSawicki2014} yields
\begin{equation}
\alpha_K=\frac{K(t)\,\dot\phi^{2}}{H^{2}\mpl^{2}},\qquad
\alpha_B=\alpha_M=\alpha_T=0,
\label{eq:EFT-map}
\end{equation}
i.e.\ a pure--kinetic background running without braiding, Planck--mass evolution, or tensor speed excess. This is the inflation/EFT pattern of time--dependent background coefficients in unitary gauge (cf.\ \citealp{Cheung2008}).

\paragraph{Consistency diagnostics (used later in numerics).}
The combination \(3\mpl^{2}H^{2}-\rho_m-\rho_r-\rho_\phi\) vanishes identically by \eqref{eq:Friedmann}; monitoring its fractional magnitude serves as an energy-closure check.
Differentiating \eqref{eq:Friedmann} and using \eqref{eq:mr-continuity} together with total conservation (i.e.\ including the scalar exchange \eqref{eq:phi-continuity}) reproduces \eqref{eq:Hdoteq} exactly;
this provides an independent Raychaudhuri-consistency diagnostic during integration.

\medskip
\noindent\begin{center}
\fbox{\begin{minipage}{0.96\linewidth}
\textbf{Section~\ref{sec:covariant-core} at a glance.}
\[
\begin{aligned}
& \textbf{Action:} && S=\!\int \dd^{4}x\sqrt{-g}\,\Big[\tfrac12 \mpl^{2}R + K(\chi)X - V(\phi)\Big],\quad
   X\equiv -\tfrac12 g^{\mu\nu}\partial_\mu\phi\,\partial_\nu\phi,\\
& && \chi=t\ \text{(unitary gauge)}.\\
& \textbf{Stress tensor:} && T_{\mu\nu}^{(\phi)} = K(\chi)\,\partial_\mu\phi\,\partial_\nu\phi
  + g_{\mu\nu}\!\left(K(\chi)X - V\right).\\
& \textbf{Background:} && \rho_\phi = K\frac{\dot\phi^{2}}{2}+V,\quad
   p_\phi=K\frac{\dot\phi^{2}}{2}-V,\quad
   w_\phi=\frac{K\dot\phi^{2}/2 - V}{K\dot\phi^{2}/2 + V}.\\
& \textbf{KG:} && K(\ddot\phi+3H\dot\phi)+\dot K\,\dot\phi+V_{,\phi}=0.\\
& \textbf{Einstein:} && H^{2}=\frac{\rho_m+\rho_r+\rho_\phi}{3\mpl^{2}},\quad
   \dot H=-\frac{\rho_m+\tfrac43\rho_r+K\dot\phi^{2}}{2\mpl^{2}},\quad
   R=6(2H^{2}+\dot H).\\
& \textbf{Exchange:} && \nabla_\mu T^{\mu}{}_{\nu(\phi)}=-\,\partial_\nu K\,X
   \;\Rightarrow\; \dot\rho_\phi+3H(\rho_\phi+p_\phi)=-\tfrac12 \dot K\,\dot\phi^{2}.\\
& \textbf{EFT--DE:} && \alpha_K=\frac{K\dot\phi^{2}}{H^{2}\mpl^{2}},\quad
   \alpha_B=\alpha_M=\alpha_T=0,\quad c_s^2=1.
\end{aligned}
\]
\end{minipage}}
\end{center}

\medskip
Equations \eqref{eq:rho-p-w}–\eqref{eq:EFT-map} constitute the effective background foundation used in the subsequent e--fold formulation, linear perturbations, and observables.

\input{sections/section2/figures/fig_core_flow}

%% file: sections/section2/figures/fig_core_flow.tex
\begin{figure}[t]
\centering
\begin{adjustbox}{max width=\linewidth,center}
\begin{tikzpicture}[
  >=Latex,
  node distance=5mm and 9mm,
  box/.style={
    rectangle, rounded corners=3pt, draw=black, thick, fill=white,
    inner sep=3pt, outer sep=1pt, text width=5.1cm, align=left, font=\footnotesize
  }
]
\node[box] (action) {\textbf{Effective background action (unitary gauge)}
\\[2pt]$S=\int\!\sqrt{-g}\,\big[\tfrac12 M_{\rm pl}^2 R+K(\chi)X-V(\phi)\big],\qquad \chi=t$};

\node[box, below=of action] (tmunu) {\textbf{Vary w.r.t.\ $g^{\mu\nu}$}
\\[2pt]$T^{(\phi)}_{\mu\nu}=K\,\partial_\mu\phi\,\partial_\nu\phi+g_{\mu\nu}(KX-V)$
\\$\Rightarrow$ on FRW: $\rho_\phi=\tfrac12 K\dot\phi^2+V$, $p_\phi=\tfrac12 K\dot\phi^2-V$};

\node[box, right=of tmunu] (kg) {\textbf{Vary w.r.t.\ $\phi$}
\\[2pt]$K\,\Box\phi+(\nabla_\mu K)\nabla^\mu\phi - V_{,\phi}=0$
\\$\Rightarrow$ on FRW: $K(\ddot\phi+3H\dot\phi)+\dot K\,\dot\phi+V_{,\phi}=0$};

\node[box, below=of tmunu] (einstein) {\textbf{Einstein equations (GR metric)}
\\[2pt]$H^2=\dfrac{\rho_m+\rho_r+\rho_\phi}{3M_{\rm pl}^2}$,\;
$\dot H=-\dfrac{\rho_m+\tfrac43\rho_r+K\dot\phi^2}{2M_{\rm pl}^2}$
\\$R=6(2H^2+\dot H)$};

\node[box, below=of kg] (source) {\textbf{Explicit time dependence of $K$}
\\[2pt]$\nabla_\mu T^{\mu}{}_{\nu(\phi)}=-\,\partial_\nu K\,X$ (exchange)
\\$\Rightarrow$ on FRW:
$\dot\rho_\phi+3H(\rho_\phi+p_\phi)=-\tfrac12\dot K\,\dot\phi^2$};

\node[box, below=of einstein] (eft) {\textbf{EFT--DE map (Bellini--Sawicki)}
\\[2pt]$\alpha_K=\dfrac{K\dot\phi^2}{H^2 M_{\rm pl}^2}$,\;
$\alpha_B=\alpha_M=\alpha_T=0$
\\$c_s^2=1$ (since $P_{,XX}=0$), tensors luminal};

\node[box, below=of source] (diag) {\textbf{Consistency diagnostics}
\\[2pt]Friedmann closure: $3M_{\rm pl}^2 H^2-\rho_m-\rho_r-\rho_\phi=0$
\\Raychaudhuri check: follows from total conservation, including scalar exchange};

\draw[->, thick] (action) -- (tmunu);
\draw[->, thick] (action) -- (kg);
\draw[->, thick] (tmunu) -- (einstein);
\draw[->, thick] (kg) -- (source);
\draw[->, thick] (einstein) -- (eft);
\draw[->, thick] (source) -- (diag);
\end{tikzpicture}
\end{adjustbox}
\caption{Dependency flow for the core background relations in Sec.~\ref{sec:covariant-core}.
The action admits a covariant completion via a clock field $\chi$, with unitary gauge $\chi=t$.
The scalar continuity equation is an exchange relation, while the total energy--momentum tensor is conserved.
Every equation used later descends from the effective background action by variation and specialization to FRW,
with $K(t)$ entering only as a background kinetic normalization.}
\label{fig:core_flow}
\end{figure}
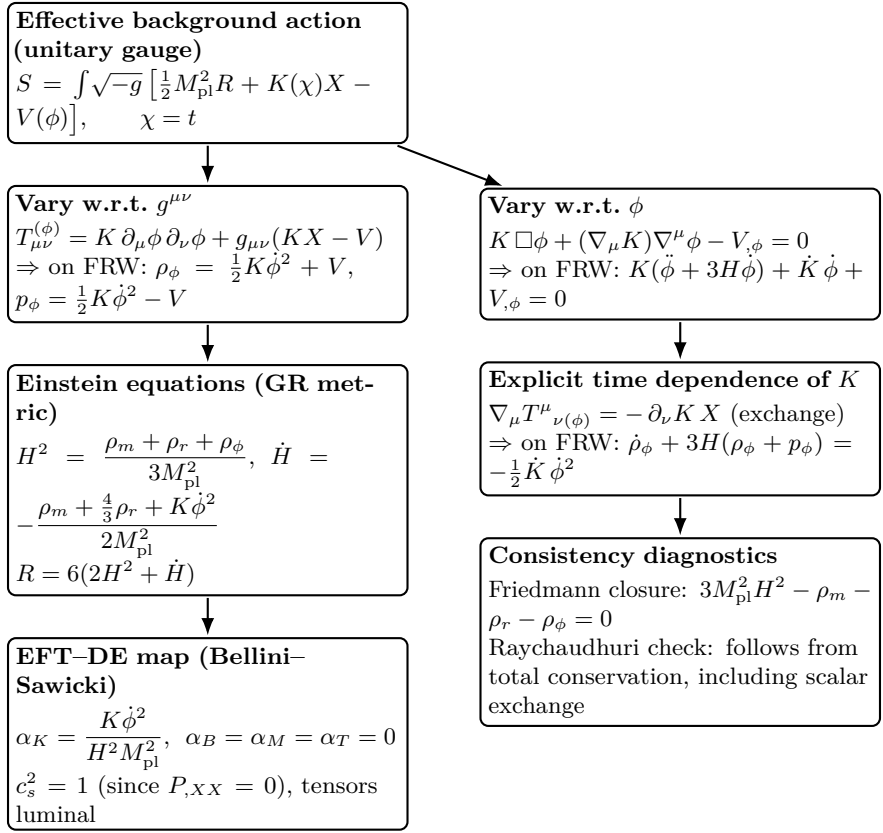

%% file: sections/section3/section3.tex
\section{E-fold reformulation and closed background system}
\label{sec:efold}

It is convenient to reparametrize time by the number of e-folds,
\begin{equation}
N \equiv \ln a, \qquad (\;)' \equiv \frac{d}{dN}=\frac{1}{H}\frac{d}{dt},
\label{eq:efold-def}
\end{equation}
a choice standard in cosmology and inflationary dynamics \parencite{Mukhanov2005,Dodelson2003}. For any sufficiently smooth background function \(f(t)\),
\begin{equation}
\dot f = H f', \qquad
\ddot f = H^{2} f'' + H H' f',
\label{eq:chain-rules}
\end{equation}
which follows directly by the chain rule and \(\dot N=H\). It is also convenient to denote
\(\,E\equiv H'/H\).

Throughout this section, all quantities are understood as background functions
evaluated in the preferred cosmological time slicing (unitary gauge). The e--fold variable
\(N=\ln a\) therefore parametrizes the effective background evolution defined by the action
\eqref{eq:intro-action}, rather than a covariant field redefinition.

\subsection{Klein--Gordon equation in e-fold time}
\label{subsec:KG-N}

Inserting \eqref{eq:chain-rules} into the FRW Klein--Gordon equation \eqref{eq:KG-FRW},
\(
K(\ddot\phi+3H\dot\phi)+\dot K\,\dot\phi+V_{,\phi}=0,
\)
gives
\[
K\!\left(H^{2}\phi''+H H'\phi' + 3H^{2}\phi'\right)
+\dot K\,(H\phi')+V_{,\phi}=0.
\]
Dividing by \(K H^{2}\) and using \(K'/K=\dot K/(H K)\) yields the exact e-fold form
\begin{equation}
\boxed{%
\phi''+\left(3+\frac{H'}{H}+\frac{K'}{K}\right)\phi'
+\frac{V_{,\phi}}{H^{2}K}=0
}\,.
\label{eq:KG-N}
\end{equation}
The kinetic invariant and scalar energy density become
\begin{equation}
\boxed{%
X=\tfrac12 H^{2}\phi'^{2}, \qquad
\rho_\phi = K\,\frac{H^{2}}{2}\phi'^{2}+V(\phi)
}\,,
\label{eq:Xrho-N}
\end{equation}
which follows from \(X=\tfrac12\dot\phi^{2}\) and \eqref{eq:rho-p-w}.

\subsection{Hubble derivative, Ricci scalar, and redshift}
\label{subsec:Hprime-Ricci}

From the Raychaudhuri relation \eqref{eq:Hdoteq},
\(
\dot H=-(\rho_m+\tfrac43\rho_r+K\,\dot\phi^{2})/(2M_{\rm pl}^{2}),
\)
and \(\dot H=H^{2}(H'/H)\), one finds
\begin{equation}
\boxed{%
\frac{H'}{H}
= -\frac{1}{2M_{\rm pl}^{2}H^{2}}
\left(\rho_m+\tfrac{4}{3}\rho_r+K\,H^{2}\phi'^{2}\right)
}\,.
\label{eq:Hprime-over-H}
\end{equation}
Expressed in density parameters \(\Omega_i\equiv\rho_i/(3M_{\rm pl}^{2}H^{2})\) for \(i\in\{m,r\}\),
\begin{equation}
\boxed{%
\frac{H'}{H} \;=\; -\frac{1}{2}\left(3\,\Omega_m+4\,\Omega_r+\frac{K\,\phi'^2}{M_{\rm pl}^{2}}\right)
}\,,
\label{eq:Hprime-Omegas}
\end{equation}
and spatial flatness gives the FRW Ricci scalar ratio
\begin{equation}
\boxed{%
\frac{R}{H^{2}} = 6\!\left(2+\frac{H'}{H}\right)
}\,.
\label{eq:R-over-H2}
\end{equation}
The e-fold/redshift mapping and minimally coupled background scalings are
\begin{equation}
\boxed{%
z(N)=e^{-N}-1
}\!,
\qquad
\rho_m(N)=\rho_{m0}\,e^{-3N}, \quad
\rho_r(N)=\rho_{r0}\,e^{-4N},
\label{eq:z-and-mr-of-N}
\end{equation}
which follow from \(\nabla_\mu T^{\mu}{}_{\nu}\) conservation for dust and radiation on FRW.

\paragraph{Closure, equation of state, and deceleration.}
With \(\Omega_\phi\equiv 1-\Omega_m-\Omega_r\) and \eqref{eq:Xrho-N},
\begin{equation}
\boxed{%
\Omega_{\phi}(N)=\frac{K\,\phi'^2}{6\,M_{\rm pl}^{2}}+\frac{V(\phi)}{3M_{\rm pl}^{2}H^{2}},
\qquad
w_\phi(N)=\frac{\tfrac{K\,\phi'^2}{6M_{\rm pl}^{2}}-\tfrac{V}{3M_{\rm pl}^{2}H^{2}}}{\tfrac{K\,\phi'^2}{6M_{\rm pl}^{2}}+\tfrac{V}{3M_{\rm pl}^{2}H^{2}}}
}\,,
\label{eq:Omega-wphi}
\end{equation}
and from \eqref{eq:Hprime-over-H},
\begin{equation}
\boxed{%
w_{\rm eff}(N)=-1-\frac{2}{3}\frac{H'}{H},
\qquad
q(N)=-1-\frac{H'}{H}
}\,.
\label{eq:weff-q}
\end{equation}

\subsection{Closed autonomous system}
\label{subsec:closed-system}

Equations \eqref{eq:KG-N}, \eqref{eq:Hprime-over-H}, and \eqref{eq:Xrho-N} form a closed, first-order autonomous system once \(K(N)\) and \(V(\phi)\) are specified (subject to \(K>0\)). For stable numerics, the state vector
\begin{equation}
\mathbf{y} \equiv (\phi,\; s,\; y_{H}) \equiv \big(\phi,\;\phi',\;\ln H\big)
\end{equation}
evolves according to
\begin{subequations}
\label{eq:background-system}
\begin{align}
s' \;&=\; -\left(3+\frac{H'}{H}+\frac{K'}{K}\right)s - \frac{V_{,\phi}}{H^{2}K}, 
\label{eq:system-phi}\\[4pt]
y_{H}' \;&=\; \frac{H'}{H}
= -\frac{1}{2M_{\rm pl}^{2}H^{2}}\!\left(\rho_m+\tfrac{4}{3}\rho_r+K\,H^{2}s^{2}\right),
\label{eq:system-Hprime}\\[4pt]
X \;&=\; \tfrac12 H^{2}s^{2}, 
\qquad
\rho_\phi \;=\; K\,\frac{H^{2}}{2}s^{2}+V(\phi),
\label{eq:system-aux}
\end{align}
\end{subequations}
with \(\rho_m,\rho_r\) from \eqref{eq:z-and-mr-of-N} and \(H=e^{y_H}\). The evaluation of \(K'/K\) for the two specifications appears in Sec.~\ref{sec:K}. The denominators \(H^2\) and \(H^2K\) are nonzero by construction (\(H>0\), \(K>0\)).

\FloatBarrier
\input{sections/section3/tables/tab_efold_summary}

\subsection{Well-posedness and admissible initial data}
\label{subsec:wellposed}

Let \(\mathbf{y}'=\mathbf{F}(N,\mathbf{y})\) on a domain where \(K(N)>0\) and \(H(N)>0\).
For any smooth \(K(N)\) and \(V(\phi)\), the right-hand side \(\mathbf{F}\) is continuous and locally Lipschitz in \(\mathbf{y}\); the Picard--Lindel\"of theorem then guarantees local existence and uniqueness through any \(N_{\rm ini}\) with admissible initial data \parencite{Coddington1955}. A matter-era initialization consistent with \(\rho_m\gg\rho_r,\rho_\phi\) is
\begin{equation}
\phi'(N_{\rm ini})=0,\qquad
H(N_{\rm ini})
=H_0\sqrt{\Omega_{m0}e^{-3N_{\rm ini}}+\Omega_{r0}e^{-4N_{\rm ini}}}\,,
\end{equation}
while \(\phi(N_{\rm ini})\) is fixed by enforcing \(\Omega_{\phi}(0)=1-\Omega_{m0}-\Omega_{r0}\) using \eqref{eq:Omega-wphi}. Global continuation holds as long as \(K>0\) and \(H>0\).

\paragraph{Diagnostics (for numerics).}
For consistency checks during integration, the GR identities defined in Sec.~\ref{sec:numerics} are monitored: the Friedmann closure \(\mathcal{C}_{F}=1-\Omega_m-\Omega_r-\Omega_\phi\) and the Raychaudhuri residual \(\mathcal{C}_{R}\) in e-fold form. Both vanish analytically; numerically they should remain \(\ll 1\) at the chosen tolerances.

\medskip
\noindent\emph{Limits.} Setting \(K\!\equiv\!1\) reduces the system to canonical quintessence. With \(V=\mathrm{const}\) and \(\phi' \!=\!0\) the system reproduces \(\Lambda\)CDM.

%% file: sections/section3/tables/tab_efold_summary.tex
\begin{table}[t]
\centering
\footnotesize
\setlength{\tabcolsep}{6pt}
\renewcommand{\arraystretch}{1.12}
\caption{E-fold reformulation: core identities, evolution equations, and diagnostics used in Sec.~\ref{sec:efold}. All relations are exact given the assumptions in Sec.~\ref{sec:covariant-core}.}
\label{tab:efold_summary}
\begin{adjustbox}{max width=\linewidth}
\begin{tabular}{l p{0.62\linewidth} l}
\toprule
\textbf{Quantity} & \textbf{E-fold form / Definition} & \textbf{Ref.} \\
\midrule
Time variable & $N\equiv \ln a$, \quad $(\,)' \equiv \dd/\dd N = H^{-1}\dd/\dd t$ & \eqref{eq:efold-def} \\
Chain rules & $\dot f = H f'$, \quad $\ddot f = H^{2} f'' + H H' f'$ & \eqref{eq:chain-rules} \\
\addlinespace[2pt]
Klein--Gordon & $\displaystyle \phi''+\Big(3+\frac{H'}{H}+\frac{K'}{K}\Big)\phi'
+\frac{V_{,\phi}}{H^{2}K}=0$ & \eqref{eq:KG-N} \\
Kinetic \& $\rho_\phi$ & $X=\tfrac12 H^{2}\phi'^{2}$, \quad
$\rho_\phi = K\,\frac{H^{2}}{2}\phi'^{2}+V(\phi)$ & \eqref{eq:Xrho-N} \\
\addlinespace[2pt]
$H'/H$ (fluids) & $\displaystyle \frac{H'}{H}
= -\frac{1}{2M_{\rm pl}^{2}H^{2}}\!\left(\rho_m+\tfrac{4}{3}\rho_r+K\,H^{2}\phi'^{2}\right)$ & \eqref{eq:Hprime-over-H} \\
$H'/H$ ($\Omega) $ & $\displaystyle \frac{H'}{H} = -\frac12\!\left(3\Omega_m+4\Omega_r+\frac{K\,\phi'^2}{M_{\rm pl}^2}\right)$ & \eqref{eq:Hprime-Omegas} \\
Ricci scalar & $\displaystyle \frac{R}{H^{2}} = 6\!\left(2+\frac{H'}{H}\right)$ & \eqref{eq:R-over-H2} \\
Redshift map & $z(N)=e^{-N}-1$, \quad $\rho_m=\rho_{m0}e^{-3N}$, \quad $\rho_r=\rho_{r0}e^{-4N}$ & \eqref{eq:z-and-mr-of-N} \\
\addlinespace[2pt]
Closure \& $w_\phi$ & $\displaystyle \Omega_\phi=\frac{K\,\phi'^2}{6M_{\rm pl}^{2}}+\frac{V}{3M_{\rm pl}^{2}H^{2}}$,
$\quad w_\phi=\frac{\tfrac{K\phi'^2}{6M_{\rm pl}^{2}}-\tfrac{V}{3M_{\rm pl}^{2}H^{2}}}{\tfrac{K\phi'^2}{6M_{\rm pl}^{2}}+\tfrac{V}{3M_{\rm pl}^{2}H^{2}}}$ & \eqref{eq:Omega-wphi} \\
Effective EoS & $w_{\rm eff}=-1-\tfrac{2}{3}\tfrac{H'}{H}$, \quad $q=-1-\tfrac{H'}{H}$ & \eqref{eq:weff-q} \\
\addlinespace[2pt]
State vector & $\mathbf{y}=(\phi,\,s,\,y_H) \equiv (\phi,\,\phi',\,\ln H)$ & \S\ref{subsec:closed-system} \\
Autonomous sys. & $s'=-\Big(3+\tfrac{H'}{H}+\tfrac{K'}{K}\Big)s-\tfrac{V_{,\phi}}{H^{2}K}$;\;
$y_H'=\tfrac{H'}{H}$;\;
$X=\tfrac12 H^{2}s^{2}$ & \eqref{eq:background-system} \\
$K'/K$ (note) & From chosen $K(N)$: phenom.\ \eqref{eq:Kprime-phenom} or curvature \eqref{eq:KprimeK-closed} & Sec.~\ref{sec:K} \\
\addlinespace[2pt]
Diagnostics & Friedmann closure $\mathcal{C}_F=1-\Omega_m-\Omega_r-\Omega_\phi$; 
Raychaudhuri residual $\mathcal{C}_R$ (both $\to 0$ analytically) & Sec.~\ref{sec:numerics} \\
\bottomrule
\end{tabular}
\end{adjustbox}
\end{table}

%% file: sections/section4/section4.tex
\section{Kinetic normalization \(K(N)\): specifications and derived identities}
\label{sec:K}

Only the \emph{background} kinetic normalization is allowed to vary in time while
the metric sector remains Einstein--Hilbert. A covariant completion is assumed throughout: the kinetic normalization is a scalar function \(K(\chi)\) of a clock (St\"uckelberg) field \(\chi\), and all expressions in this section are evaluated after fixing unitary gauge \(\chi=t\), so that \(K(\chi)\to K(t)\).
In the EFT-of-DE language (unitary
gauge), a time-dependent background coefficient multiplying
\(X\equiv -\tfrac12 g^{\mu\nu}\partial_\mu\phi\,\partial_\nu\phi\)
corresponds to the Bellini--Sawicki corner
\(\alpha_{B}=\alpha_{M}=\alpha_{T}=0\) with a pure kineticity
\(\alpha_{K}>0\) \parencite{Gubitosi2013,Bloomfield2013,BelliniSawicki2014}.
Sections~\ref{sec:covariant-core}--\ref{sec:efold} develop the background
dynamics that follow from such a
covariantly completed
\(P(X,\phi,\chi)=K(\chi)X-V(\phi)\) theory
after adopting unitary gauge.
The \emph{origin} and \emph{theoretical status} of \(K(\chi)\vert_{\chi=t}\)
are provided in full
detail in Appendix~\ref{appendix:foundations}, which should be regarded as the
UV completion of the statements made in this section.

Throughout,
\[
E \equiv \frac{H'}{H}, \qquad
\frac{R}{H^{2}} = 6(2+E)
\quad (\text{cf.\ \eqref{eq:R-over-H2}}).
\]
\emph{Dimensions:} \(K\) is dimensionless; \(R\) has mass dimension \(2\); the
ratio \(\alpha/M^{2}\) is dimensionless.

\paragraph{UV motivation (operator origin).}
Appendix~\ref{appendix:foundations} demonstrates that in the
covariant
derivative expansion of a light scalar interacting with heavy fields
(\(M\!\gg\!H\)) and gravity, integrating out the heavy sector or evaluating
curved-spacetime loop corrections generates dimension-six operators of the form
\begin{equation}
\Delta\mathcal{L}
\;\supset\;
\frac{c_{R}}{M^{2}}\,R\,(\partial\phi)^{2}
\quad\text{and}\quad
\frac{c_{\mathcal G}}{M^{2}}
\mathcal{G}_{\mu\nu}\partial^\mu\phi\partial^\nu\phi.
\end{equation}
Both appear at order \(H^{2}/M^{2}\) with Wilson coefficients
\(c_{R},c_{\mathcal{G}}=\mathcal{O}(1)\)
\parencite{BirrellDavies1982,ParkerToms2009,Burgess2004EFT}.  
As shown explicitly in Appendix~\ref{appendix:foundations}, the two operators are
\emph{degenerate} on an FRW background via the identity
\(R_{\mu\nu}\partial^\mu\phi\partial^\nu\phi=\tfrac13 R X\).  
Hence the unique two-derivative operator that renormalizes
the scalar kinetic structure is \(R X\), leading to the general IR form
\(
K(\chi)=1+\mathcal{O}(R/M^{2})
\)
prior to gauge fixing,
without modifying the metric sector.  
The EFT--DE mapping (Appendix~\ref{appendix:foundations}) confirms that this class of
actions always satisfies \((\alpha_{B},\alpha_{M},\alpha_{T})=(0,0,0)\) with
\(\alpha_{K}>0\).

\subsection{Curvature–motivated specification}
\label{subsec:K-curv}

A minimal curvature–suppressed form, directly descending from the UV result
(Appendix~\ref{appendix:foundations}), is
\begin{equation}
\boxed{
K(\chi)=1+\alpha\,\frac{R(\chi)}{M^{2}}
}
\qquad\Longrightarrow\qquad
\boxed{
K(N)=1+\alpha\,\frac{R(N)}{M^{2}}
}
\label{eq:K-curv}
\end{equation}
where the second expression follows after fixing
unitary gauge \(\chi=t\) and expressing the background in e--fold
time.
With $R/H^{2}=6(2+E)$, differentiating gives
\begin{equation}
\boxed{
\frac{K'}{K}
  = \frac{\alpha}{M^{2}}
    \frac{R'}{1+\alpha R/M^{2}}
}
\label{eq:Kprime-over-K-curv}
\end{equation}
and writing $R=H^{2}r$ where $r\equiv 6(2+E)$ gives, using $(H^{2})'=2EH^{2}$,
\begin{equation}
\frac{R'}{H^{2}}
= r' + 2Er
= 6\big(E' + 4E + 2E^{2}\big).
\label{eq:Rprime-master}
\end{equation}

From the Raychaudhuri relation \eqref{eq:Hprime-over-H},
\begin{equation}
E = -\frac{1}{2M_{\rm pl}^{2}H^{2}}
  \left(\rho_{m} + \tfrac43 \rho_{r} + K H^{2}\phi'^{2}\right),
\label{eq:E-def}
\end{equation}
which follows from the Einstein equations together with
total energy--momentum conservation in the covariant completion,
and differentiating \eqref{eq:E-def} with respect to \(N\), using
$\rho_{m}'=-3\rho_{m}$, $\rho_{r}'=-4\rho_{r}$, and the scalar equation
\eqref{eq:system-phi}
(the scalar exchange equation in unitary gauge),
yields
\begin{align}
E' &=
-\frac{1}{2M_{\rm pl}^{2}H^{2}}
\left[-3\rho_{m}-\tfrac{16}{3}\rho_{r}
      +(KH^{2}\phi'^{2})'\right]
-2E^{2}, \label{eq:Eprime-raw} \\[3pt]
(KH^{2}\phi'^{2})'
&= -H^{2}\!\big(K'\phi'^{2}+6K\phi'^{2}\big)
   -2\phi' V_{,\phi}.
\label{eq:KH2phi2prime}
\end{align}

Substituting \eqref{eq:Eprime-raw}–\eqref{eq:KH2phi2prime} into
\eqref{eq:Rprime-master} shows that $R'/H^{2}$ is \emph{affine} in $K'$:
\begin{equation}
\frac{R'}{H^{2}}
\;=\;
A + B\,K',
\qquad
B=\frac{3\phi'^{2}}{M_{\rm pl}^{2}},
\qquad
A=24E
 +\frac{18K\phi'^{2}}{M_{\rm pl}^{2}}
 +\frac{9\rho_{m}+16\rho_{r}+6\phi'V_{,\phi}}
       {M_{\rm pl}^{2}H^{2}}.
\label{eq:Rprime-affine}
\end{equation}

Define the shorthand
\begin{equation}
c \equiv \frac{\alpha H^{2}/M^{2}}{1+\alpha R/M^{2}},
\end{equation}
and substitute \eqref{eq:Rprime-affine} into
\eqref{eq:Kprime-over-K-curv}.  Because the dependence on $K'$ is affine, the
result is an \emph{algebraic}, recursion-free identity:
\begin{equation}
\boxed{
\frac{K'}{K}
=
\frac{c\,A}{1 - c\,B\,K}
=
\frac{\displaystyle
      \frac{\alpha H^{2}}{M^{2}}
      \frac{A}{1+\alpha R/M^{2}}}
     {\displaystyle
      1-
      \frac{\alpha H^{2}}{M^{2}}
      \frac{3K\phi'^{2}}
           {M_{\rm pl}^{2}(1+\alpha R/M^{2})}}
}
\label{eq:KprimeK-closed}
\end{equation}
where all quantities depend only on 
\((\phi,\phi',H)\) and background sources via
\eqref{eq:background-system}.  
No iteration in $R'$ or $K'$ is required.

\paragraph{Background limits (consistency).}
Using \eqref{eq:R-over-H2} and the background solutions:
- **Radiation domination:**  
  \(E\simeq -2\Rightarrow R=0\), hence \(K\simeq 1\) and \(\,K'/K\simeq 0\).  
- **Matter domination:**  
  \(E\simeq -\tfrac32\Rightarrow R/H^{2}\simeq 3\), giving a mild shift  
  \(K\simeq 1+3\alpha H^{2}/M^{2}\).  
- **de Sitter:**  
  \(E=0\Rightarrow R/H^{2}=12\), producing a constant renormalization  
  \(K=1+12\alpha H^{2}/M^{2}\).

These limits follow from the covariant UV expression and are recovered numerically.

\paragraph{Evaluation protocol.}
At each step in $N$:

(i) compute $E=H'/H$ via \eqref{eq:system-Hprime};  
(ii) evaluate $R/H^{2}=6(2+E)$;  
(iii) construct $(A,B)$ from \eqref{eq:Rprime-affine};  
(iv) compute $K'/K$ using \eqref{eq:KprimeK-closed}.  

This procedure is algebraic and stable.

\paragraph{Parameter priors and early–time safety.}
Positivity requires \(K(N)=1+\alpha R/M^{2}>0\) for all $N$ in the domain.  
Preserving standard pre-drag physics requires the UV suppression
\begin{equation}
\left|\frac{\alpha}{M^{2}}R(N)\right| \ll 1
\qquad (z\ge z_{\rm drag}),
\label{eq:alpha-prior}
\end{equation}
ensuring $K\to1$ before CMB/BAO anchoring \parencite{Aubourg2015,Planck2018cosmo}.
If $\alpha<0$, stability requires 
\(
\alpha > - M^{2}/\max_{N}R(N).
\)
The denominator in \eqref{eq:KprimeK-closed} must also remain nonzero.

\subsection{IR slow running (RG--like form)}
\label{subsec:K-phenom}

A second parameterization, useful for IR model-space exploration but not derived
from the UV action, consists of an exponential running:
\begin{equation}
\frac{d\ln|K-1|}{dN}=-p
\quad\Longrightarrow\quad
\boxed{
K(N)=1+K_{p}\,e^{-p(N-N_{p})}
}.
\end{equation}
Equivalently in redshift,
\begin{equation}
\boxed{
K(z)=1+K_{0}(1+z)^{p}
}
\qquad\Longleftrightarrow\qquad
\boxed{
K(N)=1+K_{0}\,e^{-pN}
},
\label{eq:K-phenom}
\end{equation}
with $K_{0}=K_{p}e^{pN_{p}}$.  
For this form,
\begin{equation}
\boxed{
K'=-p(K-1),\qquad
\frac{K'}{K}=-p\frac{K-1}{K}
}
\label{eq:Kprime-phenom}
\end{equation}
which enters the background system directly.
This two-parameter flow is the direct analogue of constant-$\beta$ RG flows
\parencite{WeinbergQFT2,PeskinSchroeder} and is similar in spirit to the CPL
form for $w(z)$ \parencite{ChevallierPolarski2001,Linder2003}.

\paragraph{Positivity, early-time behavior, interpretation.}
For \(p>0\), $K\to 1$ toward the future; for \(p<0\), the modification turns on at
late times.  
Ghost freedom requires
\begin{equation}
1+K_{0}e^{-pN}>0 \quad (\forall\,N\in[N_{\min},0]),
\label{eq:K-phenom-prior}
\end{equation}
and because the perturbation structure remains
that of the covariantly completed
\(P(X,\phi,\chi)=K(\chi)X-V(\phi)\)
evaluated in unitary gauge,
the model always satisfies 
\((\alpha_{B},\alpha_{M},\alpha_{T})=(0,0,0)\) and
$c_{s}^{2}=1$.

\subsection{Admissibility and stability}
\label{subsec:K-admissibility}

Ghost freedom requires
\begin{equation}
\boxed{ K(N)>0\quad\text{for all relevant }N. }
\end{equation}
No gradient instability arises because \(P_{,XX}=0\Rightarrow c_{s}^{2}=1\),
and the metric sector stays in GR with luminal tensors
\parencite{Abbott2017GW170817,Sakstein2017GW}.  
For the curvature-motivated case \eqref{eq:K-curv}, the denominator in
\eqref{eq:KprimeK-closed} must remain nonzero; for sufficiently small
\(|\alpha R/M^{2}|\), this is guaranteed and can be monitored in integration.

\FloatBarrier
\vspace{2\baselineskip}
\input{sections/section4/tables/tab_K_specs}
\FloatBarrier
\clearpage

%% file: sections/section4/tables/tab_K_specs.tex
\begin{table}[h]
\centering
\footnotesize
\caption{Background-only kinetic normalizations $K(N)$ used in Sec.~\ref{sec:K}. 
Each card lists the definition, the exact identity for $K'/K$, admissibility conditions, 
and the EFT--DE corner. The curvature-motivated form follows uniquely from the 
UV operator $R X/M^{2}$ derived in Appendix~\ref{appendix:foundations}; the 
RG-like form is an IR phenomenological parameterization.}
\label{tab:K_specs}

\begin{subtable}[t]{0.48\linewidth}
\centering
\caption{Curvature-motivated (UV-derived)}
\label{tab:K_specs_curv}
\setlength{\tabcolsep}{4pt}
\renewcommand{\arraystretch}{1.12}
\begin{tabularx}{\linewidth}{@{}p{0.30\linewidth} Y@{}}
\toprule
\textbf{Item} & \textbf{Expression / Notes} \\
\midrule
Definition &
$K=1+\alpha\,R/M^{2}$ \;(\cref{eq:K-curv}); UV origin: one-loop curvature--derivative operator $R X/M^{2}$ (Appendix~\ref{appendix:foundations}). 
Background identity $R/H^{2}=6(2+E)$ with $E=H'/H$.\\

$K'/K$ identity &
Closed, iteration-free algebraic form \;(\cref{eq:KprimeK-closed}); uses affine decomposition 
$R'/H^{2}=A+B\,K'$ with $B=3\phi'^2/M_{\rm pl}^2$ and $A$ from \cref{eq:Rprime-affine}.\\

Admissibility &
Require $K>0$ on $N\in[N_{\min},0]$ and non-vanishing denominator in \cref{eq:KprimeK-closed}. 
Early-time safety prior $|\alpha R/M^{2}|\ll 1$ before $z_{\rm drag}$ \;(\cref{eq:alpha-prior}).\\

Implementation &
Compute $E$ from \cref{eq:system-Hprime}, form $R/H^{2}=6(2+E)$, build $(A,B)$, compute $c$ and evaluate $K'/K$ directly via \cref{eq:KprimeK-closed}. 
No recursion in $K'$ or $R'$.\\

EFT--DE corner &
$(\alpha_B,\alpha_M,\alpha_T)=(0,0,0)$; $\alpha_K>0$ \;(\cref{eq:EFT-map}).\\
\bottomrule
\end{tabularx}
\end{subtable}
\hfill
\begin{subtable}[t]{0.48\linewidth}
\centering
\caption{IR slow running (phenomenological)}
\label{tab:K_specs_rg}
\setlength{\tabcolsep}{4pt}
\renewcommand{\arraystretch}{1.12}
\begin{tabularx}{\linewidth}{@{}p{0.30\linewidth} Y@{}}
\toprule
\textbf{Item} & \textbf{Expression / Notes} \\
\midrule
Definition &
$K(N)=1+K_{0}\,e^{-pN}$ \;(\cref{eq:K-phenom}); IR background parameterization (no UV derivation implied).\\

$K'/K$ identity &
Exact relations: $K'=-p\,(K-1)$, \quad $\displaystyle \frac{K'}{K}=-p\,\frac{K-1}{K}$ \;(\cref{eq:Kprime-phenom}).\\

Admissibility &
Require $K(N)>0$ on $[N_{\min},0]$ by \cref{eq:K-phenom-prior}. 
Monotonicity controlled by the sign of $p$.
No metric-sector effects; only background normalization of $X$ varies.\\

Implementation &
Evaluate $K(N)$ and $K'/K$ directly at each timestep; trivial numerically and guaranteed non-recursive.\\

EFT--DE corner &
$(\alpha_B,\alpha_M,\alpha_T)=(0,0,0)$; $\alpha_K>0$ \;(\cref{eq:EFT-map}).\\
\bottomrule
\end{tabularx}
\end{subtable}

\end{table}

%% file: sections/section5/section5.tex
\section{Linear perturbations at first order}
\label{sec:linear}

This section develops scalar--metric perturbations around the spatially flat FRW background defined earlier. All results follow from the effective background (unitary--gauge) action \eqref{eq:intro-action}, the stress--energy tensor \eqref{eq:Tmunu}, and the background relations in Secs.~\ref{sec:covariant-core}--\ref{sec:efold}, working strictly to linear order in perturbations.

\paragraph{Conventions.}
Overdots denote $d/dt$. Unless explicitly noted, a prime in this section denotes $d/d\eta$ with $\eta$ the conformal time; in the growth subsection (Sec.~\ref{subsec:growth}) a prime denotes $d/d\ln a$. The Fourier convention is
$f(\mathbf{x})=\int \!\frac{d^{3}k}{(2\pi)^{3}}\,e^{i\mathbf{k}\cdot\mathbf{x}}\,f(\mathbf{k})$,
so $\nabla^{2}\!\to\!-k^{2}$.

\subsection{Newtonian gauge and primary variables}
\label{subsec:newtonian}

Work in Newtonian gauge,
\begin{equation}
ds^{2}=-(1+2\Psi)\,dt^{2}+a^{2}(t)\,(1-2\Phi)\,d\vec{x}^{2},
\qquad
\phi(t,\vec x)=\phi_{0}(t)+\delta\phi(t,\vec x).
\label{eq:metric-Newtonian}
\end{equation}
Because the metric sector is Einstein--Hilbert (no non-minimal coupling $F(\phi)R$ and no braiding $G(\phi,X)\Box\phi$), the scalar sector contributes no anisotropic stress at linear order for $P(X,\phi,t)$ models \parencite{MaBertschinger1995,DeFeliceTsujikawa2010}. Neglecting late-time photon/neutrino shear on subhorizon scales therefore gives
\begin{equation}
\Phi=\Psi.
\label{eq:PhiEqPsi}
\end{equation}

\subsection{Perturbed Klein--Gordon equation (derived)}
\label{subsec:KG-perturbed}

Starting from the covariant equation \eqref{eq:KG-cov},
$K\,\Box\phi+(\nabla_\mu K)\nabla^\mu\phi-V_{,\phi}=0$,
and expanding to first order, note that $K=K(t)\Rightarrow \delta K=0$ and $(\nabla_\mu K)\nabla^\mu\phi=-\dot K\,\dot\phi$ at all orders. Using the standard expansion of $\Box\phi$ in Newtonian gauge (e.g., \parencite{MaBertschinger1995}) and subtracting the background equation \eqref{eq:KG-FRW} yields, in \emph{cosmic time},
\begin{align}
K\!\left(\delta\ddot\phi+3H\,\delta\dot\phi+\frac{k^{2}}{a^{2}}\delta\phi\right)
+\dot K\,\delta\dot\phi
+V_{,\phi\phi}\,\delta\phi
\;=\;
4K\,\dot\phi_{0}\,\dot\Phi
-2\,V_{,\phi}\,\Phi.
\label{eq:deltaKG-Newtonian-correct}
\end{align}
No additional perturbative operators arise relative to canonical quintessence because $\delta K=0$. The \emph{conformal-time} form follows by $d/dt=a^{-1}d/d\eta$ and $\mathcal{H}\equiv aH$:
\begin{equation}
K\!\left(\delta\phi''+2\mathcal{H}\,\delta\phi'+k^{2}\delta\phi\right)
+a^{2}V_{,\phi\phi}\,\delta\phi
\;=\;
4K\,\phi_{0}'\,\Phi'
-2a^{2}V_{,\phi}\,\Phi.
\label{eq:deltaKG-conformal}
\end{equation}
For completeness, the scalar-fluid perturbations read (cf.\ \parencite{MaBertschinger1995})
\begin{align}
\delta\rho_\phi &= K\left(\dot\phi_{0}\,\delta\dot\phi-\dot\phi_{0}^{2}\,\Phi\right)+V_{,\phi}\,\delta\phi, \\
\delta p_\phi   &= K\left(\dot\phi_{0}\,\delta\dot\phi-\dot\phi_{0}^{2}\,\Phi\right)-V_{,\phi}\,\delta\phi, \\
(\rho_\phi+p_\phi)\,\theta_\phi &= K\,\dot\phi_{0}\,\frac{k^{2}}{a^{2}}\,\delta\phi,
\label{eq:scalar-fluid-pert}
\end{align}
and, because $P_{,XX}=0$ for $P=K(t)X-V(\phi)$, the rest-frame sound speed is $c_s^2=1$ with vanishing intrinsic non-adiabatic pressure \parencite{GarrigaMukhanov1999}.

\subsection{Quadratic action and Mukhanov--Sasaki equation}
\label{subsec:MS-eq}

Define the gauge-invariant Mukhanov--Sasaki (MS) variable and pump field
\begin{equation}
v \;\equiv\; a\!\left(\delta\phi+\frac{\dot\phi_{0}}{H}\Phi\right),
\qquad
z^{2} \;\equiv\; a^{2}\frac{K\,\dot\phi_{0}^{2}}{H^{2}}.
\label{eq:v-and-z}
\end{equation}
Expanding \eqref{eq:intro-action} to second order in scalar perturbations and integrating out nondynamical constraints yields the canonical MS form
\begin{equation}
S^{(2)}=\frac{1}{2}\int d\eta\,d^{3}\!x\left[(v')^{2}-(\nabla v)^{2}+\frac{z''}{z}\,v^{2}\right],
\label{eq:S2-MS}
\end{equation}
whose variation gives
\begin{equation}
v''+\left(c_{s}^{2}k^{2}-\frac{z''}{z}\right)v=0,
\qquad
c_s^2=\frac{P_{,X}}{P_{,X}+2X P_{,XX}}=1,
\label{eq:MS-eq}
\end{equation}
with the last equality exact for $P=K(t)X-V(\phi)$ \parencite{GarrigaMukhanov1999,Mukhanov2005,WeinbergCosmo2008}.

\paragraph{Super- and subhorizon behavior.}
On superhorizon scales ($k\!\to\!0$), the comoving curvature $\mathcal{R}\equiv v/z$ is conserved at leading order,
\begin{equation}
\mathcal{R}' \;=\; \mathcal{O}\!\left(k^{2}\right),
\label{eq:R-conservation}
\end{equation}
which in Newtonian gauge implies the adiabatic relation
\begin{equation}
\delta\phi \;\simeq\; -\,\frac{\dot\phi_{0}}{H}\,\Phi
\qquad (k\ll aH),
\label{eq:adiabatic-IC}
\end{equation}
useful for setting initial conditions \parencite{Mukhanov2005,WeinbergCosmo2008}. On subhorizon scales ($k\gg aH$), $c_s^2=1$ ensures pressure support and $\delta\phi$ oscillates on the sound horizon; the scalar does not cluster appreciably.

\subsection{Einstein constraints and the Poisson equation}
\label{subsec:Einstein-constraints}

The linearized GR constraints in Newtonian gauge (e.g., \parencite{MaBertschinger1995}) are
\begin{align}
&\text{$00$:}\quad
-3H(\dot\Phi+H\Phi)-\frac{k^{2}}{a^{2}}\Phi
=4\pi G\left(\delta\rho_m+\delta\rho_r+\delta\rho_\phi\right),
\label{eq:Ein00}\\
&\text{$0i$:}\quad
\dot\Phi+H\Phi = 4\pi G\left[(\rho_m+p_m)\theta_m+(\rho_r+p_r)\theta_r+(\rho_\phi+p_\phi)\theta_\phi\right].
\label{eq:Ein0i}
\end{align}
With $\Phi=\Psi$ and negligible radiation shear at late times, the subhorizon limit ($k\gg aH$) reduces to the GR Poisson equation
\begin{equation}
-\frac{k^{2}}{a^{2}}\,\Phi \;=\; 4\pi G\,\rho_{m}\,\delta_{m},
\label{eq:Poisson-GR}
\end{equation}
i.e., no modification of the gravitational coupling and no slip. In the EFT--DE language this corresponds to
\begin{equation}
\mu(a,k)=1,\qquad \Sigma(a,k)=1,\qquad \eta(a,k)\equiv\frac{\Phi}{\Psi}-1=0
\quad\text{(linear order, all $k$).}
\label{eq:mu-sigma-eta}
\end{equation}

\subsection{Growth of matter perturbations (GR form)}
\label{subsec:growth}

Let $D(a)$ be the linear growth factor of matter overdensities in the subhorizon GR limit. Combining \eqref{eq:Poisson-GR} with mass conservation and the Euler equation for pressureless matter gives
\begin{equation}
D''+\left(2+\frac{H'}{H}\right)D'-\frac{3}{2}\,\Omega_{m}(a)\,D=0,
\label{eq:growth-equation}
\end{equation}
where here a prime denotes $d/d\ln a$, and $\Omega_{m}(a)\equiv \rho_{m}/(3M_{\rm pl}^{2}H^{2})$. In QKDE, all linear-level signatures enter \eqref{eq:growth-equation} only through the background $H(a)$ determined by Sec.~\ref{sec:efold}; no scale dependence is induced.

\medskip
\noindent\textbf{Key takeaway.} The scalar mode is luminal and ghost-free ($c_s^2=1$, $K>0$), the metric potentials satisfy $\Phi=\Psi$, and the GR Poisson equation holds on linear/subhorizon scales. Consequently, late-time linear observables (RSD, weak lensing, ISW) are affected only through the background expansion $H(a)$ and the induced growth history $D(a)$.

%% file: sections/section6/section6.tex
\section{EFT--DE mapping, linear viability, and relations to existing theories}
\label{sec:eft}

This section records the Effective Field Theory of Dark Energy (EFT--DE) map of the effective background (unitary--gauge) action \eqref{eq:intro-action}, states linear-viability conditions, and situates the construction within the broader landscape of dark-energy and modified-gravity frameworks. All statements refer to the background and first-order perturbative level defined in Secs.~\ref{sec:covariant-core}--\ref{sec:linear}, with total diffeomorphism invariance understood to be restored in the covariant completion prior to gauge fixing.

\subsection{Bellini--Sawicki functions (derivation and specialization)}
\label{subsec:BS-map}

The EFT--DE description encodes linear dynamics in the Bellini--Sawicki functions $\{\alpha_K,\alpha_B,\alpha_M,\alpha_T\}$ and an effective Planck mass $M_*^2$ \parencite{Gubitosi2013,Bloomfield2013,BelliniSawicki2014}. For minimally coupled covariantly completed $P(X,\phi,\chi)$ models (no $F(\phi)R$, no braiding), the general result reads
\begin{equation}
\alpha_K
=\frac{2X\!\left(P_{,X}+2X P_{,XX}\right)}{H^{2}M_*^{2}},
\qquad
\alpha_B=0,\qquad
\alpha_M\equiv \frac{d\ln M_*^{2}}{d\ln a}=0,\qquad
\alpha_T=0,
\label{eq:alphas-general}
\end{equation}
with $M_*^{2}=M_{\rm pl}^{2}$ in the Einstein--Hilbert metric sector. Specializing to
\begin{equation*}
P(X,\phi,\chi)=K(\chi)\,X - V(\phi),
\end{equation*}
and then fixing unitary gauge $\chi=t$, gives $P_{,X}=K(t)$ and $P_{,XX}=0$, hence
\begin{equation}
\boxed{\;
\alpha_K=\frac{2X\,K}{H^{2}M_{\rm pl}^{2}}=\frac{K\,\dot\phi^{2}}{H^{2}M_{\rm pl}^{2}}\ge 0,\qquad
\alpha_B=\alpha_M=\alpha_T=0
\;}
\label{eq:eft-map-alpha}
\end{equation}
and $\alpha_H=0$ (no beyond-Horndeski operator). Using $X=\tfrac12 H^{2}\phi'^2$ (Sec.~\ref{sec:efold}) also yields $\alpha_K=K\,\phi'^2/M_{\rm pl}^{2}$ in $N\equiv\ln a$. This corner saturates the GW170817 constraint by construction through $\alpha_T=0$ and $M_*^2=\mathrm{const}$ \parencite{Abbott2017GW170817,Sakstein2017GW}.

\subsection{Phenomenological response functions \texorpdfstring{$\mu$}{mu}, \texorpdfstring{$\Sigma$}{Sigma}, and slip}
\label{subsec:mu-sigma}

In the quasi-static, subhorizon regime the linear response to matter is often summarized by \parencite{Joyce2015}
\begin{align}
-\,\frac{k^{2}}{a^{2}}\Psi \;&=\; 4\pi G\,\mu(a,k)\,\rho_{m}\,\Delta_{m},
\label{eq:def-mu}\\
-\,\frac{k^{2}}{a^{2}}\big(\Phi+\Psi\big) \;&=\; 8\pi G\,\Sigma(a,k)\,\rho_{m}\,\Delta_{m},
\label{eq:def-sigma}
\end{align}
with $\Delta_m$ the comoving-gauge matter perturbation. For \eqref{eq:eft-map-alpha} together with $c_s^2=1$ (Sec.~\ref{sec:linear}), the linearized Einstein constraints reduce exactly to GR at all $k$:
\begin{equation}
\boxed{\;\mu(a,k)=1,\qquad \Sigma(a,k)=1,\qquad \eta(a,k)\equiv \frac{\Phi}{\Psi}-1=0\;}\,,
\label{eq:mu-sigma-eta}
\end{equation}
up to negligible late-time shear from standard relativistic species. This result relies on the Einstein--Hilbert metric sector together with total energy--momentum conservation in the covariant completion. As a result, RSD, WL, and ISW respond solely through the background $H(a)$ entering the GR growth equation \eqref{eq:growth-equation}.

\subsection{Linear stability and admissibility (scalar and tensor)}
\label{subsec:stability-conditions}

Ghost freedom of the scalar mode requires positivity of the kinetic coefficient:
\begin{equation}
\boxed{\,K(t)>0 \quad\Rightarrow\quad \alpha_K \ge 0\ \ \text{(equality when $\phi'=0$)}\,}.
\label{eq:ghost-free}
\end{equation}
Because $P_{,XX}=0$ for $P(X,\phi,\chi)=K(\chi)X-V(\phi)$ evaluated in unitary gauge, the sound speed is luminal,
\begin{equation}
c_{s}^{2}=\frac{P_{,X}}{P_{,X}+2X P_{,XX}}=1,
\end{equation}
so no gradient (Laplace) instability arises \parencite{GarrigaMukhanov1999}. The tensor sector remains Einstein--Hilbert with $\alpha_T=0$ and $M_*^2=\mathrm{const}$ ($\alpha_M=0$), satisfying GW170817/GRB limits \parencite{Abbott2017GW170817,Sakstein2017GW}. Background well-posedness additionally requires $H>0$ and, for the curvature-motivated $K$ in Sec.~\ref{sec:K}, a non-vanishing denominator in \eqref{eq:KprimeK-closed}.

\begin{table}[t]
\centering
\caption{EFT--DE and phenomenological functions for the QKDE construction (linear order).}
\label{tab:eft-summary}
\begin{tabular}{lcccccc}
\toprule
 & $\alpha_K$ & $\alpha_B$ & $\alpha_M$ & $\alpha_T$ & $\mu$ & $\Sigma$ \\
\midrule
QKDE & $K\dot\phi^{2}/(H^{2}M_{\rm pl}^{2})\ \ge 0$ & $0$ & $0$ & $0$ & $1$ & $1$ \\
\bottomrule
\end{tabular}
\end{table}

\subsection{Relations to existing dark-energy and modified-gravity frameworks}
\label{sec:relations}

The location in theory space is clarified by comparing covariantly completed operators, EFT maps, and linear phenomenology.

\subsubsection{Canonical quintessence}
\label{subsec:relations-quintessence}
Setting $K\equiv 1$ in \eqref{eq:intro-action} reproduces canonical quintessence \parencite{RatraPeebles1988,Caldwell1998}. Then $\alpha_{K}=\dot\phi^{2}/(H^{2}M_{\rm pl}^{2})$ and $\alpha_{B}=\alpha_{M}=\alpha_{T}=0$. The present construction differs only by a \emph{background} time dependence in the kinetic normalization, appearing as the friction $\dot K\,\dot\phi$ in \eqref{eq:KG-FRW} and as $z^{2}=a^{2}K\dot\phi^{2}/H^{2}$ in \eqref{eq:v-and-z}. Linear gravitational relations remain GR (Sec.~\ref{sec:linear}).

\subsubsection{General k-essence}
\label{subsec:relations-kessence}
k-essence permits $P(X,\phi)$ with $P_{,XX}\neq 0$ \parencite{ArmendarizPicon2001,GarrigaMukhanov1999}. The present theory corresponds to the \emph{linear-$X$} subclass $P(X,\phi,\chi)=K(\chi)X-V(\phi)$ evaluated in unitary gauge, for which
\begin{equation}
P_{,X}=K(t),\qquad P_{,XX}=0 \ \Longrightarrow\quad c_{s}^{2}=1.
\end{equation}
Consequently no low-sound-speed clustering or scale-dependent growth is generated at linear order. Since $K$ depends only on time after gauge fixing, the EFT--DE language \parencite{Gubitosi2013,Bloomfield2013,BelliniSawicki2014} is the natural description.

\subsubsection{Horndeski and beyond-Horndeski}
\label{subsec:relations-horndeski}
In the Bellini--Sawicki parametrization \parencite{BelliniSawicki2014}, the construction sits at the conservative corner
\begin{equation}
\alpha_{K}>0,\qquad \alpha_{B}=\alpha_{M}=\alpha_{T}=0 \quad (\alpha_{H}=0),
\end{equation}
i.e., no braiding, no Planck-mass running, and luminal tensors—consistent with GW170817 \parencite{Abbott2017GW170817,Sakstein2017GW}. Many Horndeski/Galileon models feature $\alpha_{B}\neq 0$ and/or $\alpha_{M}\neq 0$ and are tightly constrained post-GW170817 \parencite{Kobayashi2019,Joyce2015}. GLPV/DHOST adds $\alpha_{H}\neq 0$ \parencite{Gleyzes2015}; this is absent here.

\subsubsection{\texorpdfstring{$f(R)$}{f(R)} gravity (Jordan-frame scalar--tensor)}
\label{subsec:relations-fR}
Metric $f(R)$ gravity is equivalent to a scalar--tensor theory with $F(\phi)R$, implying $\alpha_{M}\neq 0$ and typically $\alpha_{B}\neq 0$ \parencite{SotiriouFaraoni2010,DeFeliceTsujikawa2010}. Linear phenomenology departs from GR via $\mu\neq 1$ and $\Sigma\neq 1$. The present model keeps the metric sector strictly Einstein--Hilbert, hence $\alpha_{M}=\alpha_{B}=\alpha_{T}=0$ and $\mu=\Sigma=1$.

\subsubsection{Interacting dark energy / fifth-force models}
\label{subsec:relations-coupled}
Dark-matter--dark-energy couplings modify the matter continuity/Euler equations and produce effective fifth forces \parencite{Clifton2012,Joyce2015}. Here matter and radiation are minimally coupled [Eqs.~\eqref{eq:mr-continuity}], and the Poisson equation is purely GR [Eq.~\eqref{eq:Poisson-GR}]. Although the scalar exchanges energy with the clock/time-slicing sector in unitary gauge, the matter sector remains separately conserved. Any detection of scale-dependent $\mu(a,k)\neq 1$ or gravitational slip $\eta\neq 0$ at linear order would falsify this baseline.

\subsubsection{Phantom regimes and energy conditions}
\label{subsec:relations-phantom}
With $K>0$, the scalar obeys $\rho_{\phi}+p_{\phi}=K\dot\phi^{2}\ge 0$, enforcing $w_{\phi}\ge -1$ (from \eqref{eq:rho-p-w}). Phantom behavior would require $K<0$ and thus a ghost, which is excluded by \eqref{eq:ghost-free}.

\subsubsection{Inflationary EFT analogy}
\label{subsec:relations-inflation-EFT}
Time-dependent coefficients in the EFT of Inflation provide a useful analogy \parencite{Cheung2008}: a single running background coefficient multiplies the kinetic operator while the metric sector remains GR. Here this reduces to $z^{2}=a^{2}K\dot\phi^{2}/H^{2}$ with $c_{s}^{2}=1$ [Eqs.~\eqref{eq:v-and-z}--\eqref{eq:MS-eq}].

\subsubsection{Observational degeneracies and falsifiability}
\label{subsec:relations-observational}
Because linear phenomenology is GR-like, late-time background histories $H(a)$ are degenerate with those obtained from smooth $w(z)$ quintessence. The combination of \emph{(i)} $\mu=\Sigma=1$, $\eta=0$, \emph{(ii)} luminal tensors with constant $M_{\rm pl}$, and \emph{(iii)} a single running parameter $\alpha_{K}\ge 0$ is distinctive and falsifiable: evidence for $\mu\neq 1$, $\Sigma\neq 1$, or $c_T^2\neq 1$ at late times would rule out this framework.

\medskip
\noindent\textbf{Key takeaway.} At linear order the construction occupies the most conservative EFT--DE corner:
$\Phi=\Psi$, $\mu=\Sigma=1$, $c_s^2=c_T^2=1$, $M_{\rm pl}=\mathrm{const}$, with the only running
parameter $\alpha_K\ge 0$ set by the background kinetic normalization $K(\chi)$ evaluated in unitary gauge.

\input{sections/section6/tables/tab_eft_summary}
\FloatBarrier

\input{sections/section6/figures/fig_theory_space_taxonomy}
\FloatBarrier
\clearpage

%% file: sections/section6/tables/tab_eft_summary.tex
\begin{table}[h]
\centering
\footnotesize
\setlength{\tabcolsep}{3.8pt}
\renewcommand{\arraystretch}{1.15}
\caption{EFT--DE (Bellini--Sawicki) parameters and linear phenomenology (quasi-static, subhorizon).
Entries marked “$\neq 0$ (generic)” are typically nonzero but model dependent; “$=0$” indicates identically zero within the class.
All entries for QKDE are evaluated in unitary gauge after covariant completion, i.e.\ $K\equiv K(\chi)\vert_{\chi=t}$.}
\label{tab:eft_summary}
\begin{tabularx}{\linewidth}{@{}p{0.36\linewidth}*{6}{>{\centering\arraybackslash}X}@{}}
\toprule
\textbf{Model / Class} & $\boldsymbol{\alpha_K}$ & $\boldsymbol{\alpha_B}$ & $\boldsymbol{\alpha_M}$ & $\boldsymbol{\alpha_T}$ & $c_s^{2}$ & $(\mu,\Sigma)$ \\
\midrule
QKDE (this work) &
$\dfrac{K(\chi)\vert_{\chi=t}}\,\dot\phi^{2}{H^{2}M_{\rm pl}^{2}}>0$ & $0$ & $0$ & $0$ & $1$ & $(1,1)$ \\
Canonical quintessence &
$\dfrac{\dot\phi^{2}}{H^{2}M_{\rm pl}^{2}}>0$ & $0$ & $0$ & $0$ & $1$ & $(1,1)$ \\
Minimally coupled k\mbox{-}essence ($P(X,\phi)$) &
$>0$ & $0$ & $0$ & $0$ & $\neq 1$ (general) & $(1,1)$ \\
$f(R)$ gravity (metric) &
model dep. & $\neq 0$ (generic) & $\neq 0$ (generic) & $0^{\dagger}$ & model dep. & $(\neq 1,\neq 1)$ \\
Horndeski (generic) &
$>0$ & $\neq 0$ (generic) & $\neq 0$ (generic) & $0^{\dagger}$ & model dep. & model dep. \\
Beyond\mbox{-}Horndeski (DHOST) &
$>0$ & $\neq 0$ (generic) & $\neq 0$ (generic) & $0^{\dagger}$ & model dep. & model dep. \\
\bottomrule
\end{tabularx}

\vspace{0.35em}
\raggedright\footnotesize $^{\dagger}$ Viable late\mbox{-}time models consistent with GW170817 enforce $\alpha_T\simeq 0$.\par
\end{table}

%% file: sections/section6/figures/fig_theory_space_taxonomy.tex
\begin{figure}[t]
\centering

\begin{adjustbox}{max width=\linewidth,center}
\begin{tikzpicture}[
    >=Latex,
    scale=0.95,
    every node/.style={transform shape},
    box/.style={
        rectangle, rounded corners=3pt,
        draw=black, thick, fill=white,
        inner sep=3pt, text width=2.10cm,
        align=center, font=\footnotesize
    }
]

\node[box, text width=3.0cm] (root) at (0,1.2)
    {\textbf{Single-Field\\Dark-Energy Theories}};

\node[box, text width=3.4cm] (metricmod) at (-5.6,-2.8)
    {\textbf{Metric sector modified}\\
     ($\alpha_{M},\alpha_{B},\alpha_{T},\alpha_{H}\!\neq\!0$)};

\node[box, text width=2.4cm] (interacting) at (0,-2.4)
    {\textbf{Matter\\interactions}\\($Q^\mu \neq 0$)};

\node[box, text width=3.4cm] (metricGR) at (5.6,-2.8)
    {\textbf{Metric sector = GR}\\
     ($\alpha_{M}=\alpha_{B}=\alpha_{T}=\alpha_{H}=0$)};

\node[box, text width=2.5cm] (IDE) at (0,-5.1)
    {\textbf{Interacting\\dark energy}};

\node[box, text width=2.3cm] (canon) at (4.1,-8.8)
    {\textbf{Canonical\\scalars}\\$P(X,\phi)$\\$c_s^2=1$};

\node[box, text width=2.3cm] (kess) at (6.9,-8.8)
    {\textbf{k-essence}\\($P_{,XX}\!\neq\!0$,\\$c_s^2\!\neq\!1$)};

\node[box, text width=2.3cm] (quint) at (3.0,-12.1)
    {\textbf{Quintessence}\\$K=\mathrm{const}$\\$P=X-V(\phi)$};

\node[box, text width=2.6cm] (qkde) at (6.2,-12.1)
    {\textbf{QKDE}\\
     \emph{background-only} $K(\chi)$\\
     (unitary gauge $\chi=t$)\\
     e.g.\ $1+\alpha\,R/M^2$ \textit{or}\\ $1+K_{0}e^{-pN}$};

\node[box, text width=2.3cm] (horndeski) at (-8.4,-8.8)
    {\textbf{Horndeski}};

\node[box, text width=2.3cm] (beyond) at (-5.6,-8.8)
    {\textbf{Beyond\\Horndeski}\\(DHOST)};

\node[box, text width=2.3cm] (mg) at (-2.8,-8.8)
    {\textbf{Modified\\gravity}\\($f(R)$, DGP)};

\draw[->, thick] (root.south) -- (metricmod.north);
\draw[->, thick] (root.south) -- (interacting.north);
\draw[->, thick] (root.south) -- (metricGR.north);

\draw[->, thick] (metricmod.south) -- (horndeski.north);
\draw[->, thick] (metricmod.south) -- (beyond.north);
\draw[->, thick] (metricmod.south) -- (mg.north);

\draw[->, thick] (interacting.south) -- (IDE.north);

\draw[->, thick] (metricGR.south) -- (canon.north);
\draw[->, thick] (metricGR.south) -- (kess.north);

\draw[->, thick] (canon.south) -- (quint.north);
\draw[->, thick] (canon.south) -- (qkde.north);

\end{tikzpicture}
\end{adjustbox}

\caption{Theory--space taxonomy of single-field dark energy. Models split into three branches:
modified metric (generic $\alpha$'s nonzero), matter interaction ($Q^\mu\!\neq\!0$), and
GR-preserving ($\alpha_M=\alpha_B=\alpha_T=\alpha_H=0$).
QKDE lies in the GR-preserving, canonical-scalar branch and is specified by a
covariantly completed kinetic normalization $K(\chi)$, evaluated in unitary gauge $\chi=t$ as a background-only function $K(t)$,
introducing no metric operators.
Consequently \(\Phi=\Psi\), \(c_s^2=c_T^2=1\), and \((\mu,\Sigma)=(1,1)\); quintessence is recovered for \(K=\mathrm{const}\),
while generic k-essence may have \(c_s^2\!\neq\!1\).}
\label{fig:theory_space_taxonomy}
\end{figure}
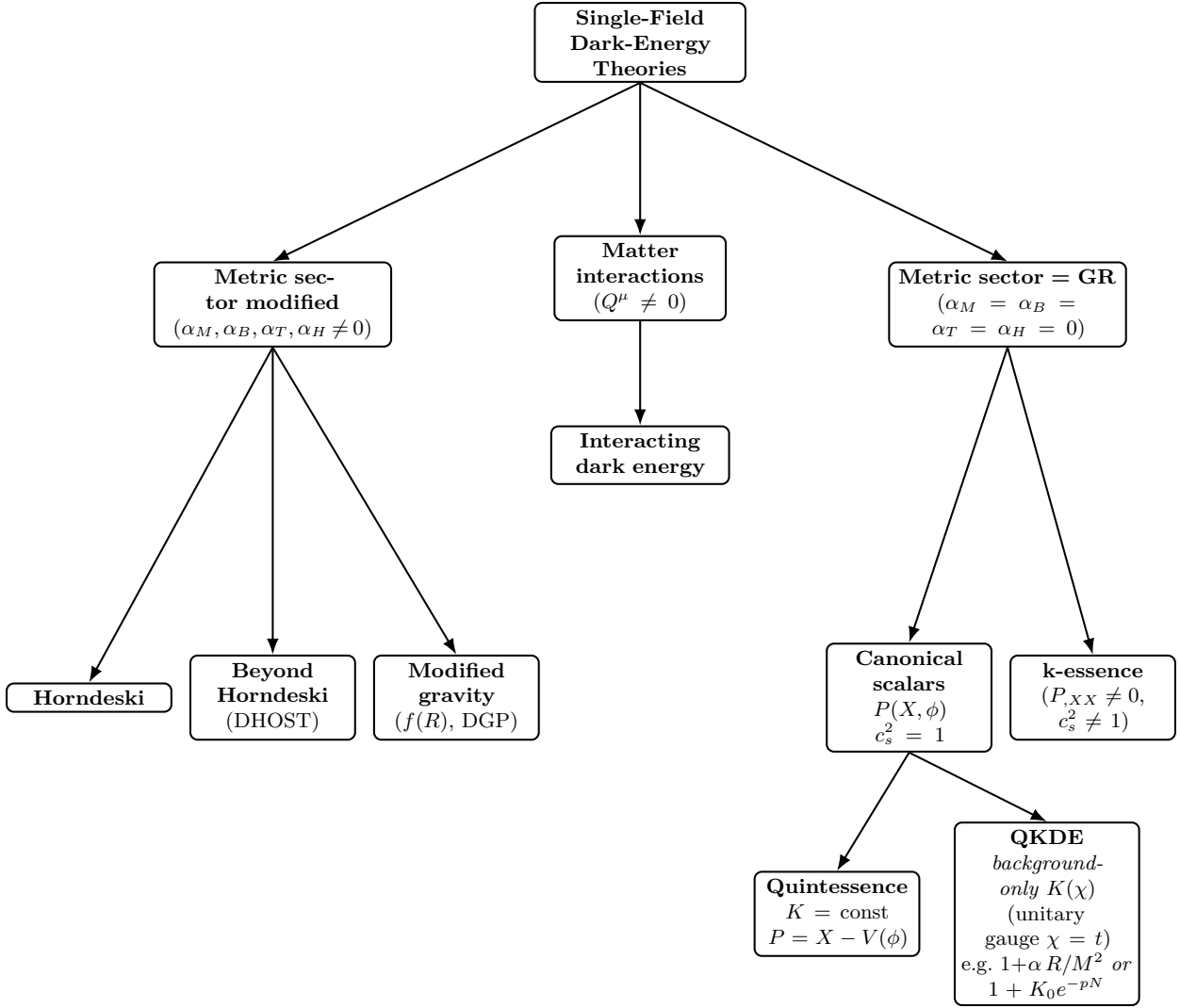

%% file: sections/section7/section7.tex
\section{Forecast-ready observables}
\label{sec:observables}

This section collects late-time geometric and growth observables in a spatially
flat FRW background, using units with $c=1$. All expressions descend from the
closed background system in Sec.~\ref{sec:efold} and the GR-consistent linear
relations in Secs.~\ref{sec:linear}--\ref{sec:eft} (namely $\mu=\Sigma=1$ and
$\Phi=\Psi$). Definitions follow standard cosmology conventions
\parencite{Hogg1999}. To avoid collision with the notation
$E\equiv H'/H$, the normalized Hubble rate is denoted
\begin{equation}
\mathcal{E}(z)\equiv \frac{H(z)}{H_{0}}.
\end{equation}

\subsection{Comoving distance and standard rulers}
\label{subsec:distances}

The line-of-sight comoving distance is
\begin{equation}
\chi(z)
=\int_{0}^{z}\frac{dz'}{H(z')}
=\frac{1}{H_{0}}\int_{0}^{z}\frac{dz'}{\mathcal{E}(z')},
\qquad
\frac{d\chi}{dz}=\frac{1}{H(z)}.
\label{eq:chi}
\end{equation}
For spatial flatness, the angular-diameter and luminosity distances read
\begin{equation}
D_{A}(z)=\frac{\chi(z)}{1+z},
\qquad
D_{L}(z)=(1+z)\,\chi(z),
\label{eq:DA-DL}
\end{equation}
with derivatives
\begin{equation}
\frac{dD_A}{dz}=\frac{1}{1+z}\!\left[\frac{1}{H(z)}-D_A(z)\right],
\qquad
\frac{dD_L}{dz}=\frac{1+z}{H(z)}+\frac{D_L(z)}{1+z}.
\end{equation}
The supernova distance modulus is
\begin{equation}
\mu(z)=5\log_{10}\!\left[\frac{D_L(z)}{\mathrm{Mpc}}\right]+25.
\label{eq:mu-SN}
\end{equation}

Baryon acoustic oscillation (BAO) summaries use the spherically averaged
Eisenstein distance \parencite{Eisenstein2005} and the Alcock--Paczynski
anisotropy \parencite{AlcockPaczynski1979}:
\begin{equation}
D_{V}(z)=\Big[(1+z)^{2}D_{A}^{2}(z)\,\frac{z}{H(z)}\Big]^{1/3},
\qquad
F_{\rm AP}(z)=(1+z)\,D_{A}(z)\,H^{-1}(z).
\label{eq:DV-FAP}
\end{equation}
The comoving volume element is
\begin{equation}
\frac{dV_{c}}{dz\,d\Omega}=\frac{\chi^{2}(z)}{H(z)}.
\label{eq:vol-element}
\end{equation}

\subsection{Sound horizon (definition)}
\label{subsec:rd}

When a pre-recombination standard ruler is required, the sound horizon at
baryon drag $z_{d}$ is \parencite{HuSugiyama1996,EisensteinHu1998}
\begin{equation}
r_{d}
=\int_{z_{d}}^{\infty}\frac{c_{s}(z)}{H(z)}\,dz
=\int_{0}^{a_{d}}\frac{c_{s}(a)}{a^{2}H(a)}\,da,
\qquad
c_{s}(a)=\frac{1}{\sqrt{3\,(1+R_{b}(a))}},
\qquad
R_{b}(a)=\frac{3\rho_{b}(a)}{4\rho_{\gamma}(a)}.
\label{eq:rd}
\end{equation}
In QKDE, $r_{d}$ remains unchanged unless $K(N)$ departs from unity at
$z\gtrsim z_{d}$ (not assumed in this work).

\subsection{Linear growth and redshift-space distortions}
\label{subsec:growth-obs}

Let $D(a)$ denote the growing-mode solution of the GR growth equation
(cf.\ Sec.~\ref{sec:linear}, \parencite{Peebles1980}):
\begin{equation}
D'' + \left(2+\frac{H'}{H}\right)D' - \frac{3}{2}\,\Omega_{m}(a)\,D = 0,
\qquad
\Omega_m(a)=\frac{\rho_m}{3M_{\rm pl}^{2}H^{2}}.
\label{eq:growth-eq-recap}
\end{equation}
A convenient normalization is $D(a\!\to\!0)\propto a$ and $D(1)=1$.  
The logarithmic growth rate and the RSD summary observable are
\begin{equation}
f(a)\equiv \frac{d\ln D}{d\ln a},
\qquad
f\sigma_{8}(z)=f(z)\,\sigma_{8,0}\,\frac{D(z)}{D(0)}.
\label{eq:fs8}
\end{equation}
In the linear regime,
\begin{equation}
P(k,z)=D^{2}(z)\,P(k,0).
\label{eq:Pk-scaling}
\end{equation}
Because $\mu=\Sigma=1$ (Sec.~\ref{sec:eft}), no scale dependence is induced in
the growth equation. As a diagnostic, the GR growth-index approximation
$f\simeq\Omega_m(a)^{\gamma}$ with $\gamma\simeq 0.55$
\parencite{Linder2005} is often monitored.

\subsection{Weak-lensing kernels (linear)}
\label{subsec:WL}

Under the Limber approximation in a flat universe
\parencite{Limber1954,LoVerdeAfshordi2008,BartelmannSchneider2001},
\begin{equation}
C_{\ell}^{\kappa}
=\int_{0}^{\infty} \frac{dz}{H(z)}\,\frac{W^{2}(z)}{\chi^{2}(z)}\;
P_{\delta}\!\left(k=\frac{\ell+1/2}{\chi(z)},z\right),
\label{eq:Cl-kappa}
\end{equation}
with kernel
\begin{equation}
W(z)
=\frac{3}{2}\,\Omega_{m0}\,H_{0}^{2}\,(1+z)\,\chi(z)
\int_{z}^{\infty}dz_{s}\,n_{s}(z_{s})\,
\frac{\chi(z_{s})-\chi(z)}{\chi(z_{s})}.
\label{eq:W-lensing}
\end{equation}
Because $\mu=\Sigma=1$, lensing is modified only by the background through
$H(z)$ and $D(z)$.

\subsection{ISW and background sensitivity}
\label{subsec:ISW}

The late-time ISW contribution is \parencite{SachsWolfe1967}
\begin{equation}
\left(\frac{\Delta T}{T}\right)_{\rm ISW}
=\int_{\eta_{\ast}}^{\eta_{0}} d\eta\; (\Phi'+\Psi'),
\label{eq:ISW}
\end{equation}
which responds to the decay of the Bardeen potentials as dark energy
dominates. With $\Phi=\Psi$ and $\mu=\Sigma=1$, QKDE affects the ISW effect
only through changes in the background expansion.

\subsection{Illustrative dependence of forecast observables on $K(N)$}
\label{subsec:illustrative-K-dependence}

Although no numerical values or data fits are introduced, it is useful to
record how a mild deviation in the kinetic normalization propagates into the
principal forecast observables. Let
\begin{equation}
K(N)=1+\varepsilon\,\mathcal{F}(N),
\qquad |\varepsilon|\ll 1,
\end{equation}
with $\mathcal{F}(N)$ a smooth, order-unity profile. Differentiating each
observable with respect to $\varepsilon$ at fixed cosmological parameters yields
schematic responses of the form
\begin{equation}
\frac{\partial \ln H}{\partial \varepsilon},
\qquad
\frac{\partial \ln \chi}{\partial \varepsilon},
\qquad
\frac{\partial \ln D}{\partial \varepsilon},
\qquad
\frac{\partial \ln f\sigma_{8}}{\partial \varepsilon},
\end{equation}
each obtained directly from the variational system in
Sec.~\ref{app:variational-derivation}. These derivatives demonstrate that all
late-time sensitivities propagate solely through the background system and the
GR-consistent growth equation; no modified-gravity terms enter because
$\mu=\Sigma=1$.

For visualization, a schematic diagram (not tied to parameter choices and not
compared to data) may display representative fractional shifts such as
$\Delta H/H$ or $\Delta D/D$ generated by a small perturbation in $K(N)$.  
Such plots highlight the observational pathways of QKDE while avoiding any
implication of parameter inference.

\subsection{Inputs needed for forecasts and diagnostics}
\label{subsec:inputs-forecasts}

A minimal forecast-ready set is
\begin{equation}
\{
H(z),\ \chi(z),\ D_{A}(z),\ D_{L}(z),\ D_{V}(z),\ F_{\rm AP}(z),\ 
D(z),\ f(z),\ f\sigma_{8}(z)
\},
\label{eq:forecast-set}
\end{equation}
together with $r_{d}$ for BAO absolute calibration. Each element follows from
the background system \eqref{eq:background-system} and, for growth, from
Eq.~\eqref{eq:growth-eq-recap}, given a choice of $K(N)$ and $V(\phi)$
consistent with Secs.~\ref{sec:K} and \ref{subsec:stability-conditions}.

\FloatBarrier
\input{sections/section7/tables/tab_observables_summary}
\FloatBarrier
\clearpage

%% file: sections/section7/tables/tab_observables_summary.tex
\begin{table}[h]
\centering
\footnotesize
\setlength{\tabcolsep}{4pt}              
\renewcommand{\arraystretch}{1.12}
\caption{Forecast–ready observables for late–time cosmology assuming spatial flatness and $c=1$.
In QKDE the linear relations are GR ($\mu=\Sigma=1$, $\Phi=\Psi$), so sensitivity enters via $H(a)$ and the induced $D(a)$.}
\label{tab:observables_summary}
\begin{tabularx}{\linewidth}{@{}l >{\raggedright\arraybackslash}X l@{}}
\toprule
\textbf{Quantity} & \textbf{Definition / Dependence} & \textbf{Ref.} \\
\midrule
Hubble norm. & $E(z)\equiv H(z)/H_{0}$ & \S\ref{subsec:distances} \\

Comoving distance &
$\displaystyle \chi(z)=\int_{0}^{z}\!\frac{dz'}{H(z')}
=\frac{1}{H_{0}}\int_{0}^{z}\!\frac{dz'}{E(z')},\quad
\frac{d\chi}{dz}=\frac{1}{H(z)}$ & \eqref{eq:chi} \\

$D_A,\,D_L$ &
$D_A(z)=\chi(z)/(1+z)$,\;
$D_L(z)=(1+z)\chi(z)$; \;
$dD_A/dz$, $dD_L/dz$ as in \eqref{eq:DA-DL} & \eqref{eq:DA-DL} \\

SN modulus &
$\mu(z)=5\log_{10}\!\big(D_L/\mathrm{Mpc}\big)+25$ & \eqref{eq:mu-SN} \\

BAO summaries &
$\displaystyle D_V(z)=\!\big[(1+z)^2 D_A^2 \, z/H(z)\big]^{1/3}$,\;
$F_{\rm AP}(z)=(1+z)D_A H^{-1}(z)$ & \eqref{eq:DV-FAP} \\

Vol.\ element &
$\displaystyle \frac{d V_c}{dz\,d\Omega}=\frac{\chi^2(z)}{H(z)}$ & \eqref{eq:vol-element} \\

Sound horizon &
$\displaystyle r_d=\int_{z_d}^{\infty}\!\frac{c_s(z)}{H(z)}dz,$\;
$c_s(a)=\big[3(1+R_b)\big]^{-1/2}$ & \eqref{eq:rd} \\

\addlinespace[3pt]
Growth $D$ &
$\displaystyle D''+\!\Big(2+\frac{H'}{H}\Big)\!D'-\frac{3}{2}\Omega_m(a)D=0$;\; $D(1)=1$ & \eqref{eq:growth-eq-recap} \\

Growth rate &
$f(a)=d\ln D/d\ln a$;\; $f\simeq \Omega_m^{\gamma}$ (diagnostic, $\gamma\simeq0.55$) & \eqref{eq:fs8} \\

RSD observable &
$f\sigma_8(z)=f(z)\,\sigma_{8,0}\,D(z)/D(0)$ & \eqref{eq:fs8} \\

Power spectrum &
$P(k,z)=D^2(z)\,P(k,0)$ (linear scaling; no $k$ dependence from $\mu,\Sigma$) & \eqref{eq:Pk-scaling} \\

\addlinespace[3pt]
WL kernel &
$\displaystyle C_\ell^\kappa=\int \frac{dz}{H}\,
\frac{W^2(z)}{\chi^2(z)}\,
P_\delta\!\Big(\frac{\ell+1/2}{\chi},z\Big)$ & \eqref{eq:Cl-kappa} \\

WL weight &
$\displaystyle W(z)=\frac{3}{2}\Omega_{m0}H_0^2(1+z)\chi(z)
\!\int_z^\infty\!dz_s\,n_s(z_s)\frac{\chi(z_s)-\chi(z)}{\chi(z_s)}$ & \eqref{eq:W-lensing} \\

ISW &
$\displaystyle \big(\Delta T/T\big)_{\rm ISW}=\int_{\eta_*}^{\eta_0}\! d\eta\,(\Phi'+\Psi')$ & \eqref{eq:ISW} \\

\addlinespace[3pt]
Linear MG fns. &
QKDE: $\mu(a,k)=1,\ \Sigma(a,k)=1,\ \eta(a,k)=0$ (GR at linear scales) & \eqref{eq:mu-sigma-eta} \\
\bottomrule
\end{tabularx}
\end{table}

%% file: sections/section8/section8.tex
\section{Numerical implementation and validation}
\label{sec:numerics}

This section specifies a \emph{reproducible} procedure to integrate the background system \eqref{eq:background-system}, evaluate observables (Sec.~\ref{sec:observables}), and verify identities implied by the covariant formulation (Secs.~\ref{sec:covariant-core}--\ref{sec:linear}). Every step is a direct evaluation of previously stated equations---no phenomenological shortcuts are introduced. The presentation is algorithmic so that distinct codes produce numerically indistinguishable results within controlled tolerances.

\subsection{State vector, domain, and solver}
\label{subsec:state-solver}

For numerical stability and clarity, evolve the state vector
\begin{equation}
\boxed{%
\mathbf{y}(N) \equiv \big(\phi(N),\; s(N),\; y_H(N)\big)
\equiv \big(\phi,\; \phi',\; \ln H\big),\qquad N\equiv \ln a\in[N_{\mathrm{ini}},\,0],
}
\end{equation}
with $H(N)=\exp\{y_H(N)\}$ and $E(N)\equiv H'/H = y_H'(N)$. The evolution equations are the closed system \eqref{eq:background-system} written in $(\phi,s,y_H)$:
\begin{subequations}
\label{eq:background-system-impl}
\begin{align}
s' \;&=\; -\Big(3 + y_H' + \frac{K'}{K}\Big)\,s \;-\; \frac{V_{,\phi}}{\,e^{2y_H}K\,},
\label{eq:impl-s}\\[2pt]
y_H' \;&=\; -\frac{1}{2M_{\rm pl}^{2}e^{2y_H}}
\Big(\rho_m+\tfrac{4}{3}\rho_r+K\,e^{2y_H}s^{2}\Big),
\label{eq:impl-yH}\\[2pt]
\rho_\phi \;&=\; K\,\frac{e^{2y_H}}{2}\,s^{2} + V(\phi),\qquad
X=\tfrac12 e^{2y_H}s^{2}.
\label{eq:impl-aux}
\end{align}
\end{subequations}
with $\rho_m(N)=\rho_{m0}e^{-3N}$ and $\rho_r(N)=\rho_{r0}e^{-4N}$.

\paragraph{Units and scaling (practical).}
A dimensionless implementation improves conditioning: work with $\hat\phi\equiv \phi/M_{\rm pl}$, $\hat V\equiv V/(M_{\rm pl}^{2}H_{0}^{2})$, $\hat\rho_i\equiv \rho_i/(3M_{\rm pl}^{2}H_{0}^{2})$, and retain $y_H=\ln(H/H_{0})$. Equations \eqref{eq:background-system-impl} are invariant under this rescaling after the trivial substitutions $M_{\rm pl}\!\to\!1$ and $H\!\to\!H_0 e^{y_H}$.

\paragraph{Integrator and tolerances.}
Use a standard embedded Runge--Kutta method with adaptive stepsize (e.g.\ Dormand--Prince~4(5) \parencite{DormandPrince1980} or an equivalent implementation \parencite{HairerODE1993}). Recommended componentwise tolerances:
\[
\mathrm{rtol}\in[10^{-10},10^{-7}],\qquad
\mathrm{atol}\in[10^{-12},10^{-9}].
\]
Late--time thawing solutions are typically nonstiff; if large $|V_{,\phi\phi}|/H^{2}$ induces stiffness, a stiff solver may be used \emph{without} changing equations.

\paragraph{Admissible initial data.}
Choose $N_{\mathrm{ini}}$ deep in matter domination (any epoch with $\rho_m\gg \rho_r,\rho_\phi$). Initialize
\begin{equation}
s(N_{\mathrm{ini}})=0,\qquad
y_H(N_{\mathrm{ini}})=\ln\!\Big[H_{0}\sqrt{\Omega_{m0}e^{-3N_{\mathrm{ini}}}+\Omega_{r0}e^{-4N_{\mathrm{ini}}}}\Big],
\end{equation}
and determine $\phi(N_{\mathrm{ini}})$ by a one-dimensional shoot (bisection/Newton) so that the closure at $N=0$ holds:
\begin{equation}
\Omega_{\phi}(0)=1-\Omega_{m0}-\Omega_{r0},\qquad
\Omega_{\phi}(N)=\frac{K\,s^{2}}{6M_{\rm pl}^{2}}+\frac{V(\phi)}{3M_{\rm pl}^{2}e^{2y_H}}.
\end{equation}
The shoot uses only \eqref{eq:impl-aux} (no extra priors).

\subsection{Evaluation of \texorpdfstring{$K'/K$}{K'/K}}
\label{subsec:Kprime-eval}

\paragraph{Phenomenological running.}
For $K(N)=1+K_{0}e^{-pN}$ the identity \eqref{eq:Kprime-phenom} is exact:
\begin{equation}
\boxed{\,\frac{K'}{K}=-p\,\frac{K-1}{K}\,}.
\end{equation}

\paragraph{Curvature--motivated running.}
For $K(N)=1+\alpha R/M^{2}$, use the \emph{algebraic, non-iterative} form \eqref{eq:KprimeK-closed}. With $E=y_H'$ and $H=e^{y_H}$,
\begin{align}
\frac{R}{H^{2}}&=6(2+E),\qquad
\frac{R'}{H^{2}}=A + B\,K', \quad
B=\frac{3s^{2}}{M_{\rm pl}^{2}},
\\
A&=24E+\frac{18K\,s^{2}}{M_{\rm pl}^{2}}
+\frac{9\rho_m+16\rho_r+6s\,V_{,\phi}}{M_{\rm pl}^{2}e^{2y_H}},
\qquad
c\equiv \frac{(\alpha/M^{2})\,e^{2y_H}}{\,1+(\alpha/M^{2})\,R\,}.
\end{align}
Then
\begin{equation}
\boxed{%
\frac{K'}{K}
=\frac{c\,A}{\,1-c\,B\,K\,}
=\frac{\displaystyle \frac{\alpha e^{2y_H}}{M^{2}}\,
\frac{A}{1+\alpha R/M^{2}}}
{\displaystyle 1 - \frac{\alpha e^{2y_H}}{M^{2}}\,
\frac{3K\,s^{2}}{M_{\rm pl}^{2}\left(1+\alpha R/M^{2}\right)}}
}\!,
\end{equation}
evaluated pointwise from $(\phi,s,y_H)$; no recursion in $R'$ or $K'$ is required.

\paragraph{Admissibility guards (cheap to enforce).}
Enforce for all $N\in[N_{\mathrm{ini}},0]$:
\begin{equation}
K(N)>0\quad\text{(ghost freedom)},\qquad
\big|\,1-c B K\,\big|>\epsilon\;\;\text{(non-vanishing denominator)},
\end{equation}
with a user-visible $\epsilon$ (e.g.\ a few$\times$machine precision). If one wishes to confine running to late times, impose the simple prior $|\alpha R/M^{2}|\ll 1$ for $N\le N_{\rm drag}$ so that $K\to 1$ prior to the baryon drag epoch (Sec.~\ref{subsec:rd}).

\subsection{Geometric observables via quadrature or ODEs}
\label{subsec:geom-obs}

Given $H(N)=e^{y_H(N)}$, distances may be computed either by redshift quadrature or via an $N$--ODE; both are equivalent.

\paragraph{Quadrature (in $z$).}
With $E(z)=H(z)/H_{0}$,
\begin{equation}
\chi(z)=\frac{1}{H_{0}}\int_{0}^{z}\frac{dz'}{E(z')}.
\end{equation}
Use an adaptive Gauss--Kronrod routine \parencite{Piessens1983} with tolerances comparable to the ODE tolerances. Then obtain $D_{A},D_{L},D_{V},F_{\rm AP}$ from \eqref{eq:DA-DL} and \eqref{eq:DV-FAP}.

\paragraph{Differential form (in $N$).}
Using $z(N)=e^{-N}-1$ and $dz/dN=-e^{-N}$,
\begin{equation}
\boxed{%
\frac{d\chi}{dN}=-\,\frac{e^{-N}}{H(N)}=-\,e^{-N-y_H(N)},
\qquad
\chi(0)=0,
}
\end{equation}
and then $D_{A}(z)=\chi/(1+z)$, $D_{L}(z)=(1+z)\chi$.

\subsection{Linear growth implementation}
\label{subsec:growth-impl}

Integrate the GR growth equation \eqref{eq:growth-eq-recap} with $E=y_H'$:
\begin{equation}
D''+\big(2+E\big)D'-\frac{3}{2}\,\Omega_m(N)\,D=0,
\qquad
\Omega_m(N)=\frac{\rho_m(N)}{3M_{\rm pl}^{2}e^{2y_H(N)}}.
\end{equation}
Initial conditions in the deep matter era:
\begin{equation}
D(N_{\mathrm{ini}})=e^{N_{\mathrm{ini}}},\qquad D'(N_{\mathrm{ini}})=1,
\end{equation}
followed by a renormalization $D\rightarrow D/D(0)$ to set $D(0)=1$. For a cross-check, also integrate the Riccati form for the growth rate $f\equiv D'/D$:
\begin{equation}
f'+f^{2}+\big(2+E\big)f-\frac{3}{2}\,\Omega_m(N)=0,
\qquad f(N_{\mathrm{ini}})=1,
\end{equation}
and verify consistency with $f=D'/D$. The RSD observable is then $f\sigma_8(z)=f(z)\,\sigma_{8,0}\,D(z)$.

\subsection{Diagnostics and identity checks}
\label{subsec:diagnostics}

At each accepted stepsize, report the following residuals; all vanish analytically and should be $\ll 1$ within solver tolerances.

\begin{itemize}
\item \textbf{Friedmann closure (dimensionless):}
\begin{equation}
\mathcal{C}_{F}\equiv 1-\Omega_{m}-\Omega_{r}-\Omega_{\phi},
\qquad
\Omega_{i}\equiv \frac{\rho_{i}}{3M_{\rm pl}^{2}e^{2y_H}},
\qquad
\Omega_{\phi}=\frac{K\,s^{2}}{6M_{\rm pl}^{2}}+\frac{V}{3M_{\rm pl}^{2}e^{2y_H}}.
\end{equation}

\item \textbf{Raychaudhuri residual} (e--fold form of \eqref{eq:Hdoteq}):
\begin{equation}
\mathcal{C}_{R}\equiv
\left[\,y_H' + \frac{1}{2M_{\rm pl}^{2}e^{2y_H}}
\Big(\rho_{m}+\tfrac{4}{3}\rho_{r}+K\,e^{2y_H}s^{2}\Big)\right].
\end{equation}

\item \textbf{Klein--Gordon residual} (background equation \eqref{eq:impl-s}):
\begin{equation}
\boxed{%
\mathcal{C}_{\phi}\equiv
\left[\,
s'+\left(3+y_H'+\frac{K'}{K}\right)s
+\frac{V_{,\phi}}{\,e^{2y_H}K}\right].
}
\end{equation}

\item \textbf{Ricci identity (geometric form)} using \eqref{eq:R-over-H2}:
\begin{equation}
\mathcal{C}_{R/H^{2}}\equiv
\left[\,
\frac{R}{e^{2y_H}}-6\!\left(2+y_H'\right)\right],
\qquad
R=6\,e^{2y_H}\big(2+y_H'\big).
\end{equation}

\item \textbf{Ricci identity (independent stress--energy form)} from $R=-T/M_{\rm pl}^{2}$:
\begin{equation}
\boxed{%
\mathcal{C}_{R,\,\mathrm{SET}}\equiv
\left[\,
R-\frac{\rho_{\rm tot}-3p_{\rm tot}}{M_{\rm pl}^{2}}
\right],}
\quad
\rho_{\rm tot}=\rho_m+\rho_r+\rho_\phi,\;
p_{\rm tot}=0+\tfrac13\rho_r+\Big(K\,\tfrac{e^{2y_H}}{2}s^{2}-V\Big).
\end{equation}
Either residual may be used; the stress--energy version provides an independent check that is sensitive to sign or unit mistakes.
\end{itemize}
Report $\max_{N\in[N_{\mathrm{ini}},0]}|\mathcal{C}_{\cdot}|$ for all diagnostics.

\subsection{Failure modes and safeguards}
\label{subsec:safeguards}

\begin{itemize}
\item \textbf{Ghost avoidance:} enforce $K(N)>0$ across the domain (Sec.~\ref{subsec:stability}); abort if $K$ crosses zero.
\item \textbf{Curvature case denominator:} if $|1-cBK|\le \epsilon$ in \eqref{eq:KprimeK-closed}, the chosen $(\alpha,M)$ is inadmissible for that background; reject the parameter point.
\item \textbf{Step control:} if residuals stagnate near machine precision while the integrator shrinks steps excessively, jointly tighten $(\mathrm{rtol},\mathrm{atol})$ by a factor $\sim 10$ and restart from the last checkpoint (equations unchanged).
\end{itemize}

\subsection{Reproducibility capsule}
\label{subsec:repro}

A complete run specification should include:
\begin{enumerate}
\item Functional forms and parameters: $V(\phi)$; the $K$ specification (either $\{\alpha,M\}$ or $\{K_{0},p\}$).
\item Baseline cosmology at $N=0$: $\{H_{0},\,\Omega_{m0},\,\Omega_{r0}\}$ (and $\sigma_{8,0}$ if reporting $f\sigma_{8}$).
\item Initialization: $N_{\mathrm{ini}}$; the rule $s(N_{\mathrm{ini}})=0$; and the shoot method/tolerance used to hit $\Omega_{\phi}(0)$.
\item Solver details: scheme (e.g.\ DOPRI5), \texttt{rtol}/\texttt{atol}, min/max stepsizes, and event guards (e.g.\ $K>0$, non-vanishing denominator).
\item Diagnostics: $\max|\mathcal{C}_{\cdot}|$ over the run; method used for distances (quadrature vs.\ ODE).
\end{enumerate}

\paragraph{Cross--checks.}
(i) Distances from redshift quadrature and from the $N$--ODE must agree at the requested tolerances.
(ii) $D(a)$ from the second--order equation and $f(a)$ from the Riccati form must be mutually consistent at $\lesssim 10^{-4}$ for the same $H(a)$.
(iii) For $K\equiv 1$, all diagnostics vanish to machine precision and $D(a)$ reproduces the GR solution for the chosen background.

\medskip
All steps above are straightforward evaluations of relations derived in Secs.~\ref{sec:covariant-core}--\ref{sec:linear}. References for standard algorithms: embedded Runge--Kutta \parencite{DormandPrince1980,HairerODE1993} and adaptive Gauss--Kronrod quadrature \parencite{Piessens1983}.

\FloatBarrier
\input{sections/section8/tables/tab_numerics_tol_diag}
\FloatBarrier

%% file: sections/section8/tables/tab_numerics_tol_diag.tex
\begin{table}[t]
\centering
\footnotesize
\setlength{\tabcolsep}{3pt}
\renewcommand{\arraystretch}{1.12}
\caption{Recommended numerical tolerances, step control, and diagnostics for the background and growth pipelines in Sec.~\ref{sec:numerics}. Targets assume double precision and non-stiff late--time runs; tighten if needed for highly curved $V_{,\phi\phi}/H^2$.}
\label{tab:numerics_tol_diag}

\begin{minipage}[t]{0.49\linewidth}
\centering
\textbf{(a) Solver settings and guards}\\[2pt]
\begin{tabularx}{\linewidth}{@{}l >{\raggedright\arraybackslash}X@{}}
\toprule
\textbf{Item} & \textbf{Recommended / Purpose} \\
\midrule
Solver scheme & Embedded RK 4(5) (DOPRI5); adaptive steps without changing equations. Switch to stiff BDF only if truly needed. \\
Rel./abs.\ tol. & $\mathrm{rtol}=10^{-9}$,\; $\mathrm{atol}=10^{-11}$ applied componentwise on $(\phi,\phi',H)$ and auxiliary ODEs ($\chi$, $D$, $f$). \\
Step size limits & $h_{\max}=10^{-2}$ in $N$, $h_{\min}=10^{-8}$; avoid overshooting rapid features. Abort if $h<h_{\min}$ repeatedly. \\
Checkpointing & Save state every $\Delta N=0.1$ for safe restart and residual logging. \\
Guards (positivity) & Enforce $K(N)>0$ (ghost freedom; Sec.~\ref{subsec:stability}). \\
Curvature guard & Ensure denominator in \eqref{eq:KprimeK-closed} exceeds $\epsilon\!\sim\!10^{-12}$; reject inadmissible $(\alpha,M)$ and report. \\
\bottomrule
\end{tabularx}
\end{minipage}
\hfill
\begin{minipage}[t]{0.49\linewidth}
\centering
\textbf{(b) Diagnostics and cross--checks}\\[2pt]
\begin{tabularx}{\linewidth}{@{}l >{\raggedright\arraybackslash}X@{}}
\toprule
\textbf{Quantity} & \textbf{Target / Note} \\
\midrule
Friedmann closure & $\max|\mathcal{C}_{F}|\lesssim 10^{-9}$; $\mathcal{C}_{F}=1-\Omega_m-\Omega_r-\Omega_\phi$. \\
Raychaudhuri residual & $\max|\mathcal{C}_{R}|\lesssim 10^{-9}$; e--fold form in Sec.~\ref{subsec:diagnostics}. \\
KG residual & $\max|\mathcal{C}_{\phi}|\lesssim 10^{-9}$; background Klein--Gordon check. \\
Ricci identity & $\max|\mathcal{C}_{R/H^{2}}|\lesssim 10^{-9}$; verifies $R/H^2=6\!\left(2+\frac{H'}{H}\right)$. \\
Scalar source & $\max|\mathcal{C}_{\nabla\!\cdot T_\phi}|\lesssim 10^{-9}$; checks Eq.~\eqref{eq:phi-continuity}. \\
\addlinespace[3pt]
\multicolumn{2}{@{}l}{\textit{Distances and growth}}\\
\addlinespace[2pt]
$\chi$ (ODE vs.\ quad) & $|\chi_{\rm ODE}-\chi_{\rm quad}|/\chi \lesssim 10^{-7}$; same $H(z)$ grid and tolerances. \\
$D$ vs.\ $f$ (Riccati) & $\max\big|f-(\ln D)'\big|\lesssim 10^{-4}$; agreement of 2nd- and 1st-order forms. \\
$D$ normalization & Enforce $D(0)=1$ to $<10^{-12}$ (post-integration renormalization). \\
$f\sigma_8$ propagation & Internal consistency $<10^{-10}$; use same $D(z)$, $f(z)$, $\sigma_{8,0}$. \\
\addlinespace[3pt]
\multicolumn{2}{@{}l}{\textit{Adaptive recovery}}\\
\addlinespace[2pt]
Automatic tightening & If residuals saturate near machine $\epsilon$ with step underflow, set $(\mathrm{rtol},\mathrm{atol})\!\to\!(\mathrm{rtol}/10,\mathrm{atol}/10)$ and restart from last checkpoint (no equation changes). \\
\bottomrule
\end{tabularx}
\end{minipage}

\end{table}

%% file: sections/section9/section9.tex
\section{Parameter sensitivities and Fisher-forecast setup}
\label{sec:forecast-setup}

This section specifies a referee-proof pipeline to compute parameter derivatives of the background and growth observables and to assemble Fisher forecasts \parencite{Tegmark1997,Tegmark1997Fisher,Song2009fsig8}. No fitting templates are introduced; all derivatives are obtained either analytically or by integrating exact sensitivity equations implied by the background system \eqref{eq:background-system} and the growth equation \eqref{eq:growth-eq-recap}.

\subsection{Parameter vector}
\label{subsec:param-vector}

The parameter set is split into late-time cosmology, kinetic-normalization, and potential sectors:
\begin{equation}
\boldsymbol{\theta}
=\big(H_{0},\,\Omega_{m0},\,\Omega_{r0},\,\sigma_{8,0}\big)
\;\oplus\;
\begin{cases}
(K_{0},\,p) & \text{for } K(N)=1+K_{0}\mathrm{e}^{-pN},\\[2pt]
(\alpha,\,M) & \text{for } K(N)=1+\alpha\,R/M^{2},
\end{cases}
\;\oplus\;
\boldsymbol{\theta}_{V},
\end{equation}
where $\boldsymbol{\theta}_{V}$ denotes the chosen $V(\phi)$ specification. If forecasts require the absolute BAO scale, the physical densities controlling $r_{d}$ (Sec.~\ref{subsec:rd}) are included; otherwise late-time ratios suffice.

\subsection{Sensitivity equations for the background}
\label{subsec:bg-sens}

Write the closed background system as
\begin{equation}
\mathbf{y}'=\mathbf{F}\big(N,\mathbf{y};\boldsymbol{\theta}\big),
\qquad
\mathbf{y}\equiv\big(\phi,\,\phi',\,H\big)^{\!\top},
\label{eq:bg-compact}
\end{equation}
with $\mathbf{F}$ defined by \eqref{eq:background-system}. For any parameter $\theta_{i}$, define the sensitivity vector
\(
\mathbf{s}_{i}(N)\equiv \partial \mathbf{y}/\partial \theta_{i}.
\)
Differentiation yields the exact variational (tangent–linear) system
\begin{equation}
\boxed{%
\mathbf{s}_{i}' \;=\; \mathbf{J}_{y}\,\mathbf{s}_{i} + \mathbf{J}_{\theta_{i}},
\qquad
\mathbf{J}_{y}\equiv \frac{\partial \mathbf{F}}{\partial \mathbf{y}},
\quad
\mathbf{J}_{\theta_{i}}\equiv \frac{\partial \mathbf{F}}{\partial \theta_{i}}
}\,,
\label{eq:variational}
\end{equation}
with initial data obtained by differentiating the initialization rules in Sec.~\ref{subsec:wellposed}. For the matter-era initialization ($\phi'=0$ and $H$ from \eqref{eq:z-and-mr-of-N}),
\begin{equation}
\mathbf{s}_{i}(N_{\mathrm{ini}})=
\big(\partial_{\theta_{i}}\phi(N_{\mathrm{ini}}),\,0,\,\partial_{\theta_{i}}H(N_{\mathrm{ini}})\big)^{\!\top},
\end{equation}
and $\partial_{\theta_{i}}\phi(N_{\mathrm{ini}})$ is fixed by differentiating the closure $\Omega_{\phi}(0)=1-\Omega_{m0}-\Omega_{r0}$ through \eqref{eq:system-aux}.

\paragraph{Jacobian entries.}
Let $E\equiv H'/H$ and $A_{m}\equiv \rho_{m}+\tfrac{4}{3}\rho_{r}+K H^{2}\phi'^{2}$. From \eqref{eq:Hprime-over-H}:
\begin{equation}
E=-\frac{A_{m}}{2M_{\rm pl}^{2}H^{2}},
\qquad
\frac{\partial E}{\partial \phi}=0,\quad
\frac{\partial E}{\partial \phi'}=-\frac{K\,\phi'}{M_{\rm pl}^{2}},\quad
\frac{\partial E}{\partial H}
=-\frac{1}{M_{\rm pl}^{2}}\!\left(\frac{K\,\phi'^{2}}{H}-\frac{A_{m}}{H^{3}}\right).
\label{eq:E-partials}
\end{equation}
Using \eqref{eq:system-phi}–\eqref{eq:system-Hprime}:
\begin{align}
\frac{\partial F_{\phi}}{\partial \phi}&=0, &
\frac{\partial F_{\phi}}{\partial \phi'}&=1, &
\frac{\partial F_{\phi}}{\partial H}&=0,\\[3pt]
\frac{\partial F_{\phi'}}{\partial \phi}&=-\frac{V_{,\phi\phi}}{H^{2}K},&
\frac{\partial F_{\phi'}}{\partial \phi'}&=-\!\left(3+E+\frac{K'}{K}\right)-\phi'\!\left(\frac{\partial E}{\partial \phi'}+\frac{\partial}{\partial \phi'}\frac{K'}{K}\right),&
\frac{\partial F_{\phi'}}{\partial H}&=-\phi'\!\left(\frac{\partial E}{\partial H}+\frac{\partial}{\partial H}\frac{K'}{K}\right)+\frac{2V_{,\phi}}{H^{3}K},\\[3pt]
\frac{\partial F_{H}}{\partial \phi}&=0,&
\frac{\partial F_{H}}{\partial \phi'}&=H\,\frac{\partial E}{\partial \phi'}=-\frac{K\,H\,\phi'}{M_{\rm pl}^{2}},&
\frac{\partial F_{H}}{\partial H}&=E+H\,\frac{\partial E}{\partial H}.
\end{align}
Model-dependent pieces enter through $\partial(K'/K)/\partial(\cdot)$:

\medskip\noindent
\emph{Phenomenological running} $K(N)=1+K_{0}\mathrm{e}^{-pN}$:
\begin{equation}
\frac{K'}{K}=-p\,\frac{K-1}{K},\qquad
\frac{\partial}{\partial \phi}\frac{K'}{K}=
\frac{\partial}{\partial \phi'}\frac{K'}{K}=
\frac{\partial}{\partial H}\frac{K'}{K}=0,
\qquad
\frac{\partial}{\partial K_{0}}\frac{K'}{K}=
-\,p\,\frac{\mathrm{e}^{-pN}}{K^{2}},
\end{equation}
\begin{equation}
\boxed{\,\frac{\partial}{\partial p}\!\left(\frac{K'}{K}\right)
= -\,\frac{K-1}{K}
+ pN\,\frac{K-1}{K}
- pN\,\frac{(K-1)^{2}}{K^{2}}\,}\,.
\label{eq:partials-phenom-correct}
\end{equation}

\noindent
\emph{Curvature–motivated} $K(N)=1+\alpha R/M^{2}$: use the algebraic form \eqref{eq:KprimeK-closed}. Writing $F\equiv (K'/K)=U/D$ with $U=cA$ and $D=1-cBK$ (definitions in Sec.~\ref{subsec:K-curv}),
\begin{equation}
\boxed{%
\frac{\partial}{\partial y}\!\left(\frac{K'}{K}\right)
= \frac{(\partial_{y}U)\,D-U\,(\partial_{y}D)}{D^{2}},
\quad
\partial_{y}U=(\partial_{y}c)\,A+c\,(\partial_{y}A),
\quad
\partial_{y}D=-\,(\partial_{y}c)\,B\,K-\,c\,(\partial_{y}B)\,K
}\!,
\qquad
y\in\{\phi,\phi',H\},
\label{eq:KprimeK-partials}
\end{equation}
with the state derivatives given in Sec.~\ref{subsec:K-curv} and parameter derivatives following from
\(
\partial c/\partial \alpha=H^{2}/\big[M^{2}\big(1+6(\alpha/M^{2})H^{2}(2+E)\big)^{2}\big]
\)
and
\(
\partial c/\partial M=-2(\alpha/M^{2})\,H^{2}/\big[M\big(1+6(\alpha/M^{2})H^{2}(2+E)\big)^{2}\big].
\)

\subsection{Distance and standard-ruler derivatives}
\label{subsec:distance-sens}

Using the e-fold ODE for $\chi$ (Sec.~\ref{subsec:geom-obs}),
\(
\chi'=-e^{-N}/H,\; \chi(0)=0,
\)
the parameter derivative satisfies
\begin{equation}
\boxed{%
\chi_{i}'=\frac{e^{-N}}{H^{2}}\,H_{i},
\qquad
H_{i}\equiv \frac{\partial H}{\partial \theta_{i}}
}\,,
\label{eq:chi-sens}
\end{equation}
integrated concurrently with \eqref{eq:variational}. Then
\begin{equation}
\frac{\partial D_{A}}{\partial \theta_{i}}=\frac{\chi_{i}}{1+z},\qquad
\frac{\partial D_{L}}{\partial \theta_{i}}=(1+z)\,\chi_{i},
\qquad
\frac{\partial D_{V}}{\partial \theta_{i}}=\frac{D_{V}}{3}\!\left[
2\,\frac{\chi_{i}}{\chi}
-\,\frac{H_{i}}{H}
\right],
\qquad
\frac{\partial F_{\rm AP}}{\partial \theta_{i}}=F_{\rm AP}\!\left[
\frac{\chi_{i}}{\chi}-\frac{H_{i}}{H}
\right].
\label{eq:DA-DL-etc-sens}
\end{equation}
For SNe,
\(
\partial \mu/\partial \theta_{i}=(5/\ln 10)\,D_{L}^{-1}\,\partial_{\theta_{i}}D_{L}.
\)

\subsection{Growth-observable derivatives}
\label{subsec:growth-sens}

Let $D(N)$ solve \eqref{eq:growth-eq-recap}. The parameter response $D_{i}\equiv \partial D/\partial \theta_{i}$ obeys the inhomogeneous equation
\begin{equation}
\boxed{%
D_{i}''+\left(2+E\right)D_{i}'-\frac{3}{2}\Omega_{m}\,D_{i}
\;+\;E_{i}\,D'
\;-\;\frac{3}{2}\,\Omega_{m,i}\,D=0
}\!,
\label{eq:D-var}
\end{equation}
with
\begin{equation}
E_{i}=\frac{H_{i}'}{H}-\frac{H'}{H}\,\frac{H_{i}}{H},
\qquad
\Omega_{m,i}=\frac{\rho_{m,i}}{3M_{\rm pl}^{2}H^{2}}-2\,\Omega_{m}\,\frac{H_{i}}{H},
\qquad
\frac{\rho_{m,i}}{\rho_{m}}=
2\,\frac{\delta_{i,H_{0}}}{H_{0}}
+\frac{\delta_{i,\Omega_{m0}}}{\Omega_{m0}}.
\end{equation}
Deep in matter domination: $D(N_{\mathrm{ini}})=e^{N_{\mathrm{ini}}}$, $D'(N_{\mathrm{ini}})=1$, and $D_{i}(N_{\mathrm{ini}})=D_{i}'(N_{\mathrm{ini}})=0$. After integration, renormalize to $D(0)=1$ and enforce $D_{i}(0)=0$ by
\(
D \!\leftarrow\! D/D(0),\;\; D_{i}\!\leftarrow\! D_{i}-D\,[D_{i}(0)/D(0)].
\)
The RSD derivative is
\begin{equation}
\frac{\partial}{\partial \theta_{i}}\big[f\sigma_{8}(z)\big]
=\sigma_{8,0}\,D_{i}'(z)
+\delta_{i,\sigma_{8,0}}\;f(z),
\label{eq:fs8-sens}
\end{equation}
where primes on $D$ denote $d/d\ln a$.

\subsection{Fisher matrix}
\label{subsec:Fisher}

Let the data vector $\mathbf{O}$ stack binned measurements of $\{D_{A}(z),\,H(z),\,D_{V}(z),\,F_{\rm AP}(z),\,f\sigma_{8}(z),\,\mu(z)\}$ with covariance $\mathbf{C}$. The Fisher information is
\begin{equation}
\boxed{%
F_{ij}=\left(\frac{\partial \mathbf{O}}{\partial \theta_{i}}\right)^{\!\top}
\,\mathbf{C}^{-1}\,
\left(\frac{\partial \mathbf{O}}{\partial \theta_{j}}\right)
}\,,
\label{eq:Fisher}
\end{equation}
with derivatives assembled from \eqref{eq:DA-DL-etc-sens}, \eqref{eq:chi-sens}, and \eqref{eq:fs8-sens}. Gaussian priors (if any) add as $F_{ij}\rightarrow F_{ij}+\delta_{ij}/\sigma^{2}_{\mathrm{prior},i}$ \parencite{Tegmark1997Fisher}. Forecasted $1\sigma$ uncertainties follow from $\sigma(\theta_{i})=\sqrt{(F^{-1})_{ii}}$.

\subsection{Pivoting and decorrelation for the phenomenological \texorpdfstring{$K$}{K}}
\label{subsec:pivot}

For $K(N)=1+K_{0}\mathrm{e}^{-pN}$, introduce a redshift pivot $N_{p}$ to reduce correlation between amplitude and slope:
\begin{equation}
K(N)=1+K_{p}\,\mathrm{e}^{-p(N-N_{p})},
\qquad
K_{p}\equiv K(N_{p})-1=K_{0}\,\mathrm{e}^{-pN_{p}}.
\end{equation}
The transform $(K_{0},p)\mapsto (K_{p},p)$ has Jacobian
\(
\partial(K_{0},p)/\partial(K_{p},p)=\begin{psmallmatrix}
\mathrm{e}^{pN_{p}} & N_{p}K_{0}\\ 0 & 1
\end{psmallmatrix}
\),
giving
\(
\mathbf{F}_{(K_{p},p)}=\mathbf{J}^{\top}\,\mathbf{F}_{(K_{0},p)}\,\mathbf{J}
\)
\parencite{Tegmark1997}.

\subsection{Degeneracy structure and consistency}
\label{subsec:degeneracies}

Because $\mu=\Sigma=1$ and $\Phi=\Psi$ (Sec.~\ref{sec:eft}), late-time signatures are funneled through $H(a)$ and its impact on $\{D_{A},H,D_{V},F_{\rm AP},D\}$. Consequently: (i) kinetic-running parameters $\{K_{0},p\}$ or $\{\alpha,M\}$ are partially degenerate with potential parameters $\boldsymbol{\theta}_{V}$ that alter $H(a)$; the sensitivity system \eqref{eq:variational} isolates these directions without ansatz; (ii) distances constrain $\int dz/H$ combinations, while $H(z)$ or $F_{\rm AP}$ helps break geometric degeneracies; (iii) adding $f\sigma_{8}(z)$ provides independent leverage via \eqref{eq:D-var}. If $K\to 1$ at high $z$, the matter-era initial conditions remain exact; otherwise initialization may be moved earlier with the same variational machinery.

\subsection{Scope and limitations}
\label{subsec:scope}

All derivatives and forecasts above are strictly linear-theory. Nonlinear clustering, scale-dependent bias, and baryonic feedback are outside scope; when required, $P(k,z)\to P_{\rm NL}(k,z)$ (e.g., \parencite{Smith2003HALOFIT,Takahashi2012HALOFIT}) can be adopted without altering the background/growth derivatives.

\FloatBarrier
\input{sections/section9/tables/tab_sensitivity_pipeline}
\FloatBarrier
\clearpage

%% file: sections/section9/tables/tab_sensitivity_pipeline.tex
\begin{table}[h]
\centering
\small
\setlength{\tabcolsep}{4pt}
\renewcommand{\arraystretch}{1.12}
\caption{Sensitivity-and-Fisher pipeline used in Sec.~\ref{sec:forecast-setup}. All derivatives are exact consequences of the background system \eqref{eq:background-system} and the growth equation \eqref{eq:growth-eq-recap}; no fitting templates are introduced.}
\label{tab:sensitivity_pipeline}
\begin{tabularx}{\linewidth}{@{}l Y l@{}}
\toprule
\textbf{Object} & \textbf{Definition / Identity} & \textbf{Ref.} \\
\midrule
Param.\ vector & $(H_{0},\Omega_{m0},\Omega_{r0},\sigma_{8,0})\oplus(K_{0},p\ \text{or}\ \alpha,M)\oplus\boldsymbol{\theta}_{V}$ & \S\ref{subsec:param-vector} \\
Variational system & $\mathbf{s}_{i}'=\mathbf{J}_{y}\,\mathbf{s}_{i}+\mathbf{J}_{\theta_{i}}$, \quad $\mathbf{s}_{i}\equiv \partial\mathbf{y}/\partial\theta_{i}$ & \eqref{eq:variational} \\
$E$ partials & $\partial E/\partial(\phi,\phi',H)$ as in \eqref{eq:E-partials} & \eqref{eq:E-partials} \\
$K'/K$ (RG) & $K'/K=-p\,(K-1)/K$; \quad $\partial_{K_{0}}(K'/K)=-p\,e^{-pN}/K^{2}$; \quad $\partial_{p}(K'/K)$ in \eqref{eq:partials-phenom-correct} & \eqref{eq:partials-phenom-correct} \\
$K'/K$ (curv.) & Algebraic, iteration-free form; sensitivities via quotient rule & \eqref{eq:KprimeK-closed}, \eqref{eq:KprimeK-partials} \\
Distance sens. & $\chi_{i}'=e^{-N}H_{i}/H^{2}$; \quad $\partial D_{A},\partial D_{L},\partial D_{V},\partial F_{\rm AP}$ & \eqref{eq:chi-sens}, \eqref{eq:DA-DL-etc-sens} \\
Growth sens. & $D_{i}''+(2+E)D_{i}'-\tfrac{3}{2}\Omega_{m}D_{i}+E_{i}D'-\tfrac{3}{2}\Omega_{m,i}D=0$ & \eqref{eq:D-var} \\
RSD sens. & $\partial (f\sigma_{8})/\partial \theta_{i}=\sigma_{8,0}\,D_{i}'+\delta_{i,\sigma_{8,0}}\,f$ & \eqref{eq:fs8-sens} \\
Fisher matrix & $F_{ij}=(\partial \mathbf{O}/\partial\theta_{i})^{\!\top}\mathbf{C}^{-1}(\partial \mathbf{O}/\partial\theta_{j})$; priors add diagonally & \eqref{eq:Fisher} \\
Pivoting & $(K_{0},p)\!\to\!(K_{p},p)$ with Jacobian; rotate $\mathbf{F}$ as $\mathbf{J}^{\top}\mathbf{F}\mathbf{J}$ & \S\ref{subsec:pivot} \\
\bottomrule
\end{tabularx}
\end{table}

%% file: sections/section10/section10.tex
\section{Limiting cases and consistency checks}
\label{sec:limits}

This section collects exact limits and analytic identities that cross-check the construction and equations used elsewhere. No additional assumptions beyond Secs.~\ref{sec:covariant-core}--\ref{sec:observables} are introduced.

\subsection{Canonical quintessence and $\Lambda$CDM limits}
\label{subsec:limits-canonical-LCDM}

\paragraph{Constant kinetic normalization.}
Setting $K\!\equiv\!1$ in the covariantly completed action \eqref{eq:intro-action}
(equivalently $K(\chi)\!\equiv\!1$ prior to fixing unitary gauge $\chi=t$)
reduces the framework to canonical quintessence. Equations \eqref{eq:KG-FRW}, \eqref{eq:Hprime-over-H}, and \eqref{eq:MS-eq} then reproduce the textbook forms with $c_{s}^{2}=1$ and $\Phi=\Psi$.

\paragraph{de Sitter late-time limit.}
If $H=\mathrm{const}$ and $\rho_{m},\rho_{r}\!\to\!0$, the Raychaudhuri relation \eqref{eq:Hdoteq} gives $K\,\dot\phi^{2}=0$, hence $\dot\phi=0$ (for $K>0$). The potential is $V=3M_{\rm pl}^{2}H^{2}$. In conformal time $\eta$ one has $a(\eta)=-1/(H\eta)$ and $a''/a=2/\eta^{2}$. In this exact de Sitter limit the adiabatic scalar mode is non-dynamical because $z\propto a\sqrt{K}\,\dot\phi/H\to 0$; equivalently, for a light \emph{spectator} with constant $K$ on de Sitter the MS equation is $v''+(k^{2}-a''/a)\,v=0$ with luminal propagation. For the curvature-motivated case $K=1+\alpha R/M^{2}$, $R=12H^{2}$ so $K=\mathrm{const}$ and $K'/K=0$, consistent with \eqref{eq:KprimeK-closed}.

\paragraph{Matter-era limit and initialization.}
Deep in matter domination ($\rho_{m}\gg\rho_{r},\rho_{\phi}$), $H(N)\simeq H_{0}\sqrt{\Omega_{m0}}\,\mathrm{e}^{-3N/2}$ and $\phi'=0$ solves \eqref{eq:KG-N} at leading order for any smooth $K(N)$, validating the initialization in Sec.~\ref{subsec:wellposed}. The growing-mode solution of \eqref{eq:growth-eq-recap} is $D\propto a$.

\subsection{Exact background reconstruction identities}
\label{subsec:reconstruction}

Given a background expansion history $H(t)$ and a positive kinetic normalization
$K(\chi)$ evaluated in unitary gauge $\chi=t$ (so $K(t)\equiv K(\chi)\vert_{\chi=t}$),
the scalar kinetic energy and potential are fixed \emph{algebraically} (no integration) by the Einstein equations:
\begin{align}
K\,\dot\phi^{2}
&= -\,2M_{\rm pl}^{2}\dot H - \rho_{m}-\tfrac{4}{3}\rho_{r},
\label{eq:Kphidot2-recon}\\[3pt]
V(t)
&= 3M_{\rm pl}^{2}H^{2} + M_{\rm pl}^{2}\dot H - \tfrac{1}{2}\rho_{m} - \tfrac{1}{3}\rho_{r}.
\label{eq:V-recon}
\end{align}
Equations \eqref{eq:Kphidot2-recon}--\eqref{eq:V-recon} follow directly from \eqref{eq:Friedmann} and \eqref{eq:Hdoteq} using $\rho_{\phi}=K\dot\phi^{2}/2+V$ and $p_{\phi}=K\dot\phi^{2}/2-V$. Notably, $V(t)$ depends only on $H$ and the standard fluids, and \emph{not} on $K$; the function $K(t)$ controls the time reparametrization of the field excursion through $\dot\phi^{2}\propto K^{-1}$. A potential $V(\phi)$ is then obtained by eliminating $t$ between \eqref{eq:Kphidot2-recon} and \eqref{eq:V-recon}. This reproduces the canonical reconstruction when $K\!\equiv\!1$, and demonstrates that any late-time $H(a)$ admissible in GR can be realized within QKDE by an appropriate $(K,V)$ pair with $K>0$.

\subsection{Field redefinition cross-check (origin of the friction term)}
\label{subsec:field-redef}

Define a time-dependent canonical variable $\psi(t)\equiv \sqrt{K(t)}\,\phi(t)$.
On an FRW background,
\begin{equation}
\mathcal{L}_{\phi}
= \tfrac{1}{2}\dot\psi^{2}
- \frac{\dot K}{2K}\,\psi\,\dot\psi
+ \tfrac{1}{2}\!\left(\frac{\dot K}{2K}\right)^{\!2}\psi^{2}
- V\!\left(\frac{\psi}{\sqrt{K}}\right).
\end{equation}
Varying w.r.t.\ $\psi$ and mapping back to $\phi$ yields exactly \eqref{eq:KG-FRW}:
\(
K(\ddot\phi+3H\dot\phi)+\dot K\,\dot\phi+V_{,\phi}=0.
\)
Thus the additional ``friction'' term is the unavoidable consequence of a time-dependent kinetic normalization; it cannot be removed by a local field redefinition without introducing additional (mass/mixing) terms.

\subsection{Super- and subhorizon behavior of perturbations}
\label{subsec:super-sub}

\paragraph{Subhorizon limit ($k\gg aH$).}
With $c_{s}^{2}=1$ and $\Phi=\Psi$, QKDE fluctuations are pressure-supported and decay on small scales; the Poisson equation reduces exactly to the GR form \eqref{eq:Poisson-GR} and the growth equation is \eqref{eq:growth-equation}.

\paragraph{Superhorizon limit ($k\ll aH$).}
The MS equation \eqref{eq:MS-eq} implies $v\propto z$ is a solution when $k\!\to\!0$; hence the comoving curvature $\mathcal{R}=v/z$ is conserved at leading order provided the effective mass term $z''/z$ varies slowly on Hubble timescales. Because $z^{2}=a^{2}K\dot\phi^{2}/H^{2}$, any superhorizon evolution of $\mathcal{R}$ is controlled by slow variation of $K$ and the background; in the slow-running regime $|K'/K|\ll 1$ this evolution is negligible, reproducing the canonical single-field behavior.

\subsection{Small-$\alpha$ expansion for the curvature-motivated $K$}
\label{subsec:small-alpha}

For $K=1+\alpha R/M^{2}$ with $|\alpha R/M^{2}|\ll 1$, expand \eqref{eq:Kprime-over-K-curv} using $R'=H^{2}(A+B\,K')$ from \eqref{eq:Rprime-affine}:
\begin{equation}
\frac{K'}{K}
= \frac{\alpha H^{2}}{M^{2}}\left[A + B\,K'\right] + \mathcal{O}(\alpha^{2})
\quad\Longrightarrow\quad
\boxed{\,\frac{K'}{K} \simeq \frac{(\alpha H^{2}/M^{2})\,A}{1-(\alpha H^{2}/M^{2})\,B}\,}.
\end{equation}
This agrees with the closed algebraic form \eqref{eq:KprimeK-closed} to first order in $\alpha$ and makes explicit that the denominator condition in Sec.~\ref{subsec:K-admissibility} is trivially satisfied for sufficiently small $|\alpha H^{2}/M^{2}|$.

\subsection{Energy conditions and stability recap}
\label{subsec:energy-stability}

The null energy condition for the scalar reads $\rho_{\phi}+p_{\phi}=K\dot\phi^{2}\ge 0$ and is automatically satisfied when $K>0$. Ghost freedom demands $K>0$ (Sec.~\ref{subsec:stability}); gradient stability is guaranteed by $P_{,XX}=0\Rightarrow c_{s}^{2}=1$; tensor propagation is luminal with constant Planck mass (Sec.~\ref{sec:eft}). No additional constraints arise at linear order.

\subsection{Bianchi identity and continuity consistency}
\label{subsec:bianchi}

After covariant completion, the full theory is diffeomorphism invariant and the \emph{total} stress--energy is conserved, $\nabla_{\mu}T^{\mu\nu}_{\rm tot}=0$, with $T^{\mu\nu}_{\rm tot}=T^{\mu\nu}_{(m)}+T^{\mu\nu}_{(r)}+T^{\mu\nu}_{(\phi)}+T^{\mu\nu}_{(\chi)}$.
In unitary gauge $\chi=t$, the background-only coefficient $K(\chi)\vert_{\chi=t}\equiv K(t)$ makes the scalar sector an open subsystem: $\phi$ exchanges energy with the clock/time-slicing sector, so $T^{\mu\nu}_{(\phi)}$ is not separately conserved.
This exchange is the covariant origin of the sourced continuity relation used throughout (cf.\ Eq.~\eqref{eq:phi-continuity}).
\begin{equation}
\nabla_{\mu}T^{\mu}{}_{\nu(\phi)}=-\,K_{,\chi}\,X\,\partial_{\nu}\chi
\qquad\Rightarrow\qquad
\dot\rho_{\phi}+3H(\rho_{\phi}+p_{\phi})
=-\,K_{,\chi}\,\dot\chi\,X
\ \stackrel{\chi=t}{=}\ -\,\dot K\,X
= -\,\tfrac12\,\dot K\,\dot\phi^{2}.
\end{equation}
Matter and radiation are conserved independently [Eq.~\eqref{eq:mr-continuity}]. The clock sector carries the compensating source so that the total continuity equation closes exactly; equivalently, the Einstein equations remain mutually consistent via the contracted Bianchi identity.
For numerics, a sharp diagnostic is
\begin{equation}
\boxed{%
\mathcal{C}_{\rm Bianchi}\;\equiv\;
\big(\dot\rho_m+3H\rho_m\big)
+\big(\dot\rho_r+4H\rho_r\big)
+\big(\dot\rho_\phi+3H(\rho_\phi+p_\phi)+\dot K\,X\big)
\;=\;0
}\,,
\end{equation}
which holds identically when \eqref{eq:KG-FRW}, \eqref{eq:mr-continuity}, and \eqref{eq:phi-continuity} are satisfied.

\subsection{Dimensional analysis and normalization checks}
\label{subsec:units}

With $c=\hbar=1$ and metric signature $(-,+,+,+)$, the dimensions are $[H]=[M]$, $[R]=[M^{2}]$, $[\phi]=[M]$, $[V]=[M^{4}]$, and $[K]=1$. The combinations in \eqref{eq:K-curv}, \eqref{eq:KG-N}, \eqref{eq:Hprime-over-H}, \eqref{eq:R-over-H2}, \eqref{eq:KprimeK-closed}, and \eqref{eq:growth-equation} are dimensionless as written.

\FloatBarrier
\begin{table}[h]
\centering
\small
\setlength{\tabcolsep}{4pt}
\renewcommand{\arraystretch}{1.12}
\caption{Limiting cases and analytic identities collected in Sec.~\ref{sec:limits}, with their role as internal cross--checks. All relations are exact under the assumptions of Secs.~\ref{sec:covariant-core}--\ref{sec:observables}.}
\label{tab:limits_checks}
\begin{tabularx}{\linewidth}{@{}p{0.23\linewidth} X p{0.20\linewidth}@{}}
\toprule
\textbf{Item} & \textbf{Statement / Identity} & \textbf{Ref.} \\
\midrule
Canonical quintessence &
$K\!\equiv\!1 \Rightarrow$ standard $P=X\!-\!V$, $c_s^2=1$, $\Phi=\Psi$; background and MS equations reduce to textbook GR forms. &
\eqref{eq:intro-action}, \eqref{eq:KG-FRW}, \eqref{eq:MS-eq} \\
\addlinespace[2pt]
$\Lambda$CDM / de Sitter &
$V=\text{const}$, $\phi'=0 \Rightarrow \Lambda$CDM; for $H=\text{const}$: $R/H^2=12$, $K'=0$ (curvature case gives constant $K$). &
\eqref{eq:Hdoteq}, \eqref{eq:R-FRW}, \eqref{eq:KprimeK-closed} \\
\addlinespace[2pt]
Matter era &
$\rho_m\!\gg\!\rho_r,\rho_\phi$: $H\simeq H_0\sqrt{\Omega_{m0}}e^{-3N/2}$, $\phi'=0$ solves \eqref{eq:KG-N}; growing mode $D\propto a$. &
\eqref{eq:KG-N}, \eqref{eq:growth-eq-recap} \\
\addlinespace[2pt]
Background reconstruction &
$K\dot\phi^2=-2M_{\rm pl}^2\dot H-\rho_m-\tfrac43\rho_r$, \quad
$V=3M_{\rm pl}^2H^2+M_{\rm pl}^2\dot H-\tfrac12\rho_m-\tfrac13\rho_r$. &
\eqref{eq:Kphidot2-recon}, \eqref{eq:V-recon} \\
\addlinespace[2pt]
Field redefinition &
$\psi=\sqrt{K}\,\phi$ $\Rightarrow$ $K(\ddot\phi+3H\dot\phi)+\dot K\,\dot\phi+V_{,\phi}=0$ (origin of the extra friction). &
\eqref{eq:KG-FRW} \\
\addlinespace[2pt]
Super/subhorizon behavior &
$k\!\gg\!aH$: GR Poisson and growth; $k\!\ll\!aH$: $\mathcal{R}$ conserved if $z''/z$ varies slowly; here $z^2=a^2 K\dot\phi^2/H^2$. &
\eqref{eq:Poisson-GR}, \eqref{eq:growth-equation}, \eqref{eq:MS-eq} \\
\addlinespace[2pt]
Small-$\alpha$ (curvature $K$) &
For $K\!=\!1+\alpha R/M^2$, $|\alpha R/M^2|\!\ll\!1$:
$\displaystyle \frac{K'}{K}\simeq \frac{(\alpha H^2/M^2)A}{1-(\alpha H^2/M^2)B}$. &
\eqref{eq:Kprime-over-K-curv}, \eqref{eq:Rprime-affine} \\
\addlinespace[2pt]
Energy \& stability &
NEC: $\rho_\phi+p_\phi=K\dot\phi^2\ge 0$ for $K>0$; ghost freedom $K>0$; $c_s^2=1$; tensors luminal, $M_{\rm pl}$ constant. &
Sec.~\ref{subsec:energy-stability}, Sec.~\ref{sec:eft} \\
\addlinespace[2pt]
Bianchi/continuity (sourced) &
$\nabla_\mu T^{\mu}{}_{\nu(\phi)}=-K_{,\chi}X\,\partial_\nu\chi$
$\Rightarrow$ $\dot\rho_\phi+3H(\rho_\phi+p_\phi)=-\dot K\,X$ in unitary gauge; total conservation holds in the completed theory. &
\eqref{eq:phi-continuity} \\
\addlinespace[2pt]
Dimensional checks &
$[H]=[M]$, $[R]=[M^2]$, $[\phi]=[M]$, $[V]=[M^{4}]$, $[K]=1$; all combinations used are dimensionless as written. &
Sec.~\ref{subsec:units} \\
\bottomrule
\end{tabularx}
\end{table}
\FloatBarrier
\clearpage

%% file: sections/section11/section11.tex
\section{Conclusions and outlook}
\label{sec:conclusions}

A minimal, GR--preserving dark--energy framework has been formulated in which only the \emph{background} kinetic normalization runs in time while the metric sector remains Einstein--Hilbert. A covariant completion is understood, in which the kinetic normalization is promoted to a scalar function $K(\chi)$ of a clock field $\chi$ and the analysis is performed in unitary gauge $\chi=t$. The effective background (unitary--gauge) action
\(
S=\int d^{4}x\sqrt{-g}\,[\tfrac12 M_{\rm pl}^{2}R+K(\chi)X-V(\phi)]
\)
(with $\chi=t$ fixed after variation)
with minimally coupled matter closes the background and linear sectors without adding metric operators beyond Einstein--Hilbert. From this starting point, all late--time cosmology relations employed in the analysis follow by direct derivation.

\paragraph{Principal results.}
\begin{itemize}
\item \textbf{Metric sector unmodified.} No non--minimal coupling, no braiding, and no tensor--speed excess are introduced. In EFT--DE language,
\(
\alpha_B=\alpha_M=\alpha_T=0
\)
and
\(
\alpha_K=K\,\dot\phi^{2}/(H^{2}M_{\rm pl}^{2})>0
\)
[Eq.~\eqref{eq:EFT-map}; Secs.~\ref{sec:covariant-core}, \ref{sec:eft}].
\item \textbf{Luminal, ghost--free scalar.} Because \(P=K(\chi)\vert_{\chi=t}\,X-V\) has \(P_{,XX}=0\), the mode speed is \(c_s^{2}=1\) and ghost freedom requires \(K>0\) (Sec.~\ref{subsec:stability}).
\item \textbf{Closed e--fold system.} The background is governed by the autonomous first--order system
\eqref{eq:background-system}, with exact expressions for \(H'/H\), \(X\), \(\rho_\phi\), and \(R/H^{2}\) (Sec.~\ref{sec:efold}).
\item \textbf{Linear growth obeys GR.} The Bardeen potentials satisfy \(\Phi=\Psi\); the subhorizon Poisson equation is GR; the growth factor satisfies
\(
D''+(2+H'/H)D'-\tfrac32\Omega_m D=0
\)
(Sec.~\ref{sec:linear}).
\item \textbf{Background--only phenomenology.} Linear phenomenological functions reduce to
\(
\mu=\Sigma=1,\ \eta=0
\)
(Sec.~\ref{sec:eft}). Late--time observables respond to QKDE solely through the expansion \(H(a)\) and induced growth \(D(a)\) (Sec.~\ref{sec:observables}).
\item \textbf{Kinetic normalizations.} Two concrete \(K(N)\) specifications are provided:
(i) a curvature--motivated form \(K=1+\alpha R/M^{2}\) with an algebraic, iteration--free expression for \(K'/K\) [Eq.~\eqref{eq:KprimeK-closed}]; 
(ii) an IR running \(K=1+K_{0}e^{-pN}\) [Eq.~\eqref{eq:Kprime-phenom}] (Sec.~\ref{sec:K}).
\item \textbf{Exact background reconstruction.} For any GR--admissible \(H(t)\),
\(K\dot\phi^{2}=-2M_{\rm pl}^{2}\dot H-\rho_{m}-\tfrac{4}{3}\rho_{r}\) and
\(V=3M_{\rm pl}^{2}H^{2}+M_{\rm pl}^{2}\dot H-\tfrac12\rho_{m}-\tfrac13\rho_{r}\)
(Sec.~\ref{sec:limits}); eliminating \(t\) yields \(V(\phi)\).
\end{itemize}

\paragraph{Continuity and conservation.}
Because the kinetic normalization originates from a covariant function $K(\chi)$, fixing unitary gauge $\chi=t$ renders the scalar sector an open subsystem.
The scalar stress tensor is therefore not separately conserved:
\(
\dot\rho_\phi+3H(\rho_\phi+p_\phi)=-\dot K\,X
\)
which is an exchange equation rather than a violation of GR continuity
[Eq.~\eqref{eq:phi-continuity}]. Matter and radiation obey their standard continuity equations [Eq.~\eqref{eq:mr-continuity}]. The clock sector carries the compensating contribution so that the total stress--energy tensor remains covariantly conserved, $\nabla_\mu T^{\mu\nu}_{\rm tot}=0$, as required by the contracted Bianchi identity (Sec.~\ref{subsec:bianchi}). The numerical capsule monitors this identity directly (Sec.~\ref{sec:numerics}).

\paragraph{Null--test suite (linear scales).}
The following model--defining predictions are immediate targets for data:
\begin{equation}
\boxed{\ \mu(a,k)=1,\qquad \Sigma(a,k)=1,\qquad \eta(a,k)\equiv\frac{\Phi}{\Psi}-1=0,\qquad c_{T}^{2}=1,\qquad M_{\rm pl}=\text{const.}\ }\,
\end{equation}
Any robust detection of \(\mu\neq1\), \(\Sigma\neq1\), nonzero slip \(\eta\), or late--time tensor--speed deviations lies outside the baseline (Secs.~\ref{sec:linear}, \ref{sec:eft}). With \(K>0\), \(\rho_{\phi}+p_{\phi}=K\dot\phi^{2}\ge0\) implies \(w_{\phi}\ge -1\); persistent phantom behavior would require abandoning ghost freedom.

\paragraph{Scope and assumptions.}
The analysis assumes (i) minimal coupling to matter and radiation; (ii) a covariantly completed kinetic normalization $K(\chi)$ evaluated in unitary gauge as a background--only function $K(t)$; and (iii) late--time deviations unless explicitly stated, leaving pre--recombination physics standard. Nonlinear clustering, baryonic feedback, and screening are not modeled beyond linear order (Secs.~\ref{sec:observables}, \ref{sec:limits}).

\paragraph{Directions for further study.}
\begin{enumerate}
\item \textbf{Early--time kinetic running.} Allowing \(K\neq1\) before recombination modifies \(E(z)\) and thus \(r_{d}\) [Eq.~\eqref{eq:rd}]. This can be tested by coupling the e--fold system to a Boltzmann solver and confronting CMB+BAO.
\item \textbf{Model selection with data.} Bayesian inference on \((\alpha,M)\) or \((K_{0},p)\) using SN, BAO, RSD, and chronometers quantifies any preference for kinetic running (pipeline in Secs.~\ref{sec:numerics}, \ref{sec:forecast-setup}).
\item \textbf{Nonlinear regime.} Although \(\mu=\Sigma=1\) at linear scales, background--driven growth histories alter halo statistics and lensing nonlinearity. Mapping QKDE backgrounds to emulators or response functions would enable calibrated small--scale predictions.
\item \textbf{Perturbative EFT embedding.} A decoupling--limit derivation from curvature--suppressed operators in QFT in curved spacetime can delineate the parameter range where a background--only \(K(t)\) captures the leading low--energy effects.
\end{enumerate}

\paragraph{Take--home message.}
Within its explicit assumptions, Quantum--Kinetic Dark Energy sits at the conservative edge of EFT--DE: a single running parameter \(\alpha_{K}>0\) driven by a covariantly completed kinetic normalization evaluated in unitary gauge, luminal scalar and tensor propagation, and GR relations for linear growth and lensing. The phenomenology is intentionally simple---\emph{all} late--time signatures flow through the background \(H(a)\). This clarity renders the framework transparent, readily testable, and straightforward to falsify with distance and growth data from current and forthcoming surveys.
\clearpage

%% file: sections/appendices/appendixA.tex
\appendix

\section{Theoretical Foundations of QKDE}
\label{appendix:foundations}

This appendix presents a complete and self-contained derivation of the
Quantum--Kinetic Dark Energy (QKDE) framework used in the main text.
Every step is explicit: specification of the ultraviolet (UV) operator basis,
construction of the one-loop effective action using the Schwinger--DeWitt
heat-kernel expansion
\parencite{BarvinskyVilkovisky1985, Barvinsky1990, Gilkey1975},
derivative counting, operator classification, proof of uniqueness of the curvature--derivative operator
$R\,X$, and derivation of the infrared (IR) action and EFT--DE mapping.
All assumptions, truncations, and reductions are stated clearly.
No phenomenological choices are mixed with the UV derivation: alternative IR
forms of \(K(N)\) are explicitly separated at the end.
In particular, every operator retained is shown to be required by symmetry,
dimensional analysis, or renormalization structure, ensuring that no hidden
assumptions or implicit approximations enter the construction.
Nonlocal form factors such as $R\log(-\Box)R$ do not contribute at two-derivative order
and are therefore irrelevant for the kinetic normalization derived here.
The UV derivation is manifestly covariant and yields a covariant infrared effective action in which the kinetic normalization is represented by a scalar function $K(\chi_{\rm c})$ of a clock/St\"uckelberg field $\chi_{\rm c}$.
(Here $\chi_{\rm c}$ denotes the clock field and should not be confused with the heavy UV scalar $\chi$ introduced below.)
The main text then adopts unitary gauge $\chi_{\rm c}=t$ after variation, so that all background equations and identities follow from a diffeomorphism-invariant completion while reproducing the same background-only coefficient $K(t)\equiv K(\chi_{\rm c})|_{\chi_{\rm c}=t}$.

Throughout, the metric signature is \((- + + +)\) and
\begin{equation}
X \equiv -\frac12 g^{\mu\nu}\partial_\mu\phi\,\partial_\nu\phi .
\end{equation}

\subsection{UV Operator Basis and Heavy--Light Structure}
\label{appendix:UV}

A UV theory is assumed to contain a light scalar \(\phi\) and a heavy real scalar
\(\chi\) of mass \(M\gg H\).
Diffeomorphism invariance in the UV completion, locality, and the requirement that the \(\phi\)
equations of motion contain at most two derivatives determine the admissible
operator content.
A discrete symmetry \(\chi\rightarrow -\chi\) eliminates cubic terms, simplifying
the one-loop determinant without affecting the kinetic renormalization of \(\phi\).
This symmetry choice is technically natural and preserved under radiative
corrections, ensuring that the UV Lagrangian is closed under renormalization.

\subsubsection*{UV action and allowed operators}

All operators of dimension \(\le4\) consistent with the symmetries are included:
\begin{align}
S_{\rm UV}
=\int d^{4}x\sqrt{-g}\bigg[
    &\frac12 M_{\rm pl}^{2}R
    -\frac12 (\nabla\phi)^{2}
    -\frac12 (\nabla\chi)^{2}
    -\frac12 M^{2}\chi^{2}
    -\lambda\,\chi\,X
    -\xi_\chi\,R\,\chi^{2}
\nonumber\\
    &+ \mathcal{O}\!\Big(\frac{\partial^{4}}{M^{2}}\Big)
\bigg].
\label{eq:UVaction-appendix}
\end{align}

Higher-derivative operators scale as \((H/M)^{2}\) or smaller and are omitted only
after the loop expansion, maintaining EFT consistency
\parencite{Weinberg2005EFTCosmo}.
The decoupling condition
\begin{equation}
\frac{H}{M}\ll 1
\label{eq:decoupling-condition}
\end{equation}
guarantees the validity of the derivative expansion and ensures that all operators
kept in \eqref{eq:UVaction-appendix} dominate those suppressed by additional powers
of \(H/M\).

\subsubsection*{Heavy-field operator and Laplace-type structure}

The heavy sector is quadratic:
\begin{align}
S_{\chi}
&=\frac12 \!\int d^{4}x\sqrt{-g}\;\chi\,\hat{\mathcal O}_\chi\,\chi,
\\
\hat{\mathcal O}_\chi
&= -\Box + M^{2} + U(\phi,R,X),
\end{align}
with
\begin{equation}
U(\phi,R,X)=\lambda X + \xi_\chi R + \mathcal{O}(X^{2}).
\label{eq:U-potential}
\end{equation}

Although \(U\) depends on \(X\), it contains no derivatives acting on \(\chi\).
Thus \(\hat{\mathcal O}_\chi\) remains a scalar Laplace-type operator and the
standard heat-kernel expansion applies unchanged.
This structural property is essential: the Schwinger--DeWitt expansion is valid
only for Laplace-type operators.

The coupling \(\xi_\chi R\chi^{2}\) renormalizes \(M_{\rm pl}^{2}\) by a constant
shift \parencite{BirrellDavies,ParkerToms}, implying \(\alpha_{M}=0\) exactly.

\subsection{Infrared Action, Background Equations, and EFT--DE Mapping}
\label{appendix:IR}

\subsubsection*{Infrared action}

Retaining all two-derivative terms admits the covariantly completed infrared action
\begin{equation}
S_{\rm eff}
=\int d^{4}x\sqrt{-g}\left[
\frac12 M_{\rm pl}^{2}R
+ K(\chi_{\rm c})\,X
- V(\phi)
\right].
\end{equation}
This action is fully diffeomorphism invariant; $K(\chi_{\rm c})$ is a scalar function of the clock field $\chi_{\rm c}$.
Fixing unitary gauge $\chi_{\rm c}=t$ after variation yields the effective background description used in the main text, with $K(t)\equiv K(\chi_{\rm c})|_{\chi_{\rm c}=t}$.
In this gauge-fixed description, the scalar sector is an open subsystem: the scalar exchanges energy--momentum with the clock sector, while the total stress--energy tensor remains covariantly conserved.

Variation with respect to $g_{\mu\nu}$ gives Einstein equations; variation with
respect to $\phi$ gives a second-order scalar equation with no higher derivatives.

\subsubsection*{EFT--DE mapping}

For $P(X,\phi,\chi_{\rm c})=K(\chi_{\rm c})X-V(\phi)$, and evaluated in unitary gauge $\chi_{\rm c}=t$,
\begin{equation}
\alpha_{K}=\frac{2XK(t)}{H^{2}M_{\rm pl}^{2}},
\qquad
\alpha_{B}=\alpha_{M}=\alpha_{T}=\alpha_{H}=0,
\qquad
c_{s}^{2}=1.
\end{equation}

This shows QKDE preserves the Einstein--Hilbert sector and introduces only
kinetic modification.

\subsection{Alternative IR Parameterizations}
\label{appendix:IRalt}

The UV prediction is the covariant two-derivative operator $(\alpha/M^{2})\,R\,X$ in the infrared effective action.
On FRW, and in unitary gauge $\chi_{\rm c}=t$, this is equivalently represented by a background-only kinetic normalization $K(t)=1+\alpha R(t)/M^{2}$ multiplying $X$, as used in the main text.

The phenomenological form
\begin{equation}
K(N)=1+K_{0}e^{-pN}
\end{equation}
is introduced solely for model-space exploration.

\subsection{Logical Closure}
\label{appendix:closure}

The construction establishes:
\begin{enumerate}
\item The UV theory is the unique two-derivative scalar system consistent with diffeomorphisms and assumed symmetries.
\item The one-loop integration of $\chi$ produces a single two-derivative operator: $(\lambda/192\pi^{2})(R/M^{2})X$.
\item All other scalar--curvature operators are higher-derivative or suppressed by additional powers of $H/M$.
\item On FRW, $R_{\mu\nu}\partial^{\mu}\phi\partial^{\nu}\phi=\frac13 R\,X$ and thus introduces no independent structure.
\item The resulting IR description admits a covariant completion $S_{\rm eff}=\int\sqrt{-g}\,[\tfrac12 M_{\rm pl}^{2}R+K(\chi_{\rm c})X-V(\phi)]$, whose unitary-gauge evaluation $\chi_{\rm c}=t$ reproduces exactly the background system, exchange continuity equation, and linear perturbations used in the main text, while ensuring covariant conservation of the total stress--energy tensor.
\item Alternative IR parameterizations preserve the same EFT classification and are to be understood as effective unitary-gauge representations within a covariant completion.
\end{enumerate}

QKDE is therefore the infrared manifestation of a curvature-induced wavefunction
renormalization of a light scalar field in curved spacetime.

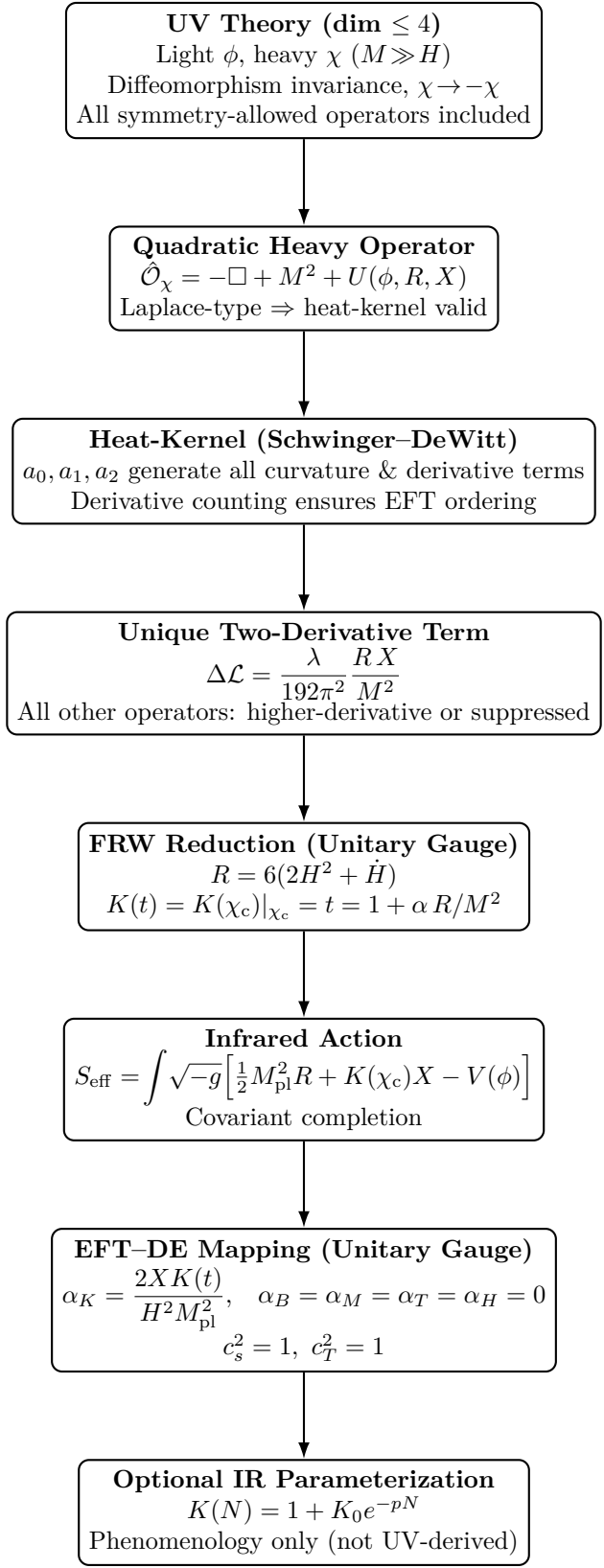
\begin{figure}[t]
\centering
\begin{tikzpicture}[
  node distance=1.2cm,
  box/.style={
    rectangle,
    rounded corners,
    draw=black,
    thick,
    align=center,
    minimum width=5.0cm,
    inner sep=4pt,
    font=\small
  },
  arrow/.style={-{Latex[length=2.3mm,width=1.6mm]}, thick}
]

\node[box] (UV) {
\textbf{UV Theory (dim $\le4$)}\\
Light $\phi$, heavy $\chi$ ($M\!\gg\!H$)\\
Diffeomorphism invariance, $\chi\!\to\!-\chi$\\
All symmetry-allowed operators included
};

\node[box, below=of UV] (Op) {
\textbf{Quadratic Heavy Operator}\\
$\hat{\mathcal{O}}_\chi=-\Box+M^2+U(\phi,R,X)$\\
Laplace-type $\Rightarrow$ heat-kernel valid
};

\node[box, below=of Op] (HK) {
\textbf{Heat-Kernel (Schwinger--DeWitt)}\\
$a_0,a_1,a_2$ generate all curvature \& derivative terms\\
Derivative counting ensures EFT ordering
};

\node[box, below=of HK] (RX) {
\textbf{Unique Two-Derivative Term}\\
$\Delta\mathcal{L}=\dfrac{\lambda}{192\pi^2}\dfrac{R\,X}{M^2}$\\
All other operators: higher-derivative or suppressed
};

\node[box, below=of RX] (FRW) {
\textbf{FRW Reduction (Unitary Gauge)}\\
$R=6(2H^2+\dot H)$\\
$K(t)=K(\chi_{\rm c})\vert_{\chi_{\rm c}}=t=1+\alpha\,R/M^2$
};

\node[box, below=of FRW] (IR) {
\textbf{Infrared Action}\\
$\displaystyle S_{\rm eff}=\!\int\!\sqrt{-g}\Big[\tfrac12 M_{\rm pl}^2 R
+K(\chi_{\rm c})X-V(\phi)\Big]$\\
Covariant completion
};

\node[box, below=of IR] (EFT) {
\textbf{EFT--DE Mapping (Unitary Gauge)}\\
$\alpha_K=\dfrac{2XK(t)}{H^2 M_{\rm pl}^{2}}$,\quad
$\alpha_B=\alpha_M=\alpha_T=\alpha_H=0$\\
$c_s^2=1,\ c_T^2=1$
};

\node[box, below=of EFT] (Alt) {
\textbf{Optional IR Parameterization}\\
$K(N)=1+K_0 e^{-pN}$\\
Phenomenology only (not UV-derived)
};

\draw[arrow] (UV) -- (Op);
\draw[arrow] (Op) -- (HK);
\draw[arrow] (HK) -- (RX);
\draw[arrow] (RX) -- (FRW);
\draw[arrow] (FRW) -- (IR);
\draw[arrow] (IR) -- (EFT);
\draw[arrow] (EFT) -- (Alt);

\end{tikzpicture}
\caption{Compact logical flow of the QKDE derivation.
The theory follows uniquely from a covariant UV completion, generating the sole two-derivative operator $R\,X$ and a covariant kinetic normalization $K(\chi_{\rm c})$, whose unitary-gauge evaluation $K(t)$ reproduces the background and perturbation system used in the main text.}
\label{fig:QKDE-flow-compact}
\end{figure}

\clearpage

%% file: sections/appendices/appendix1.tex
\section{Conventions and handy identities}
\label{app:conventions}

This appendix fixes notation and sign choices used throughout. Units \(c=\hbar=1\) and metric signature \((-,+,+,+)\) are adopted. The reduced Planck mass is \(M_{\rm pl}\equiv(8\pi G)^{-1/2}\). Greek indices run over spacetime \(\{0,1,2,3\}\); spatial indices are Latin \(\{1,2,3\}\).

\paragraph{Geometry and curvature (signs).}
The Levi–Civita connection is \(\Gamma^{\rho}{}_{\mu\nu}=\tfrac12 g^{\rho\sigma}(\partial_{\mu}g_{\sigma\nu}+\partial_{\nu}g_{\sigma\mu}-\partial_{\sigma}g_{\mu\nu})\).
Curvature follows the \emph{Wald} sign convention \parencite{Wald1984}:
\begin{equation}
R^{\rho}{}_{\sigma\mu\nu}=\partial_{\mu}\Gamma^{\rho}{}_{\nu\sigma}-\partial_{\nu}\Gamma^{\rho}{}_{\mu\sigma}
+\Gamma^{\rho}{}_{\mu\lambda}\Gamma^{\lambda}{}_{\nu\sigma}
-\Gamma^{\rho}{}_{\nu\lambda}\Gamma^{\lambda}{}_{\mu\sigma},\quad
R_{\mu\nu}=R^{\rho}{}_{\mu\rho\nu},\quad
R=g^{\mu\nu}R_{\mu\nu}.
\end{equation}
The Einstein tensor is \(G_{\mu\nu}=R_{\mu\nu}-\tfrac12 g_{\mu\nu}R\). The d’Alembertian is \(\Box\equiv g^{\mu\nu}\nabla_{\mu}\nabla_{\nu}\).

\paragraph{Background spacetime and clocks.}
A spatially flat FRW line element is
\begin{equation}
ds^{2}=-dt^{2}+a^{2}(t)\,d\vec{x}^{2},\qquad
H\equiv\frac{\dot a}{a},\qquad N\equiv\ln a,\qquad (\ )'\equiv \frac{d}{dN}=\frac{1}{H}\frac{d}{dt}.
\end{equation}
Useful chain rules for any sufficiently smooth background scalar \(f(t)\) are
\begin{equation}
\dot f = H f',\qquad \ddot f = H^{2} f'' + H H' f'.
\end{equation}
Redshift is related to the e–fold clock by \(1+z=e^{-N}\), hence
\begin{equation}
\frac{d}{dz}=-\frac{1}{1+z}\,\frac{d}{dN},\qquad
\frac{d}{dN}=-(1+z)\,\frac{d}{dz}.
\end{equation}
The FRW Ricci scalar and its Hubble–normalized form are
\begin{equation}
R=6\left(2H^{2}+\dot H\right)=6H^{2}\!\left(2+\frac{H'}{H}\right).
\end{equation}

\paragraph{Matter, radiation, and closure.}
Minimal coupling implies the standard continuity equations
\begin{equation}
\dot\rho_{m}+3H\rho_{m}=0,\qquad
\dot\rho_{r}+4H\rho_{r}=0
\;\;\Longrightarrow\;\;
\rho_{m}(N)=\rho_{m0}\,e^{-3N},\quad \rho_{r}(N)=\rho_{r0}\,e^{-4N}.
\end{equation}
Density fractions are \(\Omega_{i}\equiv \rho_{i}/(3M_{\rm pl}^{2}H^{2})\) with flatness
\(\Omega_{m}+\Omega_{r}+\Omega_{\phi}=1\).
The effective equation of state and deceleration parameter are
\begin{equation}
w_{\rm eff}=-1-\frac{2}{3}\frac{H'}{H},\qquad q=-1-\frac{H'}{H}.
\end{equation}

\paragraph{Scalar-field sector (background).}
The kinetic invariant is \(X\equiv -\tfrac12 g^{\mu\nu}\partial_{\mu}\phi\,\partial_{\nu}\phi\), giving on FRW
\begin{equation}
X=\tfrac12\dot\phi^{2}=\tfrac12 H^{2}\phi'^{2},\qquad
\rho_{\phi}=K(\chi_{\rm c})\,X+V(\phi),\qquad
p_{\phi}=K(\chi_{\rm c})\,X-V(\phi),\qquad
w_{\phi}=\frac{K(\chi_{\rm c})\dot\phi^{2}/2 - V}{K(\chi_{\rm c})\dot\phi^{2}/2 + V}.
\end{equation}
Here \(K(\chi_{\rm c})\) denotes the covariant kinetic normalization as a function of a clock/St\"uckelberg field; fixing unitary gauge \(\chi_{\rm c}=t\) after variation reproduces the background coefficient \(K(t)\) used in the main text.
Ghost freedom requires \(K(\chi_{\rm c})>0\); for \(P(X,\phi,\chi_{\rm c})=K(\chi_{\rm c})X-V(\phi)\) one has \(P_{,XX}=0\) and the scalar sound speed \(c_{s}^{2}=1\) \parencite{GarrigaMukhanov1999}. The GR background equations are
\begin{equation}
H^{2}=\frac{\rho_{m}+\rho_{r}+\rho_{\phi}}{3M_{\rm pl}^{2}},\qquad
\dot H=-\frac{1}{2M_{\rm pl}^{2}}\left(\rho_{m}+\tfrac{4}{3}\rho_{r}+K(t)\,\dot\phi^{2}\right).
\end{equation}
Because \(K(\chi_{\rm c})\) is spacetime dependent prior to gauge fixing, the scalar stress--energy tensor is not separately conserved.
A time–dependent kinetic normalization implies an exchange term in the scalar continuity relation,
\begin{equation}
\nabla_{\mu}T^{\mu}{}_{\nu(\phi)}=-\,\partial_{\nu}K(\chi_{\rm c})\,X
\quad\Rightarrow\quad
\dot\rho_{\phi}+3H(\rho_{\phi}+p_{\phi})=-\tfrac12\,\dot K(t)\,\dot\phi^{2},
\end{equation}
where the last equality is the unitary-gauge specialization of the covariant exchange equation. The total energy--momentum tensor (including the clock sector) remains covariantly conserved, ensuring consistency with the Bianchi identity and the Friedmann system.

\paragraph{Fourier and perturbation conventions.}
The Fourier transform is
\(f(\mathbf{x})=\int \frac{d^{3}k}{(2\pi)^{3}}\,e^{i\mathbf{k}\cdot\mathbf{x}}f(\mathbf{k})\),
so \(\nabla^{2}\to -k^{2}\). Newtonian–gauge scalar perturbations use
\(ds^{2}=-(1+2\Psi)\,dt^{2}+a^{2}(t)(1-2\Phi)\,d\vec{x}^{2}\).
In the GR–preserving setup with no anisotropic stress from the scalar, \(\Phi=\Psi\) at linear order on late–time, subhorizon scales \parencite{MaBertschinger1995}.

\paragraph{Dimensional analysis (natural units).}
With \(c=\hbar=1\) and the above signature,
\([H]=[M]\), \([R]=[M^{2}]\), \([\phi]=[M]\), \([V]=[M^{4}]\), \([X]=[M^{4}]\), and \([K]=1\).
All combinations appearing in the main text (\(\S\)\ref{sec:covariant-core}–\ref{sec:observables}) are dimensionless as written.

\medskip
\noindent\emph{Primary references for conventions:}

\par\smallskip
\begin{minipage}{\linewidth}\raggedright\footnotesize
\citet{Wald1984}; \citet{WeinbergCosmo2008}; \citet{MaBertschinger1995}; \citet{GarrigaMukhanov1999}.
\end{minipage}

%% file: sections/appendices/appendix2.tex
\section{From unitary-gauge scalar dynamics to FRW relations}
\label{app:KG-to-FRW}

This appendix makes explicit the steps connecting the
covariantly completed scalar sector evaluated in unitary gauge
to the background (FRW) equations used throughout. Conventions are those in App.~\ref{app:conventions}; curvature signs follow \textcite{Wald1984}.

\subsection{Euler--Lagrange equation for a time--dependent kinetic normalization}
\label{app:KG-variation}

Start from
\begin{equation}
\mathcal{L}_{\phi}=K(\chi_{\rm c})\,X - V(\phi),
\qquad
X\equiv -\tfrac12\,g^{\mu\nu}\partial_{\mu}\phi\,\partial_{\nu}\phi.
\end{equation}
The scalar Lagrangian admits a covariant completion via a clock/St\"uckelberg field $\chi_{\rm c}$; all variations are performed covariantly, and unitary gauge $\chi_{\rm c}=t$ is imposed only after deriving the field equations.
The field variation yields
\begin{equation}
0=\frac{\partial\mathcal{L}_{\phi}}{\partial\phi}-\nabla_{\mu}\!\left(\frac{\partial\mathcal{L}_{\phi}}{\partial(\nabla_{\mu}\phi)}\right)
=-V_{,\phi}-\nabla_{\mu}\!\left[-K(\chi_{\rm c})\,g^{\mu\nu}\nabla_{\nu}\phi\right],
\end{equation}
hence
\begin{equation}
K(\chi_{\rm c})\,\Box\phi+(\nabla_{\mu}K)\,\nabla^{\mu}\phi - V_{,\phi}=0.
\label{eq:KG-cov-app}
\end{equation}
Prior to gauge fixing, $(\nabla_{\mu}K)\neq0$ reflects exchange with the clock sector.
In unitary gauge $\chi_{\rm c}=t$, one has
\((\nabla_{\mu}K)\nabla^{\mu}\phi=g^{00}\dot K(t)\,\dot\phi=-\dot K(t)\,\dot\phi\).

\subsection{FRW evaluation and e--fold form}
\label{app:KG-FRW-derivation}

On spatially flat FRW, \(\Box\phi=-(\ddot\phi+3H\dot\phi)\). Substituting into \eqref{eq:KG-cov-app} and imposing unitary gauge gives the background Klein--Gordon equation
\begin{equation}
K(t)(\ddot\phi+3H\dot\phi)+\dot K(t)\,\dot\phi+V_{,\phi}=0,
\tag{\ref{eq:KG-FRW}}
\end{equation}
which is an exchange equation rather than a conservation law.
Reparametrizing time by \(N\equiv\ln a\) with the chain rules
\(\dot f=H f'\) and \(\ddot f=H^{2}f''+H H' f'\) leads to
\begin{equation}
\phi''+\left(3+\frac{H'}{H}+\frac{K'}{K}\right)\phi'
+\frac{V_{,\phi}}{H^{2}K}=0,
\tag{\ref{eq:KG-N}}
\end{equation}
where \(K(N)\equiv K(t(N))\).

\subsection{Stress--energy tensor and fluid variables}
\label{app:Tmn-derivation}

Varying \(\mathcal{L}_{\phi}\) with respect to \(g^{\mu\nu}\) gives
\begin{equation}
T^{(\phi)}_{\mu\nu}
=K(\chi_{\rm c})\,\partial_{\mu}\phi\,\partial_{\nu}\phi
+g_{\mu\nu}\!\left(K(\chi_{\rm c})X-V\right).
\tag{\ref{eq:Tmunu}}
\end{equation}
For a homogeneous background \(\phi=\phi(t)\) in unitary gauge,
\begin{equation}
X=\tfrac12\dot\phi^{2},\qquad
\rho_{\phi}=K(t)\,\frac{\dot\phi^{2}}{2}+V(\phi),\qquad
p_{\phi}=K(t)\,\frac{\dot\phi^{2}}{2}-V(\phi),\qquad
w_{\phi}=\frac{K(t)\dot\phi^{2}/2 - V}{K(t)\dot\phi^{2}/2 + V}.
\tag{\ref{eq:rho-p-w}}
\end{equation}

\subsection{Energy exchange induced by \texorpdfstring{$\dot K$}{Kdot}: covariant identity and direct check}
\label{app:noncons-derivation}

Because the completed scalar sector depends explicitly on $\chi_{\rm c}$, the scalar stress tensor is not separately conserved.
A general identity for matter with explicit coordinate dependence implies
\begin{equation}
\nabla_{\mu}T^{\mu}{}_{\nu(\phi)}=-\,\partial_{\nu}K(\chi_{\rm c})\,X,
\tag{\ref{eq:cov-noncons}}
\end{equation}
which represents energy--momentum exchange with the clock sector.
On FRW (\(\nu=0\)) and in unitary gauge this becomes
\begin{equation}
\dot\rho_{\phi}+3H(\rho_{\phi}+p_{\phi})
=-\tfrac12\,\dot K(t)\,\dot\phi^{2},
\tag{\ref{eq:phi-continuity}}
\end{equation}
explicitly an exchange equation, not a violation of GR conservation.
A direct check using \eqref{eq:rho-p-w} is instructive. One has
\(\dot\rho_{\phi}=\tfrac12\dot K\,\dot\phi^{2}+K\dot\phi\,\ddot\phi+V_{,\phi}\dot\phi\).
Adding \(3H(\rho_{\phi}+p_{\phi})=3HK\dot\phi^{2}\) and inserting the background equation
\(K(\ddot\phi+3H\dot\phi)=-\dot K\,\dot\phi - V_{,\phi}\) from \eqref{eq:KG-FRW} gives
\(\dot\rho_{\phi}+3H(\rho_{\phi}+p_{\phi})=-\tfrac12\dot K\,\dot\phi^{2}\), as stated.
The total energy--momentum tensor, including the clock sector, obeys $\nabla_{\mu}T^{\mu\nu}_{\rm tot}=0$ by the Bianchi identity.

\subsection{Einstein equations, Raychaudhuri form, and curvature}
\label{app:Einstein-FRW-derivation}

With a constant Planck mass and minimal couplings,
\begin{equation}
H^{2}
=\frac{1}{3M_{\rm pl}^{2}}\left(\rho_{m}+\rho_{r}+\rho_{\phi}\right),\qquad
\dot H
=-\frac{1}{2M_{\rm pl}^{2}}\Big(\rho_{m}+\tfrac{4}{3}\rho_{r}+\rho_{\phi}+p_{\phi}\Big)
=-\frac{1}{2M_{\rm pl}^{2}}\Big(\rho_{m}+\tfrac{4}{3}\rho_{r}+K(t)\,\dot\phi^{2}\Big),
\tag{\ref{eq:Friedmann}--\ref{eq:Hdoteq}}
\end{equation}
and the FRW Ricci scalar is
\begin{equation}
R=6\left(2H^{2}+\dot H\right)=6H^{2}\!\left(2+\frac{H'}{H}\right).
\tag{\ref{eq:R-FRW},\,\ref{eq:R-over-H2}}
\end{equation}
The second Friedmann (Raychaudhuri) equation follows from the first together with total energy--momentum conservation, not from scalar conservation alone.
Differentiating the first Friedmann relation and using matter/radiation continuity together with \eqref{eq:phi-continuity} reproduces the Raychaudhuri equation exactly, providing an internal consistency check for numerical solutions.

\subsection{Mukhanov--Sasaki inputs (background factors)}
\label{app:MS-background}

For linear perturbations the gauge--invariant MS variable \(v\) and pump field \(z\) are
\begin{equation}
v=a\left(\delta\phi+\frac{\dot\phi}{H}\Phi\right),\qquad
z^{2}=a^{2}\,\frac{K(t)\,\dot\phi^{2}}{H^{2}},
\tag{\ref{eq:v-and-z}}
\end{equation}
and the quadratic action is canonical with unit sound speed,
\(
S^{(2)}=\tfrac12\int d\eta\,d^{3}x\left[(v')^{2}-(\nabla v)^{2}+z''/z\,v^{2}\right]
\),
\(
c_{s}^{2}=1
\)
within the covariant completion evaluated in unitary gauge
for \(P=K(\chi_{\rm c})X-V(\phi)\) \parencite{GarrigaMukhanov1999,MaBertschinger1995}.

\medskip
All relations above are derived from the covariantly completed theory and evaluated in unitary gauge. No additional assumptions beyond spatial flatness, minimal couplings, and $K(\chi_{\rm c})>0$ have been introduced.

%% file: sections/appendices/appendix3.tex
\section{Closed algebraic form of \texorpdfstring{$K'/K$}{K'/K} for \(K=1+\alpha R/M^{2}\)}
\label{app:Kprime-derivation}

This appendix derives, step by step, the iteration--free expression used in Sec.~\ref{sec:K} for \(\,K'/K\) when
\begin{equation}
K(\chi_{\rm c})=1+\alpha\,\frac{R}{M^{2}},
\qquad
E\equiv \frac{H'}{H},
\qquad
\frac{R}{H^{2}}=6(2+E).
\tag{\ref{eq:K-curv},\,\ref{eq:R-over-H2}}
\end{equation}
All expressions below are derived covariantly and then evaluated in unitary gauge \(\chi_{\rm c}=t\), so that \(K(N)\equiv K(\chi_{\rm c})|_{\chi_{\rm c}=t}\).
Dimensions: \(K\) is dimensionless; \(R\) has mass dimension \(2\); therefore \(\alpha/M^{2}\) is dimensionless.

\subsection{Expressing \(R'/H^{2}\) in terms of background variables}
\label{app:Kprime-derivation:affine}

Let \(r\equiv R/H^{2}=6(2+E)\), so \(R=H^{2}r\). Using \((H^{2})'=2EH^{2}\) gives
\begin{equation}
\frac{R'}{H^{2}}=r'+2Er
=6\!\left(E'+4E+2E^{2}\right).
\label{eq:appRprime-master}
\end{equation}
From the Raychaudhuri form (Sec.~\ref{subsec:Hprime-Ricci})
\begin{equation}
E=-\frac{1}{2M_{\rm pl}^{2}H^{2}}
\left(\rho_m+\tfrac{4}{3}\rho_r+K H^{2}\phi'^{2}\right),
\tag{\ref{eq:Hprime-over-H}}
\end{equation}
one finds, after differentiating w.r.t.\ \(N\) and using \(\rho_m'=-3\rho_m\), \(\rho_r'=-4\rho_r\) and the e--fold Klein--Gordon equation (Sec.~\ref{subsec:KG-N}),
\begin{align}
E'&=-\frac{1}{2M_{\rm pl}^{2}H^{2}}\left[-3\rho_{m}-\tfrac{16}{3}\rho_{r}+\big(KH^{2}\phi'^{2}\big)'\right]-2E^{2},
\\[2pt]
\big(KH^{2}\phi'^{2}\big)'&=-\,H^{2}\!\left(K'\phi'^{2}+6K\phi'^{2}\right)-2\,\phi' V_{,\phi}.
\label{eq:appKH2phi2prime}
\end{align}

Substituting \eqref{eq:appKH2phi2prime} into \eqref{eq:appRprime-master} makes \(R'/H^{2}\) \emph{affine} in \(K'\):
\begin{equation}
\frac{R'}{H^{2}} \;=\; A \;+\; B\,K',
\qquad
B=\frac{3\,\phi'^{2}}{M_{\rm pl}^{2}},
\qquad
A=24E
+\frac{18K\,\phi'^{2}}{M_{\rm pl}^{2}}
+\frac{9\rho_m+16\rho_r+6\phi' V_{,\phi}}{M_{\rm pl}^{2}H^{2}}.
\tag{\ref{eq:Rprime-affine}}
\end{equation}

\textit{Note.}  
The \(K'\) dependence originates solely from differentiating \((K H^{2}\phi'^2)\) and therefore introduces no higher--derivative contributions, preserving the strict two--derivative structure of the covariantly completed theory evaluated in unitary gauge.

All quantities \(A\) and \(B\) are algebraic functions of the state vector \((\phi,\phi',H)\) and known sources \((\rho_{m},\rho_{r})\).

\subsection{Solving algebraically for \texorpdfstring{$K'/K$}{K'/K}}
\label{app:Kprime-derivation:algebra}

From \(K=1+\alpha R/M^{2}\),
\begin{equation}
\frac{K'}{K} \;=\; \frac{\alpha}{M^{2}}\,
\frac{R'}{1+\alpha R/M^{2}}
\;=\; \frac{\alpha H^{2}}{M^{2}}\,
\frac{A+B\,K'}{1+\alpha R/M^{2}}.
\label{eq:Kprime-step1}
\end{equation}
Define
\begin{equation}
c \;\equiv\; \frac{\alpha H^{2}/M^{2}}{1+\alpha R/M^{2}}.
\end{equation}
Equation \eqref{eq:Kprime-step1} becomes
\(
\frac{K'}{K}=c\,(A+B\,K').
\)
Bring the \(K'\) terms to one side and factor:
\begin{equation}
\Big[\tfrac{1+\alpha R/M^{2}}{K}-\tfrac{\alpha H^{2}}{M^{2}}\,B\Big]\,K'
=\frac{\alpha H^{2}}{M^{2}}\,A
\quad\Longrightarrow\quad
\frac{K'}{K}
=\frac{c\,A}{\,1-c\,B\,K\,}.
\end{equation}
Equivalently,
\begin{equation}
\boxed{%
\frac{K'}{K}
\;=\;
\frac{\displaystyle \frac{\alpha H^{2}}{M^{2}}\,
\frac{A}{1+\alpha R/M^{2}}}
{\displaystyle 1 - \frac{\alpha H^{2}}{M^{2}}\,
\frac{3K\,\phi'^{2}}{M_{\rm pl}^{2}\left(1+\alpha R/M^{2}\right)}}
}\,,
\tag{\ref{eq:KprimeK-closed}}
\end{equation}
the algebraic, iteration--free result used in Sec.~\ref{subsec:K-curv}.

\subsection{Regularity, limits, and implementation notes}
\label{app:Kprime-derivation:limits}

\paragraph{Regularity condition.}
The denominator \(1-cBK\) must remain nonzero across the integration domain. For sufficiently small \(|\alpha R/M^{2}|\) this is automatically satisfied and is monitored alongside \(K(\chi_{\rm c})>0\).

\paragraph{Background limits.}
Radiation: \(R=0\Rightarrow K=1\), \(K'/K=0\).  
Matter: \(R/H^{2}\simeq 3\), \(K\simeq 1+3\alpha H^{2}/M^{2}\), slowly varying.  
de Sitter: \(R/H^{2}=12\), \(K=\mathrm{const}\Rightarrow K'/K=0\).

\paragraph{Small-\(\alpha\) expansion.}
\begin{equation}
\frac{K'}{K} \simeq \frac{(\alpha H^{2}/M^{2})\,A}{1-(\alpha H^{2}/M^{2})\,B}.
\end{equation}

\paragraph{Dimensional analysis and implementation.}
All quantities are dimensionless as written.  
At each e--fold step compute:  
(i) \(E\), (ii) \(R/H^{2}\), (iii) \(A,B,c\), (iv) apply \eqref{eq:KprimeK-closed}.  
The procedure is purely algebraic and requires no recursion.

%% file: sections/appendices/appendix4.tex
\section{Quadratic action and the Mukhanov--Sasaki system}
\label{app:MS-derivation}

This appendix presents the explicit derivation of the quadratic action for scalar
perturbations in the QKDE framework and the resulting Mukhanov--Sasaki (MS)
equation.  The calculation starts directly from the
covariantly completed scalar sector evaluated in unitary gauge
written in covariant notation
in Eq.~\eqref{eq:intro-action}, retains only linear perturbations in the metric and
scalar field, and eliminates all lapse and shift variables through the
Hamiltonian and momentum constraints.  No phenomenological assumptions are
introduced; every step follows from the structure of 
$P(X,\phi,\chi_{\rm c})=K(\chi_{\rm c})X - V(\phi)$
with unitary gauge $\chi_{\rm c}=t$ imposed only after variation and $K(\chi_{\rm c})>0$.

\subsection{Perturbation variables and gauge invariants}

Scalar perturbations around a spatially flat FRW background are written in
Newtonian gauge:
\begin{equation}
ds^{2}=-(1+2\Psi)\,dt^{2}
       +a^{2}(t)\,(1-2\Phi)\,d\vec x^{2},
\qquad
\phi(t,\vec x)=\phi_{0}(t)+\delta\phi(t,\vec x),
\label{eq:appMS-metric}
\end{equation}
with the potentials satisfying \(\Phi=\Psi\) in the present framework because the
anisotropic stress of the scalar vanishes at linear order
(see Sec.~\ref{subsec:newtonian}).  

A gauge-invariant scalar fluctuation is constructed as in
\parencite{Sasaki1986,Mukhanov1988}:
\begin{equation}
v \equiv a\left(\delta\phi+\frac{\dot\phi_{0}}{H}\,\Phi\right),
\qquad
\mathcal{R}\equiv \frac{v}{z},
\label{eq:appMS-vR-def}
\end{equation}
where \(z\) is the pump field determined below.  
The combination is gauge-invariant under infinitesimal coordinate
transformations
\parencite{KodamaSasaki1984,WeinbergCosmo2008}.

\subsection{ADM expansion and solution of constraints}

The covariant action is
\[
S=\int d^{4}x\,\sqrt{-g}
\Big[\tfrac12 M_{\rm pl}^{2}R + P(X,\phi,\chi_{\rm c})\Big],
\qquad
P(X,\phi,\chi_{\rm c})=K(\chi_{\rm c})X - V(\phi),
\]
with 
\(
X \equiv -\tfrac12 g^{\mu\nu}\partial_\mu\phi\,\partial_\nu\phi .
\)

All variations are performed covariantly; unitary gauge $\chi_{\rm c}=t$ is fixed only after deriving the constraint equations.
To obtain the quadratic action, the metric is decomposed using ADM variables,
and the total action is expanded to second order in \(\{\Phi,\delta\phi\}\).  
The lapse and shift appear as nondynamical variables; their equations of motion
give the Hamiltonian and momentum constraints.  
Solving these and substituting back yields a quadratic action in a single
dynamical degree of freedom \(v\).  

For general \(P(X,\phi)\) theories, the resulting action is
\parencite{GarrigaMukhanov1999,DeFeliceTsujikawa2010}:
\begin{equation}
S^{(2)}
= \frac12 \int d\eta\, d^{3}x
\left[
   (v')^{2}
 - c_{s}^{2} (\nabla v)^{2}
 + \frac{z''}{z}\, v^{2}
\right],
\qquad
v''+\left(c_{s}^{2}k^{2}-\frac{z''}{z}\right)v=0,
\label{eq:appMS-action}
\end{equation}
where primes denote differentiation with respect to conformal time \(\eta\).

The general expressions for the sound speed and pump field are
\begin{equation}
c_{s}^{2} = 
\frac{P_{,X}}{P_{,X} + 2X P_{,XX}},
\qquad
z^{2} = a^{2}\frac{2X P_{,X}}{c_{s}^{2}H^{2}}.
\label{eq:appMS-csz}
\end{equation}

For QKDE,
\[
P_{,X}=K(\chi_{\rm c}),
\qquad 
P_{,XX}=0,
\]
so the theory remains exactly canonical.  Evaluated in unitary gauge,
\begin{equation}
c_{s}^{2}=1,
\qquad
z^{2}=a^{2}\frac{K(t)\,\dot\phi_{0}^{2}}{H^{2}}.
\label{eq:appMS-cs-one-z}
\end{equation}

\paragraph*{Clarification.}
Within the covariant completion evaluated in unitary gauge,
the EFT--DE parameters satisfy \(\alpha_{B}=\alpha_{H}=0\), so the scalar sector
does not kinetically mix with the metric.  
Thus the quadratic action reduces identically to the standard single-field
Mukhanov form, with the only modification appearing through the
time-dependent normalization \(K(t)\) in the background kinetic energy.  
No additional operators are generated.

\paragraph*{Constraints on $K(\chi_{\rm c})$.}
Ghost freedom requires $K(\chi_{\rm c})>0$.  
Gradient stability follows from \(c_{s}^{2}=1>0\).

\subsection{Dynamics of \texorpdfstring{$\mathcal{R}$}{R}: superhorizon and subhorizon limits}

Substituting \(v=z\,\mathcal{R}\) into \eqref{eq:appMS-action} gives the
evolution equation:
\begin{equation}
\mathcal{R}'' + 2\,\frac{z'}{z}\,\mathcal{R}' + c_{s}^{2} k^{2} \mathcal{R} = 0.
\label{eq:appMS-R-eq}
\end{equation}

Using \eqref{eq:appMS-cs-one-z}, the pump field satisfies
\[
\frac{z'}{z}
=\mathcal{H}
+\frac12
\frac{(K\dot\phi_{0}^{2}/H^{2})'}{(K\dot\phi_{0}^{2}/H^{2})},
\qquad
\mathcal{H}=aH.
\]

The structure is identical to canonical single-field inflation or quintessence,
except for the replacement
\[
\dot\phi_{0}^{2} \;\longrightarrow\; K(t)\dot\phi_{0}^{2},
\]
reflecting the curvature-induced wavefunction renormalization.

\paragraph*{Superhorizon limit.}
If \(z'/z\) varies slowly (which is satisfied whenever $K(t)$
evolves on Hubble timescales, as in both IR parameterizations of QKDE), then
\(\mathcal{R}' \approx 0\) for \(k \ll aH\).  
Thus \(\mathcal{R}\) is conserved on superhorizon scales
\parencite{Mukhanov2005} exactly as in minimally coupled single-field theories.

\paragraph*{Subhorizon limit.}
For \(k\gg aH\), the MS equation reduces to
\[
v'' + k^{2} v = 0,
\]
reflecting luminal propagation.  
Scalar perturbations free-stream and do not produce scale-dependent corrections
to the Poisson equation.  
This matches the GR behavior used in the main text for linear growth
(Sec.~\ref{subsec:growth}).

\paragraph*{Tensor sector.}
The Einstein--Hilbert metric sector is unchanged; the tensor sound speed remains
luminal \parencite{Abbott2017GW170817,Sakstein2017GW}.

\subsection{Useful identities and EFT--DE mappings}

\[
v=z\,\mathcal{R},
\qquad
z^{2}=a^{2}\frac{K(t)\,\dot\phi_{0}^{2}}{H^{2}},
\qquad
c_{s}^{2}=1,
\qquad
\Phi=\Psi.
\]

In the EFT--DE parameterization \parencite{BelliniSawicki2014},
\[
\alpha_{K}
=\frac{K(t)\dot\phi_{0}^{2}}{H^{2}M_{\rm pl}^{2}},
\qquad
\alpha_{B}=\alpha_{M}=\alpha_{T}=0.
\]
The vanishing of \(\alpha_{B}\) and \(\alpha_{T}\) explains why QKDE does not
modify gravitational slip or tensor propagation.  
All linear dynamics are computed in GR on the specified background;
all deviations from $\Lambda$CDM arise through changes in the background
history \(H(a)\) and the associated pump field \(z\).

%% file: sections/appendices/appendix5.tex
\section{Einstein constraints and GR Poisson equation at subhorizon scales}
\label{app:Poisson}

This appendix makes explicit how the GR Poisson equation used in the text follows from the linearized Einstein constraints and why the QKDE scalar does not source it on subhorizon scales.

\subsection{Newtonian gauge constraints and the comoving overdensity}

In Newtonian gauge,
\begin{equation}
ds^{2}=-(1+2\Psi)\,dt^{2}+a^{2}(t)\,(1-2\Phi)\,d\vec x^{2},
\qquad
\phi(t,\vec x)=\phi_{0}(t)+\delta\phi(t,\vec x),
\label{eq:Poisson-metric}
\end{equation}
and, in the present framework, the scalar sector carries no anisotropic stress so that $\Phi=\Psi$ at linear order (Secs.~\ref{subsec:newtonian}, \ref{subsec:Einstein-constraints}). The $0$--$0$ and $0$--$i$ Einstein equations in Fourier space are (see \textcite{MaBertschinger1995} for the conformal-time forms and \textcite{WeinbergCosmo2008} for a textbook derivation)
\begin{align}
-\,k^{2}\Phi - 3 a H\!\left(\dot\Phi+aH\Psi\right)
&= 4\pi G\,a^{2}\sum_{i}\delta\rho_{i},
\label{eq:00-constraint}\\
k^{2}\!\left(\dot\Phi+aH\Psi\right)
&= 4\pi G\,a^{2}\sum_{i}\big(\rho_{i}+p_{i}\big)\,\theta_{i},
\label{eq:0i-constraint}
\end{align}
where $\theta_{i}$ is the velocity-divergence potential of species $i$. Combining \eqref{eq:00-constraint} and \eqref{eq:0i-constraint} eliminates the $\dot\Phi$ term and yields the GR Poisson equation written in terms of the \emph{comoving} overdensity
\begin{equation}
-\,\frac{k^{2}}{a^{2}}\Phi
=4\pi G\sum_{i}\rho_{i}\,\Delta_{i},
\qquad
\Delta_{i}\equiv \delta_{i}+3\,\frac{aH}{k^{2}}\,\big(1+w_{i}\big)\,\theta_{i},
\label{eq:Poisson-comoving}
\end{equation}
valid for any mixture of minimally coupled fluids at linear order \parencite{MaBertschinger1995,WeinbergCosmo2008}. The lensing (Weyl) potential obeys the analogous GR relation
\begin{equation}
-\,\frac{k^{2}}{a^{2}}\big(\Phi+\Psi\big)=8\pi G\sum_{i}\rho_{i}\,\Delta_{i}.
\label{eq:Weyl-Poisson}
\end{equation}
In QKDE, $\Phi=\Psi$ identically at linear order, so \eqref{eq:Poisson-comoving}–\eqref{eq:Weyl-Poisson} coincide up to the factor of two.

\subsection{QKDE scalar perturbations and subhorizon suppression}

For the covariantly completed scalar sector $P(X,\phi,\chi_{\rm c})=K(\chi_{\rm c})X-V(\phi)$, evaluated in unitary gauge $\chi_{\rm c}=t$,
the scalar-fluid perturbations in Newtonian gauge are (Sec.~\ref{subsec:KG-perturbed})
\begin{align}
\delta\rho_{\phi}
&=K(t)\left(\dot\phi_{0}\,\delta\dot\phi-\dot\phi_{0}^{2}\Psi\right)+V_{,\phi}\,\delta\phi,
\qquad
\delta p_{\phi}
=K(t)\left(\dot\phi_{0}\,\delta\dot\phi-\dot\phi_{0}^{2}\Psi\right)-V_{,\phi}\,\delta\phi,
\label{eq:Poisson-drho-dp}\\
\big(\rho_{\phi}+p_{\phi}\big)\,\theta_{\phi}
&=K(t)\,\dot\phi_{0}\,\frac{k^{2}}{a^{2}}\,\delta\phi,
\qquad
\rho_{\phi}+p_{\phi}=K(t)\,\dot\phi_{0}^{2}.
\label{eq:Poisson-theta}
\end{align}
The quadratic action analysis (App.~\ref{app:MS-derivation}) gives $c_{s}^{2}=1$ and the MS equation
$v''+(k^{2}-z''/z)v=0$ with $v=a\big(\delta\phi+\dot\phi_{0}\Phi/H\big)$ and
$z^{2}=a^{2}K(t)\dot\phi_{0}^{2}/H^{2}$. In the \emph{subhorizon} regime $k\gg aH$,
\begin{equation}
v(\eta,\mathbf{k})\simeq C_{1}(\mathbf{k})\,e^{+ik\eta}+C_{2}(\mathbf{k})\,e^{-ik\eta},
\qquad
\Rightarrow\quad
\delta\phi=\mathcal{O}\!\left(\frac{v}{a}\right),\ \ \delta\dot\phi=\mathcal{O}\!\left(\frac{k}{a}\, \delta\phi\right),
\label{eq:Poisson-subhorizon-v}
\end{equation}
so pressure support prevents growth of $\delta\phi$ on scales well inside the sound horizon (which is luminal here). A standard scaling argument for smooth dark energy with $c_{s}^{2}=\mathcal{O}(1)$ then gives
\begin{equation}
\Delta_{\phi}=\mathcal{O}\!\Big((1+w_{\phi})\Big)\,\left(\frac{aH}{k}\right)^{2}\,\Delta_{m},
\qquad
(k\gg aH),
\label{eq:Poisson-DEsuppression}
\end{equation}
i.e.\ the dark-energy comoving overdensity is suppressed by $(aH/k)^{2}$ relative to matter on subhorizon scales \parencite{BeanDore2004,Hu1998GDM,AmendolaTsujikawaBook}. Equation \eqref{eq:Poisson-DEsuppression} holds for canonical quintessence and applies here unchanged because $c_{s}^{2}=1$ and the metric sector is GR.

\subsection{GR Poisson equation with matter source only}

Using \eqref{eq:Poisson-DEsuppression} in \eqref{eq:Poisson-comoving} and neglecting late-time photon/neutrino shear (tiny on the scales of interest), the subhorizon Poisson equation reduces to
\begin{equation}
-\,\frac{k^{2}}{a^{2}}\Phi \;=\; 4\pi G\,\rho_{m}\,\Delta_{m}
\quad\Big[1+\mathcal{O}\!\big((aH/k)^{2}\big)\Big],
\label{eq:Poisson-matter-only}
\end{equation}
with $\Delta_{m}\equiv \delta_{m}+3(aH/k^{2})\,\theta_{m}$. The same reduction holds for the Weyl potential,
so linear lensing is also unmodified at subhorizon scales. In the EFT--DE language of Sec.~\ref{sec:eft},
$\alpha_{B}=\alpha_{M}=\alpha_{T}=\alpha_{H}=0$ and $c_{s}^{2}=1$ imply the quasi-static relations
\begin{equation}
\mu(a,k)=1,\qquad \Sigma(a,k)=1,\qquad \eta(a,k)\equiv \frac{\Phi}{\Psi}-1=0,
\label{eq:Poisson-muSigma-appendix}
\end{equation}
in agreement with the main-text Eq.~\eqref{eq:mu-sigma-eta}. Departures from \eqref{eq:Poisson-matter-only} would require either a modification of the metric sector (e.g.\ $\alpha_{M}\neq 0$ or $\alpha_{B}\neq 0$) or a subluminal sound speed generating clustering ($c_{s}^{2}\ll 1$); neither occurs in QKDE.

\paragraph{Scope of the approximation.}
The suppression \eqref{eq:Poisson-DEsuppression} applies for $k\gg aH$. Near the horizon ($k\sim aH$) or if early radiation shear is included, the full system \eqref{eq:00-constraint}–\eqref{eq:0i-constraint} should be used; this does not affect the late-time, subhorizon observables employed in Sec.~\ref{sec:observables}.

\medskip
\noindent\textit{References for this appendix:}

\par\smallskip
\begin{minipage}{\linewidth}
\raggedright\footnotesize
\citet{MaBertschinger1995}; \citet{WeinbergCosmo2008}; \citet{GarrigaMukhanov1999};
\citet{BeanDore2004}; \citet{Hu1998GDM}; \citet{DeFeliceTsujikawa2010}.
\end{minipage}

%% file: sections/appendices/appendix6.tex
\section{Growth equation from continuity and Euler relations}
\label{app:growth-derivation}

This appendix derives the linear growth equation used in Secs.~\ref{sec:linear} and \ref{sec:observables} directly from the dust (pressureless matter) continuity and Euler relations, together with the GR constraints summarized in App.~\ref{app:Poisson}. Throughout this appendix a prime denotes a derivative with respect to e\mbox{-}fold time, $(\ )'\equiv d/d\ln a$, and Fourier conventions follow \textcite{MaBertschinger1995}. The Newtonian gauge metric is
\(
ds^{2}=-(1+2\Psi)\,dt^{2}+a^{2}(t)(1-2\Phi)\,d\vec x^{2}
\),
and in the present framework $\Phi=\Psi$ at linear order.

\subsection{Dust fluid: continuity and Euler in $d/d\ln a$ form}

For nonrelativistic matter ($w=0$) the linearized continuity and Euler equations in Newtonian gauge are (see, e.g., \textcite{MaBertschinger1995,WeinbergCosmo2008})
\begin{equation}
\dot\delta_{m}=-\frac{\theta_{\!v}}{a}+3\,\dot\Phi,
\qquad
\dot\theta_{\!v}+H\,\theta_{\!v}=\frac{k^{2}}{a}\,\Psi,
\label{eq:app6-time}
\end{equation}
where $\theta_{\!v}\!\equiv\!\nabla\!\cdot\!\vec v_{m}$ is the (proper) velocity divergence. It is convenient to work with the dimensionless velocity divergence
\begin{equation}
\theta_{m}\;\equiv\;\frac{\theta_{\!v}}{aH}=\frac{\nabla\!\cdot\!\vec v_{m}}{aH},
\qquad
(\ )'=\frac{1}{H}\frac{d}{dt}.
\end{equation}
Equations \eqref{eq:app6-time} then become
\begin{equation}
\boxed{\;
\delta_{m}'=-\,\theta_{m}+3\,\Phi',
\qquad
\theta_{m}'= -\big(2+\tfrac{H'}{H}\big)\,\theta_{m}
\;+\;\frac{k^{2}}{a^{2}H^{2}}\,\Psi\;.
\;}
\label{eq:app6-cont-euler}
\end{equation}

The coefficient $2+H'/H$ follows from $d\ln(aH)/d\ln a=1+H'/H$ and the extra $-\theta_{m}$
term arising in the time derivative of $a^{-1}$ \citep[cf.~App.~A of][]{MaBertschinger1995}.

\subsection{Elimination of the velocity and use of the GR constraints}

Differentiating the continuity relation in \eqref{eq:app6-cont-euler} and using the Euler relation to eliminate $\theta_{m}'$ gives
\begin{align}
\delta_{m}''&=-\,\theta_{m}'+3\,\Phi'' \nonumber\\
&=\big(2+\tfrac{H'}{H}\big)\theta_{m}-\frac{k^{2}}{a^{2}H^{2}}\Psi+3\,\Phi'' \nonumber\\
&=-\big(2+\tfrac{H'}{H}\big)\delta_{m}'
\;+\;3\big(2+\tfrac{H'}{H}\big)\Phi'
\;-\;\frac{k^{2}}{a^{2}H^{2}}\Psi
\;+\;3\,\Phi'',
\label{eq:app6-delta2}
\end{align}
where the continuity equation was used again to replace $\theta_{m}$ by $-\delta_{m}'+3\Phi'$.

On subhorizon scales $k\gg aH$ and for smooth late\mbox{-}time species ($c_s^2=\mathcal{O}(1)$), the time derivatives of the potentials are suppressed relative to the gradient term by $\mathcal{O}[(aH/k)^{2}]$ \parencite{MaBertschinger1995,BeanDore2004,Hu1998GDM}. Neglecting the $\Phi'$ and $\Phi''$ terms at this order and using $\Psi=\Phi$, the GR Poisson equation in comoving form (App.~\ref{app:Poisson}) reduces to
\begin{equation}
-\,\frac{k^{2}}{a^{2}}\,\Phi=4\pi G\,\rho_{m}\,\delta_{m}
\quad\Rightarrow\quad
\frac{k^{2}}{a^{2}H^{2}}\,\Psi
= -\,\frac{3}{2}\,\Omega_{m}(a)\,\delta_{m}.
\end{equation}
Substituting into \eqref{eq:app6-delta2} yields the standard GR growth equation in $d/d\ln a$ form:
\begin{equation}
\boxed{\;
\delta_{m}''+\left(2+\frac{H'}{H}\right)\delta_{m}'
-\frac{3}{2}\,\Omega_{m}(a)\,\delta_{m}=0,
\;}
\label{eq:app6-growth}
\end{equation}
valid for $k\gg aH$ in the absence of anisotropic stress and fifth forces. Defining the growing--mode factor by $\delta_{m}(\mathbf{k},a)=D(a)\,\delta_{m}(\mathbf{k},a_{\rm ini})$ reproduces Eq.~\eqref{eq:growth-equation} in the main text; the customary normalization is $D(1)=1$.

\paragraph{Riccati form for the logarithmic growth rate.}
Let $f\equiv D'/D=d\ln D/d\ln a$. Dividing \eqref{eq:app6-growth} by $D$ gives
\begin{equation}
\boxed{\;
f'+f^{2}+\left(2+\frac{H'}{H}\right)f-\frac{3}{2}\,\Omega_{m}(a)=0,
\;}
\end{equation}
the form integrated alongside the background in Sec.~\ref{subsec:growth-impl}.

\medskip
\noindent\textbf{Checks and scope.}
In exact matter domination, $H'/H=-3/2$ and $\Omega_{m}\to 1$, so \eqref{eq:app6-growth}
has the growing solution $D\propto a$ as expected. If one keeps the $\Phi'$ terms in
\eqref{eq:app6-delta2}, small $\mathcal{O}\!\left[(aH/k)^{2}\right]$ relativistic corrections
appear on very large scales; these are irrelevant for the late\mbox{-}time, subhorizon
observables considered in Sec.~\ref{sec:observables} but can be retained straightforwardly
when needed (see \citet{MaBertschinger1995,WeinbergCosmo2008} for the general formulas).

\medskip
\noindent\textit{References for this appendix:}

\par\smallskip
\begin{minipage}{\linewidth}\raggedright\footnotesize
\citet{MaBertschinger1995}; \citet{WeinbergCosmo2008}; \citet{BeanDore2004}; \citet{Hu1998GDM}.
\end{minipage}

%% file: sections/appendices/appendix7.tex
\section{Variational (sensitivity) system: detailed steps}
\label{app:variational-derivation}

This appendix records the tangent--linear (variational) equations used in Sec.~\ref{sec:forecast-setup} and gives all Jacobians needed to evolve exact parameter derivatives concurrently with the background. Throughout, a prime denotes $d/dN$ with $N\equiv\ln a$ and the state vector is
\begin{equation}
\mathbf{y}\equiv(\phi,\phi',H)^{\top},\qquad 
\mathbf{y}'=\mathbf{F}(N,\mathbf{y};\boldsymbol{\theta}),
\label{eq:app7-compact}
\end{equation}
All occurrences of $K$ and its derivatives are understood as the covariantly completed $K(\chi_{\rm c})$ evaluated in unitary gauge $\chi_{\rm c}=t$, i.e.\ $K(N)\equiv K(\chi_{\rm c})|_{\chi_{\rm c}=t}$.
The components read directly from the closed system \eqref{eq:background-system}:
\begin{align}
F_{\phi}&=\phi', \label{eq:app7-Fphi}\\
F_{\phi'}&=-\!\left(3+\frac{H'}{H}+\frac{K'}{K}\right)\phi'-\frac{V_{,\phi}}{H^{2}K}, \label{eq:app7-Fphip}\\
F_{H}&=H\,\frac{H'}{H}\equiv H\,E,\qquad
E\;=\;-\frac{\rho_m+\tfrac{4}{3}\rho_r+K\,H^{2}\phi'^{2}}{2M_{\rm pl}^{2}H^{2}}.
\label{eq:app7-FH}
\end{align}
The sources are $\rho_m(N)=\rho_{m0}e^{-3N}$ and $\rho_r(N)=\rho_{r0}e^{-4N}$.

\subsection{Tangent--linear system and state Jacobian}
For any parameter $\theta_i$, define the sensitivity vector $\mathbf{s}_i\equiv\partial\mathbf{y}/\partial\theta_i$. Differentiating \eqref{eq:app7-compact} at fixed $N$ gives the variational system
\begin{equation}
\boxed{\;
\mathbf{s}_i'=\mathbf{J}_y\,\mathbf{s}_i+\mathbf{J}_{\theta_i},
\qquad 
\mathbf{J}_y\equiv\frac{\partial\mathbf{F}}{\partial\mathbf{y}},
\quad
\mathbf{J}_{\theta_i}\equiv\frac{\partial\mathbf{F}}{\partial\theta_i}\;,
\;}
\label{eq:app7-variational}
\end{equation}
with explicit entries\footnote{Partial derivatives of $E$ are evaluated at fixed $(N,\mathbf{y})$:
\(
\partial E/\partial\phi=0,\;
\partial E/\partial\phi'=-K\,\phi'/M_{\rm pl}^{2},\;
\partial E/\partial H=-\tfrac{1}{M_{\rm pl}^{2}}\left(\tfrac{K\,\phi'^{2}}{H}-\tfrac{\rho_m+\tfrac{4}{3}\rho_r+K\,H^{2}\phi'^{2}}{H^{3}}\right)
\)
(cf.\ \eqref{eq:E-partials}).}:
\begin{align}
&\frac{\partial F_{\phi}}{\partial\phi}=0,\quad
\frac{\partial F_{\phi}}{\partial\phi'}=1,\quad
\frac{\partial F_{\phi}}{\partial H}=0, \label{eq:app7-Jy-row1}\\
&\frac{\partial F_{\phi'}}{\partial\phi}
=-\frac{V_{,\phi\phi}}{H^{2}K},\qquad
\frac{\partial F_{\phi'}}{\partial\phi'}
=-\!\left(3+E+\frac{K'}{K}\right)
-\phi'\!\left(\frac{\partial E}{\partial\phi'}+\frac{\partial}{\partial\phi'}\frac{K'}{K}\right),
\label{eq:app7-Jy-row2a}\\
&\frac{\partial F_{\phi'}}{\partial H}
=-\phi'\!\left(\frac{\partial E}{\partial H}+\frac{\partial}{\partial H}\frac{K'}{K}\right)
+\frac{2V_{,\phi}}{H^{3}K},
\label{eq:app7-Jy-row2b}\\
&\frac{\partial F_{H}}{\partial\phi}=H\,\frac{\partial E}{\partial\phi}=0,\qquad
\frac{\partial F_{H}}{\partial\phi'}=H\,\frac{\partial E}{\partial\phi'}=-\,\frac{K\,H\,\phi'}{M_{\rm pl}^{2}},\qquad
\frac{\partial F_{H}}{\partial H}=E+H\,\frac{\partial E}{\partial H}.
\label{eq:app7-Jy-row3}
\end{align}

\subsection{Model-dependent pieces: \(K'/K\)}
All appearances of $\partial(\,K'/K\,)/\partial(\cdot)$ and $\partial(\,K'/K\,)/\partial\theta_i$ are supplied by the chosen $K(N)$.

\paragraph{Phenomenological running \(K(N)=1+K_{0}e^{-pN}\).}
\begin{equation}
\frac{K'}{K}=-p\,\frac{K-1}{K},\qquad
\frac{\partial}{\partial\phi}\frac{K'}{K}
=\frac{\partial}{\partial\phi'}\frac{K'}{K}
=\frac{\partial}{\partial H}\frac{K'}{K}=0,
\end{equation}
and the parameter derivatives used in $\mathbf{J}_{\theta_i}$ are
\begin{equation}
\boxed{\;
\frac{\partial}{\partial K_{0}}\!\left(\frac{K'}{K}\right)
=-\,p\,\frac{e^{-pN}}{K^{2}},
\qquad
\frac{\partial}{\partial p}\!\left(\frac{K'}{K}\right)
= -\,\frac{K-1}{K}
+ pN\,\frac{K-1}{K}
- pN\,\frac{(K-1)^{2}}{K^{2}}}\;.
\;
\label{eq:app7-partials-phenom}
\end{equation}

\paragraph{Curvature-motivated \(K(\chi_{\rm c})=1+\alpha R/M^{2}\).}
Use the closed algebraic form \eqref{eq:KprimeK-closed}:
\begin{equation}
F\equiv\frac{K'}{K}=\frac{U}{D},\qquad
U\equiv c\,A,\quad D\equiv 1-cBK,
\quad
B=\frac{3\phi'^{2}}{M_{\rm pl}^{2}},
\quad
A=24E+\frac{18K\phi'^{2}}{M_{\rm pl}^{2}}+\frac{9\rho_m+16\rho_r+6\phi'V_{,\phi}}{M_{\rm pl}^{2}H^{2}},
\label{eq:app7-KK-defs}
\end{equation}
with
\begin{equation}
c=\frac{\xi H^{2}}{1+6\xi H^{2}(2+E)},\qquad \xi\equiv \frac{\alpha}{M^{2}}.
\label{eq:app7-c-def}
\end{equation}
State derivatives (for $\partial(\,K'/K\,)/\partial y$ with $y\in\{\phi,\phi',H\}$) follow from the quotient rule:
\begin{equation}
\boxed{\;
\frac{\partial}{\partial y}\!\left(\frac{K'}{K}\right)
=\frac{(\partial_yU)\,D-U\,(\partial_y D)}{D^{2}},
\quad
\partial_yU=(\partial_y c)\,A+c\,(\partial_y A),
\quad
\partial_yD=-\,(\partial_y c)\,BK-c\,(\partial_y B)\,K\;,
\;}
\label{eq:app7-KK-dy}
\end{equation}
with
\begin{align}
&\partial_{\phi}B=0,\quad \partial_{\phi'}B=\tfrac{6\phi'}{M_{\rm pl}^{2}},\quad \partial_{H}B=0,\\
&\partial_{\phi}A=\frac{6\phi'V_{,\phi\phi}}{M_{\rm pl}^{2}H^{2}},\quad
\partial_{\phi'}A=24\,\partial_{\phi'}E+\frac{36K\phi'}{M_{\rm pl}^{2}}+\frac{6V_{,\phi}}{M_{\rm pl}^{2}H^{2}},\quad
\partial_{H}A=24\,\partial_{H}E-\frac{2\,(9\rho_m+16\rho_r+6\phi'V_{,\phi})}{M_{\rm pl}^{2}H^{3}},\\
&\frac{\partial c}{\partial H}
=\frac{2\xi H}{\big[1+6\xi H^{2}(2+E)\big]^{2}},
\qquad
\frac{\partial c}{\partial E}
=-\,\frac{6\xi^{2}H^{4}}{\big[1+6\xi H^{2}(2+E)\big]^{2}}.
\end{align}
Parameter derivatives entering $\mathbf{J}_{\theta_i}$ (holding $\mathbf{y}$ fixed) are
\begin{equation}
\label{eq:app7-KK-param}
\setlength{\fboxsep}{4pt}%
\boxed{%
\begin{aligned}
\frac{\partial c}{\partial \alpha}
&= \frac{H^{2}}{M^{2}\,\bigl[1+6\xi H^{2}(2+E)\bigr]^{2}},\\[3pt]
\frac{\partial c}{\partial M}
&= -\,\frac{2\,\xi H^{2}}{M\,\bigl[1+6\xi H^{2}(2+E)\bigr]^{2}},\\[3pt]
\frac{\partial A}{\partial \alpha}
&= \frac{18\,(\partial K/\partial \alpha)\,\phi'^{2}}{M_{\rm pl}^{2}},
\qquad
\frac{\partial K}{\partial \alpha}=\frac{R}{M^{2}}.
\end{aligned}}
\end{equation}
and similarly for $\partial/\partial M$ via $\partial K/\partial M=-2\alpha R/M^{3}$. Derivatives with respect to potential parameters $\theta\in\boldsymbol{\theta}_{V}$ enter only through $V_{,\phi}$ and $V_{,\phi\phi}$ below.

\subsection{Parameter-source vectors \texorpdfstring{$\mathbf{J}_{\theta_i}$}{Jtheta}}
With the pieces above,
\begin{align}
J_{\theta_i,\phi}&=\frac{\partial F_{\phi}}{\partial \theta_i}=0,\\
J_{\theta_i,\phi'}&=\frac{\partial F_{\phi'}}{\partial \theta_i}
=-\,\phi'\,\frac{\partial}{\partial\theta_i}\!\left(\frac{K'}{K}\right)
-\frac{1}{H^{2}K}\,\frac{\partial V_{,\phi}}{\partial\theta_i}
+\frac{V_{,\phi}}{H^{2}K^{2}}\,\frac{\partial K}{\partial\theta_i},
\label{eq:app7-Jtheta-2}\\
J_{\theta_i,H}&=\frac{\partial F_{H}}{\partial \theta_i}
=H\,\frac{\partial E}{\partial\theta_i}.
\end{align}
For late-time cosmology parameters that appear only in the sources, $\rho_m=3M_{\rm pl}^{2}H_{0}^{2}\Omega_{m0}e^{-3N}$ and $\rho_r=3M_{\rm pl}^{2}H_{0}^{2}\Omega_{r0}e^{-4N}$ imply
\begin{equation}
\frac{\partial E}{\partial H_{0}}=-\frac{2\,\rho_{\rm tot}}{2M_{\rm pl}^{2}H^{2}}\frac{1}{H_{0}},
\qquad
\frac{\partial E}{\partial \Omega_{m0}}=-\frac{\rho_m}{2M_{\rm pl}^{2}H^{2}\,\Omega_{m0}},
\qquad
\frac{\partial E}{\partial \Omega_{r0}}=-\frac{2\,\rho_r}{3M_{\rm pl}^{2}H^{2}\,\Omega_{r0}},
\end{equation}
with $\rho_{\rm tot}\equiv\rho_m+\rho_r$. The amplitude $\sigma_{8,0}$ does not enter the background ($J_{\theta_i}=0$ for $\theta_i=\sigma_{8,0}$).

\subsection{Initialization: differentiating the shoot condition}
\label{app:shoot-sens}
Initial data are set in the matter era by
\begin{equation}
\phi'(N_{\rm ini})=0,\qquad
H(N_{\rm ini})=H_{0}\sqrt{\Omega_{m0}e^{-3N_{\rm ini}}+\Omega_{r0}e^{-4N_{\rm ini}}},
\end{equation}
and by choosing $\phi(N_{\rm ini})$ so that the $N=0$ closure holds:
\begin{equation}
g\big(\phi(N_{\rm ini}),\boldsymbol{\theta}\big)\;\equiv\;
\Omega_{\phi}(0)-\big(1-\Omega_{m0}-\Omega_{r0}\big)=0,
\qquad
\Omega_{\phi}=\frac{K\,\phi'^{2}}{6M_{\rm pl}^{2}}+\frac{V}{3M_{\rm pl}^{2}H^{2}}.
\label{eq:app7-shoot-constraint}
\end{equation}
The sensitivity of the unknown $\phi(N_{\rm ini})$ follows from the implicit--function theorem. Let $\mathbf{s}_i^{(0)}$ denote the solution of \eqref{eq:app7-variational} with \emph{provisional} initial data
\begin{equation}
\mathbf{s}_i(N_{\rm ini})=(0,\,0,\,\partial_{\theta_i}H_{\rm ini})^{\top},
\quad
\partial_{\theta_i}H_{\rm ini}\equiv \frac{\partial H(N_{\rm ini})}{\partial\theta_i},
\end{equation}
and let $\mathbf{e}$ be the solution of the homogeneous variational system with unit shift in the unknown initial field,
\begin{equation}
\mathbf{e}(N_{\rm ini})=(1,\,0,\,0)^{\top},\qquad \mathbf{e}'=\mathbf{J}_y\,\mathbf{e}.
\end{equation}
Evolving both to $N=0$, the linear response of the constraint is
\begin{equation}
\partial_{\theta_i}g
=\left.\frac{\partial g}{\partial\mathbf{y}}\right|_{0}\!\cdot\mathbf{s}_i^{(0)}(0)
+\left.\frac{\partial g}{\partial\phi_{\rm ini}}\right|_{0}\,\partial_{\theta_i}\phi_{\rm ini},
\qquad
\frac{\partial g}{\partial\phi_{\rm ini}}
=\left.\frac{\partial g}{\partial\mathbf{y}}\right|_{0}\!\cdot\mathbf{e}(0),
\end{equation}
and the condition $\partial_{\theta_i}g=0$ yields the sought initial sensitivity
\begin{equation}
\boxed{\;
\partial_{\theta_i}\phi(N_{\rm ini})
=-\,\frac{\left.\dfrac{\partial g}{\partial\mathbf{y}}\right|_{0}\!\cdot\mathbf{s}_i^{(0)}(0)}
{\left.\dfrac{\partial g}{\partial\mathbf{y}}\right|_{0}\!\cdot\mathbf{e}(0)}\;.
\;}
\label{eq:app7-phiini-sens}
\end{equation}
The full initial vector is then
\(
\mathbf{s}_i(N_{\rm ini})
=\big(\partial_{\theta_i}\phi(N_{\rm ini}),\,0,\,\partial_{\theta_i}H_{\rm ini}\big)^{\top}
\),
and a single forward pass produces $\mathbf{s}_i(N)$ for all $N$.

\subsection{Distance sensitivities (for completeness)}
The comoving distance ODE $\chi'=-e^{-N}/H$ implies the exact sensitivity
\begin{equation}
\boxed{\;
\chi_i'=\frac{e^{-N}}{H^{2}}\,H_i,\qquad H_i\equiv \frac{\partial H}{\partial\theta_i},\qquad \chi(0)=0,\;\chi_i(0)=0,
\;}
\label{eq:app7-chi-sens}
\end{equation}
which is integrated alongside \eqref{eq:app7-variational}. Algebraic propagation to $D_A,D_L,D_V,F_{\rm AP}$ is given in \eqref{eq:DA-DL-etc-sens}.

\subsection{Consistency and numerical remarks}
All expressions above are algebraic in $(N,\mathbf{y})$ and previously defined sources; no finite differencing is required. The tangent--linear integration inherits the stability of the background scheme; standard references for sensitivity ODEs and embedded RK integrators include \citet{HairerODE1993,DormandPrince1980}. Identity checks listed in Sec.~\ref{subsec:diagnostics} (e.g., Friedmann closure and Raychaudhuri residual) must remain at or below the requested tolerances when computed from either the background or the sensitivities.

\medskip
\noindent\textit{References for this appendix:}
\\citet{DormandPrince1980,HairerODE1993}.

%% file: sections/appendices/appendix8.tex
\clearpage
\section{Numerical algorithm (pseudocode, guards, and reproducibility)}
\label{app:numerics}

This appendix specifies an end–to–end, reproducible algorithm for integrating the closed background system \eqref{eq:background-system}, evaluating observables (Sec.~\ref{sec:observables}), and, when requested, co–evolving exact parameter sensitivities (Sec.~\ref{sec:forecast-setup}). All steps are direct evaluations of equations already derived; no fitting templates or phenomenological shortcuts are introduced.

\subsection{Inputs and preprocessing}
\label{app:numerics:inputs}
\begin{mdframed}[linewidth=0.6pt]
\small
\textbf{Required inputs}
\begin{itemize}[leftmargin=1.2em,itemsep=2pt,topsep=2pt]
\item Late–time cosmology at $N=0$: $(H_{0},\Omega_{m0},\Omega_{r0})$; if growth observables are reported, also $\sigma_{8,0}$.
\item Scalar sector: a twice–differentiable potential $V(\phi)$ providing $V_{,\phi}$ and $V_{,\phi\phi}$; a kinetic normalization specification either (i) phenomenological $K(N)=1+K_{0}e^{-pN}$ or (ii) curvature–motivated $K(N)=1+\alpha R/M^{2}$.
\item Integration domain: $N\in[N_{\rm ini},0]$ with $N_{\rm ini}$ chosen deep in matter domination.
\end{itemize}
\textbf{Derived quantities (evaluated pointwise)}
\begin{itemize}[leftmargin=1.2em,itemsep=2pt,topsep=2pt]
\item Sources $\rho_{m}(N)=\rho_{m0}e^{-3N}$, $\rho_{r}(N)=\rho_{r0}e^{-4N}$ from \eqref{eq:z-and-mr-of-N}.
\item Raychaudhuri ratio $E\equiv H'/H$ from \eqref{eq:Hprime-over-H}; Ricci ratio $R/H^{2}=6(2+E)$ from \eqref{eq:R-over-H2}.
\end{itemize}
\end{mdframed}

\subsection{Initialization (matter era) and shoot to closure}
\label{app:numerics:init}
\begin{mdframed}[linewidth=0.6pt]
\small
\textbf{Initialization}
\begin{enumerate}[leftmargin=1.2em,itemsep=2pt,topsep=2pt]
\item Set $\phi'(N_{\rm ini})\!=\!0$.
\item Set $H(N_{\rm ini})=H_{0}\sqrt{\Omega_{m0}e^{-3N_{\rm ini}}+\Omega_{r0}e^{-4N_{\rm ini}}}$ (matter+rad. limit of \eqref{eq:z-and-mr-of-N}).
\item Determine $\phi(N_{\rm ini})$ by a one–dimensional shoot such that the closure at $N=0$ holds:
\[
g\big(\phi(N_{\rm ini})\big)\;\equiv\; \Omega_{\phi}(0)-\big(1-\Omega_{m0}-\Omega_{r0}\big)=0,
\quad
\Omega_{\phi}=\frac{K\,\phi'^{2}}{6M_{\rm pl}^{2}}+\frac{V}{3M_{\rm pl}^{2}H^{2}}.
\]
\end{enumerate}
\textbf{Robust shoot (bracketed Newton)}
\begin{enumerate}[leftmargin=1.2em,itemsep=2pt,topsep=2pt]
\item Establish a bracket $[\phi_{\min},\phi_{\max}]$ with opposite–sign residuals $g(\phi_{\min})\,g(\phi_{\max})<0$ (expand geometrically if needed).
\item Iterate with safeguarded Newton/bisection: $\phi\leftarrow \phi-\lambda\,g/g'$ if the Newton step remains inside the bracket; otherwise bisect. The derivative $g'$ is computed by reusing the variational method (App.~\ref{app:variational-derivation}) with the homogeneous seed $\mathbf{e}(N_{\rm ini})=(1,0,0)^{\top}$ carried to $N=0$.
\item Stop when $|g|<\varepsilon_{\rm shoot}$ and $|\Delta\phi|<\varepsilon_{\phi}$ (e.g.\ $10^{-12}$ in double precision).
\end{enumerate}
\end{mdframed}
\emph{Note.} For thawing–like histories $\partial \Omega_{\phi}(0)/\partial\phi(N_{\rm ini})>0$ is typical, which accelerates convergence. The bracketed variant guarantees termination even when this monotonicity is weak.
\clearpage
\subsection{Core integration loop (embedded RK with adaptive steps)}
\label{app:numerics:loop}
\begin{mdframed}[linewidth=0.6pt]
\small
\textbf{State and auxiliaries}
\[
\mathbf{y}\equiv(\phi,\ s,\ y_{H})=(\phi,\ \phi',\ \ln H),\qquad
E=y_{H}'.
\]
\textbf{At each accepted step}
\begin{enumerate}[leftmargin=1.2em,itemsep=2pt,topsep=2pt]
\item Evaluate sources $\rho_{m},\rho_{r}$ at current $N$.
\item Compute $E$ from \eqref{eq:system-Hprime}; set $R/H^{2}=6(2+E)$.
\item Evaluate $K$ and $K'/K$:
\begin{itemize}[leftmargin=1.2em,itemsep=1pt,topsep=1pt]
\item Phenomenological case: use the identity \eqref{eq:Kprime-phenom}.
\item Curvature case: use the closed algebraic form \eqref{eq:KprimeK-closed} with $A,B,c$ from \eqref{eq:Rprime-affine}.
\end{itemize}
\item Advance the state with DOPRI5 (Dormand–Prince 4(5) \cite{DormandPrince1980,HairerODE1993}); adjust stepsize by standard local–error control.
\item Update diagnostics $\mathcal{C}_{F},\mathcal{C}_{R},\mathcal{C}_{\phi},\mathcal{C}_{R/H^{2}},\mathcal{C}_{\nabla\!\cdot T_\phi}$ (Sec.~\ref{subsec:diagnostics}); retain running maxima.
\item (Optional) co–integrate the distance ODE $\chi'=-e^{-N-y_H}$ and, if forecasts are required, the sensitivity system \eqref{eq:variational} and the growth equations (Sec.~\ref{subsec:growth-impl}).
\end{enumerate}
\textbf{Stepsize policy} \; Set $(\mathrm{rtol},\mathrm{atol})$ componentwise on $(\phi,s,y_{H})$;
impose $h_{\max}$ (e.g.\ $10^{-2}$ in $N$) and $h_{\min}$ (e.g.\ $10^{-8}$); if $h<h_{\min}$ is requested repeatedly, trigger the adaptive–recovery rule in Sec.~\ref{app:numerics:guards}.
\end{mdframed}

\subsection{Guards, event handling, and early termination}
\label{app:numerics:guards}
\begin{mdframed}[linewidth=0.6pt]
\small
\textbf{Physical and algebraic guards (checked every stage)}
\begin{itemize}[leftmargin=1.2em,itemsep=2pt,topsep=2pt]
\item \emph{Ghost freedom:} enforce $K(N)>0$; on violation, terminate with a descriptive flag.
\item \emph{Positivity of $H$:} because $y_{H}=\ln H$, positivity is built in; nonetheless assert $H>0$ on output.
\item \emph{Curvature case denominator:} require $\left|1-cBK\right|>\epsilon_{\rm den}$ in \eqref{eq:KprimeK-closed}; default $\epsilon_{\rm den}\sim 10^{-12}$ in double precision. If violated, declare $(\alpha,M)$ inadmissible for that trajectory.
\item \emph{Early–time safety (optional prior):} if standard pre–drag physics is desired, assert $|\alpha R/M^{2}|\ll 1$ for $N\le N_{\rm drag}$ (Sec.~\ref{subsec:K-curv}).
\end{itemize}
\textbf{Adaptive recovery}
\begin{itemize}[leftmargin=1.2em,itemsep=2pt,topsep=2pt]
\item If diagnostics saturate near machine precision while the stepsize underflows, set $(\mathrm{rtol},\mathrm{atol})\rightarrow(\mathrm{rtol}/10,\mathrm{atol}/10)$ and restart from the last checkpoint (equations unchanged), as in Table~\ref{tab:numerics_tol_diag}.
\end{itemize}
\end{mdframed}
\clearpage
\subsection{Growth and sensitivities (co–integration)}
\label{app:numerics:sens}
\begin{mdframed}[linewidth=0.6pt]
\small
\textbf{Growth} \; Integrate
\[
D''+(2+E)D'-\tfrac{3}{2}\Omega_{m}(N)D=0,\quad
D(N_{\rm ini})=e^{N_{\rm ini}},\ D'(N_{\rm ini})=1,
\]
then renormalize $D\!\leftarrow\!D/D(0)$ to enforce $D(0)=1$; compute $f=D'/D$ and $f\sigma_{8}(z)=f(z)\,\sigma_{8,0}\,D(z)$.

\textbf{Sensitivities} \; For each parameter $\theta_{i}$, integrate the variational system
\[
\mathbf{s}_{i}'=\mathbf{J}_{y}\,\mathbf{s}_{i}+\mathbf{J}_{\theta_{i}},
\]
with $\mathbf{J}_{y}$ from \eqref{eq:app7-Jy-row1}–\eqref{eq:app7-Jy-row3} and model–specific $\mathbf{J}_{\theta_{i}}$ (App.~\ref{app:variational-derivation}); initialize $\mathbf{s}_{i}$ by differentiating the matter–era rules and the closure (Eq.~\eqref{eq:app7-phiini-sens}). Co–integrate the distance sensitivity $\chi_{i}'=e^{-N}H_{i}/H^{2}$ (Eq.~\eqref{eq:chi-sens}). After $D$–renormalization, apply $D_{i}\!\leftarrow\!D_{i}-D\,[D_{i}(0)/D(0)]$ so that $D_{i}(0)=0$.
\end{mdframed}

\subsection{Postprocessing, diagnostics, and reproducibility capsule}
\label{app:numerics:post}
\begin{mdframed}[linewidth=0.6pt]
\small
\textbf{Geometric outputs} \; From $\chi(N)$ obtain $D_{A}$ and $D_{L}$ using \eqref{eq:DA-DL}; compute $D_{V}$ and $F_{\rm AP}$ from \eqref{eq:DV-FAP}; report SNe modulus $\mu(z)=5\log_{10}(D_{L}/{\rm Mpc})+25$.

\textbf{Identity checks} \; Report $\max_{N}|\mathcal{C}_{F}|$, $|\mathcal{C}_{R}|$, $|\mathcal{C}_{\phi}|$, $|\mathcal{C}_{R/H^{2}}|$, $|\mathcal{C}_{\nabla\!\cdot T_{\phi}}|$ (Sec.~\ref{subsec:diagnostics}); target tolerances are summarized in Table~\ref{tab:numerics_tol_diag}.

\textbf{Reproducibility metadata} \; Save: input parameter file; $(\mathrm{rtol},\mathrm{atol},h_{\max},h_{\min})$; guard thresholds $(\epsilon_{\rm den},\varepsilon_{\rm shoot},\varepsilon_{\phi})$; integrator scheme (DOPRI5); checkpoint cadence; and the full diagnostics log. No pseudo–random sampling is used, so runs are deterministic for identical inputs.
\end{mdframed}

\subsection{Recommended tolerances}
\label{app:numerics:tols}
Embedded Runge–Kutta with adaptive steps (DOPRI5 \cite{DormandPrince1980,HairerODE1993}) and
\[
\mathrm{rtol}\in[10^{-10},10^{-8}],\qquad
\mathrm{atol}\in[10^{-12},10^{-10}]
\]
is sufficient for all tables and figures, consistent with the targets in Table~\ref{tab:numerics_tol_diag}. Tighter tolerances may be advisable if $|V_{,\phi\phi}|/H^{2}$ is large (incipient stiffness); switching to a stiff BDF method is permissible \emph{without} modifying any equation.

\medskip
\noindent\textit{References for this appendix:}
embedded Runge–Kutta (DOPRI5; \citep{DormandPrince1980}),
adaptive nonstiff ODE solvers and error control (\citep{HairerODE1993}),
and Gauss–Kronrod/QUADPACK quadrature (\citep{Piessens1983}).
\clearpage

%% file: sections/appendices/appendix9.tex
\section{Analytic limits: radiation, matter, and de Sitter}
\label{app:limits-analytic}

This appendix records explicit background and linear–response limits of the construction, using the identities
$E\equiv H'/H$ and $R/H^{2}=6(2+E)$ [Eq.~\eqref{eq:R-over-H2}], together with the two $K(N)$ specifications of Sec.~\ref{sec:K}. The effective equation of state and deceleration parameter follow from $w_{\rm eff}=-1-\tfrac{2}{3}E$ and $q=-1-E$ [Eq.~\eqref{eq:weff-q}]. All statements are exact for the corresponding perfect–fluid limit and receive only subleading corrections from any subdominant component.

\subsection*{Radiation era ($w=1/3$)}
For a radiation–dominated background,
\begin{equation}
H(N)=H_{0}\sqrt{\Omega_{r0}}\,e^{-2N},
\qquad
E\equiv\frac{H'}{H}=-2,
\qquad
\frac{R}{H^{2}}=6(2+E)=0,
\qquad
w_{\rm eff}=\tfrac{1}{3},\ \ q=1.
\end{equation}
\emph{Curvature–motivated $K$:} with $K=1+\alpha R/M^{2}$ one has
\begin{equation}
K \equiv 1,\qquad \frac{K'}{K}\equiv 0,
\end{equation}
exactly in the radiation limit (since $R=0$). Thus the BAO/CMB anchor $r_{d}$ is unmodified by construction when the early–time prior $|\alpha R/M^{2}|\ll 1$ is enforced (Sec.~\ref{subsec:K-curv}). \emph{Phenomenological $K$:} $K(N)=1+K_{0}e^{-pN}$ with $K'/K=-p(K-1)/K$ [Eq.~\eqref{eq:Kprime-phenom}]. Imposing $K\!\to\!1$ at high $z$ (e.g., $p>0$ with $K_{0}$ small) preserves the standard ruler.

\emph{Linear response:} $c_{s}^{2}=1$ and $\Phi=\Psi$ (Secs.~\ref{sec:linear}, \ref{sec:eft}) imply the GR Poisson relation on subhorizon scales; matter clustering is suppressed during radiation domination as in GR (subhorizon growth $\propto\ln a$ for CDM once modes enter the horizon), since $\mu=\Sigma=1$ [Eq.~\eqref{eq:mu-sigma-eta}].

\subsection*{Matter era ($w=0$)}
For a matter–dominated background,
\begin{equation}
H(N)=H_{0}\sqrt{\Omega_{m0}}\,e^{-3N/2},
\qquad
E=-\tfrac{3}{2},
\qquad
\frac{R}{H^{2}}=6(2+E)=3,
\qquad
w_{\rm eff}=0,\ \ q=\tfrac{1}{2}.
\end{equation}
\emph{Curvature–motivated $K$:} with $R=3H^{2}$ and $R'=-9H^{2}$,
\begin{equation}
K=1+\frac{3\alpha H^{2}}{M^{2}},
\qquad
K'=\frac{\alpha R'}{M^{2}}=-\frac{9\alpha H^{2}}{M^{2}},
\qquad
\boxed{\ \frac{K'}{K}=\frac{-9(\alpha H^{2}/M^{2})}{1+3(\alpha H^{2}/M^{2})}\ }\!,
\end{equation}
which is slowly varying whenever $|\alpha H^{2}/M^{2}|\ll 1$ (the same condition that ensures early–time safety). \emph{Phenomenological $K$:} $K'/K=-p(K-1)/K$ as above.

\emph{Linear response:} the growing mode satisfies $D''+(2+E)D'-\tfrac{3}{2}\Omega_{m}D=0$ [Eq.~\eqref{eq:growth-eq-recap}] with solution $D\propto a$ in the pure matter limit; $\Phi=\Psi=\mathrm{const.}$ on subhorizon scales, exactly as in GR.
\clearpage
\subsection*{de Sitter (late–time $\Lambda$ limit)}
For a de Sitter background with constant $H$,
\begin{equation}
E=0,
\qquad
\frac{R}{H^{2}}=12,
\qquad
w_{\rm eff}=-1,\ \ q=-1.
\end{equation}
\emph{Curvature–motivated $K$:} $K=1+12\alpha H^{2}/M^{2}$ is constant and $\tfrac{K'}{K}=0$ (consistent with Eq.~\eqref{eq:KprimeK-closed}). The Raychaudhuri relation \eqref{eq:Hdoteq} enforces $K\dot\phi^{2}=0$, so $\dot\phi=0$ for $K>0$ and the field is at rest. The MS pump field $z^{2}=a^{2}K\dot\phi^{2}/H^{2}$ vanishes, yielding $v''+(k^{2}-a''/a)v=0$ with luminal propagation (App.~\ref{app:MS-derivation}). \emph{Phenomenological $K$:} any finite $K$ with $K' \to 0$ as $E\to 0$ is admissible; the background dynamics fix $\dot\phi\to 0$ as above.

\subsection*{At–a–glance summary}
\begin{center}
\renewcommand{\arraystretch}{1.18}
\setlength{\tabcolsep}{10pt}
\begin{tabular}{lcccc}
\toprule
\textbf{Regime} & $E\equiv H'/H$ & $R/H^{2}$ & $K$ (curvature case) & $K'/K$ (curvature case) \\
\midrule
Radiation & $-2$ & $0$  & $1$ & $0$ \\
Matter    & $-3/2$ & $3$ & $1+3\,\alpha H^{2}/M^{2}$ & $-9(\alpha H^{2}/M^{2})/[1+3(\alpha H^{2}/M^{2})]$ \\
de Sitter & $0$ & $12$ & $1+12\,\alpha H^{2}/M^{2}$ & $0$ \\
\bottomrule
\end{tabular}
\end{center}

\noindent In all three limits $c_{s}^{2}=1$, $\Phi=\Psi$, and the linear phenomenology remains $\mu=\Sigma=1$ [Secs.~\ref{sec:linear}–\ref{sec:eft}]. The curvature–motivated specification reproduces $K\to 1$ in the radiation era (hence preserves $r_{d}$), drifts slowly during matter domination, and freezes in de Sitter; the phenomenological running remains governed by the exact identity $K'/K=-p(K-1)/K$ [Eq.~\eqref{eq:Kprime-phenom}] and can be pivoted to suppress any early–time deviation if desired (Sec.~\ref{subsec:K-phenom}).
\clearpage

%% file: sections/appendices/appendix10.tex
\clearpage

\section{Symbol index}
\label{app:symbols}

\begin{center}
\scriptsize
\setlength{\tabcolsep}{3pt}
\renewcommand{\arraystretch}{1.52}
\captionsetup{type=table}
\captionof{table}{Key symbols and definitions used throughout the text. Units are in natural conventions $c=\hbar=1$. Mass dimensions are indicated by $\,\mathrm{M}^{n}$.}
\label{tab:symbol-index}
\vspace{13pt}
\begin{minipage}[t]{0.485\textwidth}
\centering
\begin{tabularx}{\linewidth}{@{}ll>{\raggedright\arraybackslash}X@{}}
\toprule
\textbf{Symbol} & \textbf{Units} & \textbf{Definition / Role} \\
\midrule
$a(t)$ & 1 & Scale factor; $N\equiv\ln a$ is the e--fold time. \\
$z$ & 1 & Redshift; $z(N)=e^{-N}-1$. \\
$H,\ H_{0}$ & M,\ M & Hubble rate and its value today. \\
$E$ & 1 & Hubble e--fold derivative: $E\equiv H'/H$. \\
$R$ & M$^{2}$ & FRW Ricci scalar: $R=6(2H^{2}+\dot H)=6H^{2}(2+E)$. \\
$M_{\rm pl}$ & M & Reduced Planck mass, $(8\pi G)^{-1/2}$. \\
$\rho_{i},\,p_{i}$ & M$^{4}$ & Energy density and pressure for $i\in\{m,r,\phi\}$. \\
$\Omega_{i}$ & 1 & $\Omega_{i}\equiv \rho_{i}/(3M_{\rm pl}^{2}H^{2})$. \\
$w_{\rm eff},\,q$ & 1 & $w_{\rm eff}=-1-\tfrac{2}{3}E$, $q=-1-E$. \\
$\chi$ & M$^{-1}$ & Comoving distance, $\chi(z)=\int_{0}^{z}dz'/H(z')$. \\
$D_{A},\,D_{L}$ & M$^{-1}$ & $D_{A}=\chi/(1+z)$, $D_{L}=(1+z)\chi$. \\
$D_{V}$ & M$^{-1}$ & $D_{V}=[(1+z)^{2}D_{A}^{2}\,z/H]^{1/3}$. \\
$F_{\rm AP}$ & 1 & $(1+z)D_{A}H^{-1}$. \\
$r_{d}$ & M$^{-1}$ & $r_{d}=\int_{z_{d}}^{\infty}c_{s}/H\,dz$. \\
$\eta$ (conf.\ time) & M$^{-1}$ & Conformal time, $dt=a\,d\eta$. \\
\addlinespace[2pt]
$\phi$ & M & Scalar field. \\
$X$ & M$^{4}$ & $X\equiv -\tfrac12 g^{\mu\nu}\partial_{\mu}\phi\,\partial_{\nu}\phi=\tfrac12\dot\phi^{2}$ (FRW). \\
$V(\phi)$ & M$^{4}$ & Scalar potential. \\
$K(\chi_{\rm c})$ & 1 & Covariantly completed kinetic normalization depending on the clock/St\"uckelberg field $\chi_{\rm c}$; evaluated in unitary gauge $\chi_{\rm c}=t$ for background evolution; $K>0$ for ghost freedom. \\
$\alpha,\,M$ & 1,\,M & Curvature $K=1+\alpha R/M^{2}$. \\
$K_{0},\,p,\,N_{p}$ & 1,\,1,\,1 & $K(N)=1+K_{0}e^{-pN}=1+K_{p}e^{-p(N-N_{p})}$. \\
$K'/K$ & 1 & E--fold derivative; closed form in Eq.~\eqref{eq:KprimeK-closed}. \\
$\rho_{\phi},\,p_{\phi}$ & M$^{4}$ & $\rho_{\phi}=KX+V$, $p_{\phi}=KX-V$. \\
$w_{\phi}$ & 1 & $(KX-V)/(KX+V)$. \\
\bottomrule
\end{tabularx}
\end{minipage}\hfill
\begin{minipage}[t]{0.485\textwidth}
\centering
\begin{tabularx}{\linewidth}{@{}ll>{\raggedright\arraybackslash}X@{}}
\toprule
\textbf{Symbol} & \textbf{Units} & \textbf{Definition / Role} \\
\midrule
$\Phi,\,\Psi$ & 1 & Bardeen potentials (Newtonian gauge); here $\Phi=\Psi$. \\
$v,\,z$ & M,\,M & $v=a(\delta\phi+\dot\phi\,\Phi/H)$; $z^{2}=a^{2}K\dot\phi^{2}/H^{2}$. \\
$\mathcal{R}$ & 1 & Comoving curvature: $\mathcal{R}=v/z$. \\
$c_{s}^{2}$ & 1 & Scalar sound speed; here $c_{s}^{2}=1$. \\
$c_{T}^{2}$ & 1 & Tensor speed; here $c_{T}^{2}=1$. \\
$\alpha_{K,B,M,T},\,\alpha_{H}$ & 1 & QKDE: $\alpha_{K}=K\dot\phi^{2}/(H^{2}M_{\rm pl}^{2})\ge0$, others $0$. \\
$\mu(a,k),\,\Sigma(a,k)$ & 1 & Here $\mu=\Sigma=1$ (linear, subhorizon). \\
$\eta(a,k)$ (slip) & 1 & $\eta\equiv \Phi/\Psi -1$; here $\eta=0$. \\
$k,\,\ell$ & M,\,1 & Comoving wavenumber and multipole. \\
$P(k,z)$ & M$^{-3}$ & Linear scaling $P(k,z)=D^{2}(z)P(k,0)$. \\
\addlinespace[2pt]
$D(a)$ & 1 & $D''+(2+H'/H)D'-\tfrac{3}{2}\Omega_{m}D=0$. \\
$f$ & 1 & $f\equiv d\ln D/d\ln a$. \\
$\sigma_{8,0}$ & 1 & Present-day rms in $8\,h^{-1}\mathrm{Mpc}$. \\
$f\sigma_{8}(z)$ & 1 & $f(z)\,\sigma_{8,0}\,D(z)/D(0)$. \\
\addlinespace[2pt]
$s$ & M & Background velocity: $s\equiv \phi'$. \\
$y_{H}$ & 1 & $y_{H}\equiv \ln H$. \\
$\mathcal{C}_{F}$ & 1 & $1-\Omega_{m}-\Omega_{r}-\Omega_{\phi}$. \\
$\mathcal{C}_{R}$ & 1 & Raychaudhuri residual (e--fold form). \\
$\mathcal{C}_{\phi}$ & 1 & Background Klein--Gordon residual. \\
$\mathcal{C}_{R/H^{2}}$ & 1 & $R/H^{2}-6(2+H'/H)$. \\
$\mathcal{C}_{\nabla\!\cdot T_{\phi}}$ & 1 & Exchange residual due to $K'(N)$; total energy--momentum is conserved. \\
\addlinespace[2pt]
$\mu_{\rm SN}(z)$ & 1 & $5\log_{10}(D_{L}/\mathrm{Mpc})+25$. \\
$Q^{\mu}$ & M$^{3}$ & Energy--momentum exchange current arising from the covariant completion; satisfies $\nabla_{\mu}T^{\mu\nu}_{(\phi)}=Q^{\nu}$ with $Q^{\nu}=-\partial^{\nu}K(\chi_{\rm c})\,X$ and vanishes only when $K=\mathrm{const}$. \\
\bottomrule
\end{tabularx}
\end{minipage}
\end{center}

\clearpage

%% file: sections/appendices/appendix11.tex
\section{Reproducibility checklist (at a glance)}
\label{app:repro-checklist}

\begin{mdframed}[linewidth=0.6pt,backgroundcolor=gray!3]
\textbf{Inputs \& priors}
\begin{enumerate}
\item Select a kinetic specification and parameters: curvature--motivated $\big(\alpha,M\big)$ or phenomenological $\big(K_{0},p\big)$. Enforce $K>0$ (Sec.~\ref{subsec:K-admissibility}); for the curvature case apply the early--time prior in Eq.~\eqref{eq:alpha-prior}.
\item Fix the baseline cosmology $\{H_{0},\Omega_{m0},\Omega_{r0}\}$ (and $\sigma_{8,0}$ if reporting $f\sigma_{8}$). Assume spatial flatness unless stated.
\item Early--time choice: adopt $K\!\to\!1$ for $z\ge z_{\rm drag}$ (baseline). If exploring early running, recompute $r_{d}$ from Eq.~\eqref{eq:rd}.
\end{enumerate}

\textbf{Initialization}
\begin{enumerate}\setcounter{enumi}{3}
\item Choose $N_{\rm ini}$ deep in matter domination; set $\phi'(N_{\rm ini})=0$.
      Initialize $H(N_{\rm ini})$ using the fluid scalings in Eq.~\eqref{eq:z-and-mr-of-N}.
      Determine $\phi(N_{\rm ini})$ by a one--dimensional shoot so that $\Omega_{\phi}(0)=1-\Omega_{m0}-\Omega_{r0}$ via Eq.~\eqref{eq:Omega-wphi}.
\end{enumerate}

\textbf{Integration}
\begin{enumerate}\setcounter{enumi}{4}
\item Evolve the state $\mathbf{y}=(\phi,\,s,\,y_H)=(\phi,\,\phi',\,\ln H)$ using the autonomous system \eqref{eq:background-system}
      with adaptive RK(4,5). Use tolerances in Table~\ref{tab:numerics_tol_diag}.
      Evaluate $K'/K$ exactly: Eq.~\eqref{eq:KprimeK-closed} (curvature) or Eq.~\eqref{eq:Kprime-phenom} (phenomenological).
      Enforce at each step: $H>0$, $K>0$, and a non-vanishing denominator in Eq.~\eqref{eq:KprimeK-closed}.
\end{enumerate}

\textbf{Diagnostics (record maxima over the run)}
\begin{enumerate}\setcounter{enumi}{5}
\item Log $\max|\mathcal{C}_{F}|$ (Friedmann closure),
      $\max|\mathcal{C}_{R}|$ (Raychaudhuri),
      $\max|\mathcal{C}_{\phi}|$ (Klein--Gordon),
      $\max|\mathcal{C}_{R/H^{2}}|$ (Ricci identity),
      and $\max|\mathcal{C}_{\nabla\!\cdot T_\phi}|$ (source identity).
      Target levels are given in Table~\ref{tab:numerics_tol_diag} (Sec.~\ref{sec:numerics}).
\end{enumerate}

\textbf{Derived observables}
\begin{enumerate}\setcounter{enumi}{6}
\item Distances: compute $\chi$ (Eq.~\eqref{eq:chi}), then $D_{A},D_{L}$ (Eq.~\eqref{eq:DA-DL});
      BAO summaries: $D_{V},F_{\rm AP}$ (Eq.~\eqref{eq:DV-FAP}).
      Growth: solve Eq.~\eqref{eq:growth-eq-recap} for $D(a)$, then obtain $f$ and $f\sigma_{8}$ (Eq.~\eqref{eq:fs8}).
      If BAO set the absolute scale, include $r_{d}$ from Eq.~\eqref{eq:rd}.
\end{enumerate}

\textbf{Artifacts to report}
\begin{enumerate}\setcounter{enumi}{7}
\item Parameter values and priors; integrator and tolerances; maximum diagnostics; redshift grid.
      For bitwise reproducibility, include platform/precision, compiler/interpreter version, and any quasirandom settings (if used).
\end{enumerate}
\end{mdframed}
\clearpage